%\input sacmace 
%**************************** SACMACE.TEX ***********************************
%***************************************************************************
% Macros pour articles utilisant une version modifie des macros de
%``Quantum field theory  and critical phenomena" and de harvmac
% Version with 2 type of fonts: 10 pts, 10 pts magnified
% 10 pts sort en format double colonne type "lanscape" 
% J. Zinn-Justin 9/02//1996
% *************************** output macros ********************************
\input hyperbasics
\catcode`\@=11
%%% saclay A4 paper:
%\def\unredoffs{\voffset=33mm \hoffset=19truemm}%for proceedings
\def\unredoffs{\voffset=5truemm \hoffset=8truemm} %for LA database 
\def\redoffs{\voffset=-12.truemm\hoffset=-9truemm} 
\def\speclscape{}
%%% MIT paper
\def\redoffs{\voffset=-12.truemm\hoffset=-16truemm} 
%---------------------------------------------------------------------%
\newbox\leftpage \newdimen\fullhsize \newdimen\hstitle \newdimen\hsbody
\newdimen\hdim
%\tolerance=1000
\hfuzz=1pt
\ifx\hyperdef\UNd@FiNeD\def\hyperdef#1#2#3#4{#4}\def\hyperref#1#2#3#4{#4}\fi
\def\newans{y }
%\message{ new or old (y/n)? }\read-1 to\answb
\def\answb{y }
\ifx\answb\newans\message{(This uses normal fonts.)}%
\magnification=1200% suppress for proceedings
%\hoffset=6.7mm
%\voffset=13mm
% ***************************************************************************
%
\def\bigans{b }
%\message{ big or little (b/l)? }\read-1 to\answ
\def\answ{b }% for submission
\ifx\answ\bigans\message{(Format simple colonne 12pts.}
\magnification=1200 \unredoffs\hsize=147truemm\vsize=219truemm
\hsbody=\hsize \hstitle=\hsize %take default values for unreduced format
\else\message{(Format double colonne, 10pts.} \let\l@r=L
\magnification=1000 \vsize=6.68in%\vsize=177.5truemm
\redoffs%
\hstitle=120truemm\hsbody=120truemm\fullhsize=261.5truemm\hsize=\hsbody 
\output={
  \almostshipout{\leftline{\vbox{\makeheadline\pagebody\makefootline}}}
\advancepageno%
%  \fi
}
\def\almostshipout#1{\if L\l@r \count1=1 \message{[\the\count0.\the\count1]}
      \global\setbox\leftpage=#1 \global\let\l@r=R
 \else \count1=2
  \shipout\vbox{\speclscape{\hsize\fullhsize}%\makeheadline}
      \hbox to\fullhsize{\box\leftpage\hfil#1}}  \global\let\l@r=L\fi}
\fi
% ****************************************************************************
% fonts, Dirac slash

\def\sla#1{\mkern-1.5mu\raise0.4pt\hbox{$\not$}\mkern1.2mu #1\mkern 0.7mu}
\def\Dbar{\mkern-1.5mu\raise0.4pt\hbox{$\not$}\mkern-.1mu {\rm D}\mkern.1mu}
\def\Abar{\mkern1.mu\raise0.4pt\hbox{$\not$}\mkern-1.3mu A\mkern.1mu}
\def\Bbar{\mkern-0.mu\raise0.4pt\hbox{$\not$}\mkern-.3mu B\mkern 0.6mu}
\newskip\tableskipamount \tableskipamount=8pt plus 3pt minus 3pt

%****************************

\newdimen\chapskip

\font\caprm=cmr9
\font\capit=cmti9
\font\capbf=cmbx9
\font\capsl=cmsl9
\font\capmi=cmmi9
\font\capex=cmex9
\font\capsy=cmsy9
\chapskip=17.5mm
\def\makeheadline{\vbox to 0pt{\vskip-22.5pt
\line{\vbox to8.5pt{}\the\headline}\vss}\nointerlineskip}
%***************************************************
  %obsolete??
\font\tenbi=cmmib10 
\font\ninebi=cmmib9
\font\sevenbi=cmmib7 
\font\fivebi=cmmib5
\textfont4=\tenbi
\scriptfont4=\sevenbi
\scriptscriptfont4=\fivebi

%****************************
\font\sixrm=cmr6
\font\fiverm=cmr5
\font\sixmi=cmmi6
\font\fivemi=cmmi5
\font\sixsy=cmsy6
\font\fivesy=cmsy5
\font\sixbf=cmbx6
\font\fivebf=cmbx5
\skewchar\capmi='177 \skewchar\sixmi='177 \skewchar\fivemi='177
\skewchar\capsy='60 \skewchar\sixsy='60 \skewchar\fivesy='60

\def\elevenpoint{
\textfont0=\caprm \scriptfont0=\sixrm \scriptscriptfont0=\fiverm
\def\rm{\fam0\caprm}
\textfont1=\capmi \scriptfont1=\sixmi \scriptscriptfont1=\fivemi
\textfont2=\capsy \scriptfont2=\sixsy \scriptscriptfont2=\fivesy
\textfont3=\capex \scriptfont3=\capex \scriptscriptfont3=\capex
\textfont\itfam=\capit \def\it{\fam\itfam\capit} % \it is family 4
\textfont\slfam=\capsl  \def\sl{\fam\slfam\capsl} % \sl is family 5
\textfont\bffam=\capbf \scriptfont\bffam=\sixbf
\scriptscriptfont\bffam=\fivebf
\def\bf{\fam\bffam\capbf} % \bf is family 6
\textfont4=\ninebi \scriptfont4=\sevenbi \scriptscriptfont4=\fivebi
\abovedisplayskip=11pt plus 3pt minus 8pt
\belowdisplayskip=\abovedisplayskip
\smallskipamount=2.7pt plus 1pt minus 1pt
\medskipamount=5.4pt plus 2pt minus 2pt
\bigskipamount=10.8pt plus 3.6pt minus 3.6pt
\normalbaselineskip=11pt
\setbox\strutbox=\hbox{\vrule height7.8pt depth3.2pt width0pt}
\normalbaselines \rm}
% *************************************************************************

% ****************************************************************************
%		*****	  MSSYMB.TeX	*****		       20 Sept 91
%
%	This file contains the definitions for the symbols in the two
%	"extra symbols" fonts created at the American Math. Society.
%
%       The old fonts msxm et msym have been replaced by msam et msbm. 

\catcode`\@=11

\font\tenmsa=msam10
\font\sevenmsa=msam7
\font\fivemsa=msam5
\font\tenmsb=msbm10
\font\sevenmsb=msbm7
\font\fivemsb=msbm5
\newfam\msafam
\newfam\msbfam
\textfont\msafam=\tenmsa  \scriptfont\msafam=\sevenmsa
  \scriptscriptfont\msafam=\fivemsa
\textfont\msbfam=\tenmsb  \scriptfont\msbfam=\sevenmsb
  \scriptscriptfont\msbfam=\fivemsb

\def\hexnumber@#1{\ifcase#1 0\or1\or2\or3\or4\or5\or6\or7\or8\or9\or
	A\or B\or C\or D\or E\or F\fi }

%  The following 13 lines establish the use of the Euler Fraktur font.
%  To use this font, remove % from beginning of these lines.
\font\teneuf=eufm10
\font\seveneuf=eufm7
\font\fiveeuf=eufm5
\newfam\euffam
\textfont\euffam=\teneuf
\scriptfont\euffam=\seveneuf
\scriptscriptfont\euffam=\fiveeuf
\def\frak{\ifmmode\let\next\frak@\else
 \def\next{\Err@{Use \string\frak\space only in math mode}}\fi\next}
\def\goth{\ifmmode\let\next\frak@\else
 \def\next{\Err@{Use \string\goth\space only in math mode}}\fi\next}
\def\frak@#1{{\frak@@{#1}}}
\def\frak@@#1{\fam\euffam#1}
%  End definition of Euler Fraktur font.

\edef\msa@{\hexnumber@\msafam}
\edef\msb@{\hexnumber@\msbfam}

\mathchardef\boxdot="2\msa@00
\mathchardef\boxplus="2\msa@01
\mathchardef\boxtimes="2\msa@02
\mathchardef\square="0\msa@03
\mathchardef\blacksquare="0\msa@04
\mathchardef\centerdot="2\msa@05
\mathchardef\lozenge="0\msa@06
\mathchardef\blacklozenge="0\msa@07
\mathchardef\circlearrowright="3\msa@08
\mathchardef\circlearrowleft="3\msa@09
\mathchardef\rightleftharpoons="3\msa@0A
\mathchardef\leftrightharpoons="3\msa@0B
\mathchardef\boxminus="2\msa@0C
\mathchardef\Vdash="3\msa@0D
\mathchardef\Vvdash="3\msa@0E
\mathchardef\vDash="3\msa@0F
\mathchardef\twoheadrightarrow="3\msa@10
\mathchardef\twoheadleftarrow="3\msa@11
\mathchardef\leftleftarrows="3\msa@12
\mathchardef\rightrightarrows="3\msa@13
\mathchardef\upuparrows="3\msa@14
\mathchardef\downdownarrows="3\msa@15
\mathchardef\upharpoonright="3\msa@16

\mathchardef\downharpoonright="3\msa@17
\mathchardef\upharpoonleft="3\msa@18
\mathchardef\downharpoonleft="3\msa@19
\mathchardef\rightarrowtail="3\msa@1A
\mathchardef\leftarrowtail="3\msa@1B
\mathchardef\leftrightarrows="3\msa@1C
\mathchardef\rightleftarrows="3\msa@1D
\mathchardef\Lsh="3\msa@1E
\mathchardef\Rsh="3\msa@1F
\mathchardef\rightsquigarrow="3\msa@20
\mathchardef\leftrightsquigarrow="3\msa@21
\mathchardef\looparrowleft="3\msa@22
\mathchardef\looparrowright="3\msa@23
\mathchardef\circeq="3\msa@24
\mathchardef\succsim="3\msa@25
\mathchardef\gtrsim="3\msa@26
\mathchardef\gtrapprox="3\msa@27
\mathchardef\multimap="3\msa@28
\mathchardef\therefore="3\msa@29
\mathchardef\because="3\msa@2A
\mathchardef\doteqdot="3\msa@2B

\mathchardef\triangleq="3\msa@2C
\mathchardef\precsim="3\msa@2D
\mathchardef\lesssim="3\msa@2E
\mathchardef\lessapprox="3\msa@2F
\mathchardef\eqslantless="3\msa@30
\mathchardef\eqslantgtr="3\msa@31
\mathchardef\curlyeqprec="3\msa@32
\mathchardef\curlyeqsucc="3\msa@33
\mathchardef\preccurlyeq="3\msa@34
\mathchardef\leqq="3\msa@35
\mathchardef\leqslant="3\msa@36
\mathchardef\lessgtr="3\msa@37
\mathchardef\backprime="0\msa@38
\mathchardef\risingdotseq="3\msa@3A
\mathchardef\fallingdotseq="3\msa@3B
\mathchardef\succcurlyeq="3\msa@3C
\mathchardef\geqq="3\msa@3D
\mathchardef\geqslant="3\msa@3E
\mathchardef\gtrless="3\msa@3F
\mathchardef\sqsubset="3\msa@40
\mathchardef\sqsupset="3\msa@41
\mathchardef\vartriangleright="3\msa@42
\mathchardef\vartriangleleft="3\msa@43
\mathchardef\trianglerighteq="3\msa@44
\mathchardef\trianglelefteq="3\msa@45
\mathchardef\bigstar="0\msa@46
\mathchardef\between="3\msa@47
\mathchardef\blacktriangledown="0\msa@48
\mathchardef\blacktriangleright="3\msa@49
\mathchardef\blacktriangleleft="3\msa@4A
\mathchardef\vartriangle="0\msa@4D
\mathchardef\blacktriangle="0\msa@4E
\mathchardef\triangledown="0\msa@4F
\mathchardef\eqcirc="3\msa@50
\mathchardef\lesseqgtr="3\msa@51
\mathchardef\gtreqless="3\msa@52
\mathchardef\lesseqqgtr="3\msa@53
\mathchardef\gtreqqless="3\msa@54
\mathchardef\Rrightarrow="3\msa@56
\mathchardef\Lleftarrow="3\msa@57
\mathchardef\veebar="2\msa@59
\mathchardef\barwedge="2\msa@5A
\mathchardef\doublebarwedge="2\msa@5B
\mathchardef\angle="0\msa@5C
\mathchardef\measuredangle="0\msa@5D
\mathchardef\sphericalangle="0\msa@5E
\mathchardef\varpropto="3\msa@5F
\mathchardef\smallsmile="3\msa@60
\mathchardef\smallfrown="3\msa@61
\mathchardef\Subset="3\msa@62
\mathchardef\Supset="3\msa@63
\mathchardef\Cup="2\msa@64

\mathchardef\Cap="2\msa@65

\mathchardef\curlywedge="2\msa@66
\mathchardef\curlyvee="2\msa@67
\mathchardef\leftthreetimes="2\msa@68
\mathchardef\rightthreetimes="2\msa@69
\mathchardef\subseteqq="3\msa@6A
\mathchardef\supseteqq="3\msa@6B
\mathchardef\bumpeq="3\msa@6C
\mathchardef\Bumpeq="3\msa@6D
\mathchardef\lll="3\msa@6E

\mathchardef\ggg="3\msa@6F

\mathchardef\circledS="0\msa@73
\mathchardef\pitchfork="3\msa@74
\mathchardef\dotplus="2\msa@75
\mathchardef\backsim="3\msa@76
\mathchardef\backsimeq="3\msa@77
\mathchardef\complement="0\msa@7B
\mathchardef\intercal="2\msa@7C
\mathchardef\circledcirc="2\msa@7D
\mathchardef\circledast="2\msa@7E
\mathchardef\circleddash="2\msa@7F
\def\ulcorner{\delimiter"4\msa@70\msa@70 }
\def\urcorner{\delimiter"5\msa@71\msa@71 }
\def\llcorner{\delimiter"4\msa@78\msa@78 }
\def\lrcorner{\delimiter"5\msa@79\msa@79 }
\def\yen{\mathhexbox\msa@55 }
\def\checkmark{\mathhexbox\msa@58 }
\def\circledR{\mathhexbox\msa@72 }
\def\maltese{\mathhexbox\msa@7A }
\mathchardef\lvertneqq="3\msb@00
\mathchardef\gvertneqq="3\msb@01
\mathchardef\nleq="3\msb@02
\mathchardef\ngeq="3\msb@03
\mathchardef\nless="3\msb@04
\mathchardef\ngtr="3\msb@05
\mathchardef\nprec="3\msb@06
\mathchardef\nsucc="3\msb@07
\mathchardef\lneqq="3\msb@08
\mathchardef\gneqq="3\msb@09
\mathchardef\nleqslant="3\msb@0A
\mathchardef\ngeqslant="3\msb@0B
\mathchardef\lneq="3\msb@0C
\mathchardef\gneq="3\msb@0D
\mathchardef\npreceq="3\msb@0E
\mathchardef\nsucceq="3\msb@0F
\mathchardef\precnsim="3\msb@10
\mathchardef\succnsim="3\msb@11
\mathchardef\lnsim="3\msb@12
\mathchardef\gnsim="3\msb@13
\mathchardef\nleqq="3\msb@14
\mathchardef\ngeqq="3\msb@15
\mathchardef\precneqq="3\msb@16
\mathchardef\succneqq="3\msb@17
\mathchardef\precnapprox="3\msb@18
\mathchardef\succnapprox="3\msb@19
\mathchardef\lnapprox="3\msb@1A
\mathchardef\gnapprox="3\msb@1B
\mathchardef\nsim="3\msb@1C
%\mathchardef\napprox="3\msb@1D
\mathchardef\ncong="3\msb@1D

\mathchardef\varsubsetneq="3\msb@20
\mathchardef\varsupsetneq="3\msb@21
\mathchardef\nsubseteqq="3\msb@22
\mathchardef\nsupseteqq="3\msb@23
\mathchardef\subsetneqq="3\msb@24
\mathchardef\supsetneqq="3\msb@25
\mathchardef\varsubsetneqq="3\msb@26
\mathchardef\varsupsetneqq="3\msb@27
\mathchardef\subsetneq="3\msb@28
\mathchardef\supsetneq="3\msb@29
\mathchardef\nsubseteq="3\msb@2A
\mathchardef\nsupseteq="3\msb@2B
\mathchardef\nparallel="3\msb@2C
\mathchardef\nmid="3\msb@2D
\mathchardef\nshortmid="3\msb@2E
\mathchardef\nshortparallel="3\msb@2F
\mathchardef\nvdash="3\msb@30
\mathchardef\nVdash="3\msb@31
\mathchardef\nvDash="3\msb@32
\mathchardef\nVDash="3\msb@33
\mathchardef\ntrianglerighteq="3\msb@34
\mathchardef\ntrianglelefteq="3\msb@35
\mathchardef\ntriangleleft="3\msb@36
\mathchardef\ntriangleright="3\msb@37
\mathchardef\nleftarrow="3\msb@38
\mathchardef\nrightarrow="3\msb@39
\mathchardef\nLeftarrow="3\msb@3A
\mathchardef\nRightarrow="3\msb@3B
\mathchardef\nLeftrightarrow="3\msb@3C
\mathchardef\nleftrightarrow="3\msb@3D
\mathchardef\divideontimes="2\msb@3E
\mathchardef\varnothing="0\msb@3F
\mathchardef\nexists="0\msb@40
\mathchardef\mho="0\msb@66
\mathchardef\eth="0\msb@67
\mathchardef\eqsim="3\msb@68
\mathchardef\beth="0\msb@69
\mathchardef\gimel="0\msb@6A
\mathchardef\daleth="0\msb@6B
\mathchardef\lessdot="3\msb@6C
\mathchardef\gtrdot="3\msb@6D
\mathchardef\ltimes="2\msb@6E
\mathchardef\rtimes="2\msb@6F
\mathchardef\shortmid="3\msb@70
\mathchardef\shortparallel="3\msb@71
\mathchardef\smallsetminus="2\msb@72
\mathchardef\thicksim="3\msb@73
\mathchardef\thickapprox="3\msb@74
\mathchardef\approxeq="3\msb@75
\mathchardef\succapprox="3\msb@76
\mathchardef\precapprox="3\msb@77
\mathchardef\curvearrowleft="3\msb@78
\mathchardef\curvearrowright="3\msb@79
\mathchardef\digamma="0\msb@7A
\mathchardef\varkappa="0\msb@7B
\mathchardef\hslash="0\msb@7D
\mathchardef\hbar="0\msb@7E
\mathchardef\backepsilon="3\msb@7F
\def\Bbb{\ifmmode\let\next\Bbb@\else
 \def\next{\errmessage{Use \string\Bbb\space only in math mode}}\fi\next}
\def\Bbb@#1{{\Bbb@@{#1}}}
\def\Bbb@@#1{\fam\msbfam#1}
\font\sacfont=eufm10 scaled 1440
\catcode`\@=12

\def\sla#1{\mkern-1.5mu\raise0.4pt\hbox{$\not$}\mkern1.2mu #1\mkern 0.7mu}
\def\Dbar{\mkern-1.5mu\raise0.4pt\hbox{$\not$}\mkern-.1mu {\rm D}\mkern.1mu}
\def\Abar{\mkern1.mu\raise0.4pt\hbox{$\not$}\mkern-1.3mu A\mkern.1mu}
%\nopagenumbers
%\headline={\ifnum\pageno=1\hfill\else\draftdate\hfil{\headrm\folio}%
%\hfil\fi}	 
\else\message{(This uses pseudo 12pts fonts.}
\hoffset=8mm
\voffset=16mm
\input lfont12 %pour sun

\def\sla#1{\mkern-1.5mu\raise0.5pt\hbox{$\not$}\mkern1.2mu #1\mkern 0.7mu}
\def\Dbar{\mkern-1.5mu\raise0.5pt\hbox{$\not$}\mkern-.1mu {\rm D}\mkern.1mu}
\def\Abar{\mkern1.mu\raise0.5pt\hbox{$\not$}\mkern-1.3mu A\mkern.1mu}
\fi
% ****************************************************************************
%********* end ouptut macros
% ****************************************************************************
% ****** extrait de definit.tex (obsolete ?)

% ****************************************************************************
\newcount\yearltd\yearltd=\year\advance\yearltd by -1900
\newif\ifdraftmode
\draftmodefalse
\def\draft{\draftmodetrue{\count255=\time\divide\count255 by 60
\xdef\hourmin{\number\count255} 
  \multiply\count255 by-60\advance\count255 by\time
  \xdef\hourmin{\hourmin:\ifnum\count255<10 0\fi\the\count255}}}
 
\newif\iffrancmode
\francmodefalse
% ********* A few math symbols
\def\e{\mathop{\rm e}\nolimits}

\def\d{{\rm d}}
\def\ud{{\textstyle{1\over 2}}}
\def\half{\ud}
\def\tr{\mathop{\rm tr}\nolimits}
\def\det{\mathop{\rm det}\nolimits}
\def\del{\partial}

\chardef\sigmat=27
\def\n{\noindent}  
\def\rf{\par\item{}}
\def\nrf{\par\n}
\def\frac#1#2{{\textstyle{#1\over#2}}}

\def\leaderfill{\leaders\hbox to 1em{\hss.\hss}\hfill}
% *************************************************************************
\catcode`\@=11
% ************** double alignment in eqalignno style **********************
\def\deqalignno#1{\displ@y\tabskip\centering \halign to
\displaywidth{\hfil$\displaystyle{##}$\tabskip0pt&$\displaystyle{{}##}$
\hfil\tabskip0pt &\quad
\hfil$\displaystyle{##}$\tabskip0pt&$\displaystyle{{}##}$ 
\hfil\tabskip\centering& \llap{$##$}\tabskip0pt \crcr #1 \crcr}}
% ************** double eqalign ******************************************
\def\deqalign#1{\null\,\vcenter{\openup\jot\m@th\ialign{
\strut\hfil$\displaystyle{##}$&$\displaystyle{{}##}$\hfil
&&\quad\strut\hfil$\displaystyle{##}$&$\displaystyle{{}##}$
\hfil\crcr#1\crcr}}\,}
%***************************************************************************
%********* titlepage, headline, section, subsection, sub, appendix *********
%***************************************************************************
%**************** input with check of file existence ***********************
\newread\ch@ckfile
\def\cinput#1{\def\filen@me{#1}%                              DOES NOT WORK???
\immediate\openin\ch@ckfile=\filen@me
\ifeof\ch@ckfile\closein\ch@ckfile\message{<< (\filen@me\ N'EXISTE PAS)
>>}\else% 
\input\filen@me\closein \ch@ckfile\fi}
%********* introduce equation number file: for non-causal quotation
\immediate\openin\ch@ckfile=\jobname.def
\ifeof\ch@ckfile\closein\ch@ckfile\message{<< (\jobname.def N'EXISTE PAS)
>>}
% Warning macro
\def\DefWarn#1{\ifx\UNd@FiNeD#1\else
\immediate\write16{*** WARNING: the label \string#1 is already defined%
***}\fi}% 
\else% 
\def\DefWarn#1{}
\input\jobname.def\closein \ch@ckfile\fi
%**************************************************************************
\newcount\nosection
\newcount\nosubsection
\newcount\neqno
\newcount\notenumber
\newcount\nofigure
\newif\ifappmode
\def\table{\jobname.toc}
\def\equation{\jobname.equ}
\def\labeldefs{\jobname.eqd}
\newwrite\equa
\newwrite\tab 
\newwrite\eqdf
% ******************* titlepage **********************************

\newdimen\hulp
\def\maketitle#1{
\edef\oneliner##1{\centerline{##1}}
\edef\twoliner##1{\vbox{\parindent=0pt\leftskip=0pt plus 1fill\rightskip=0pt
plus 1fill 
                     \parfillskip=0pt\relax##1}} 
\setbox0=\vbox{#1}\hulp=0.5\hsize
                 \ifdim\wd0<\hulp\oneliner{#1}\else
                 \twoliner{#1}\fi}
\def\preprint#1{{\sacfont }\hfill{#1}\vskip 20mm}
% **************** beginning
\def\title#1\par{\gdef\titlename{#1}
\maketitle{\bf\uppercase\expandafter{\titlename}}
\vskip8truemm
\nosection=0
\neqno=0
\notenumber=0
\nofigure=0
\def\prefix{}
\appmodefalse
\mark{\the\nosection}
\message{#1}
%\immediate\openout\tab=\table
\immediate\openout\equa=\equation
\immediate\openout\eqdf=\labeldefs
}
\def\abstract{\bigskip%
%\iffrancmode\centerline{RESUME}\else%
%\centerline{ABSTRACT}%
%\fi \smallskip
 \begingroup\narrower
\elevenpoint\baselineskip10pt} 
\def\endabstract{\par\endgroup\bigskip
}
% ***************** input table of contents
%\def\content{\vfill\eject% A un bug dans le format double colonne
%\begingroup\centerline{\uppercase\expandafter{Table of contents}}% 
%\bigskip\elevenpoint\noindent% 
%\parindent=2em
%\openin 1=\jobname.tab
%\ifeof 1\closein1\message{<< (\jobname.tab DOES NOT EXIST. TeX again)>>}%
%\else\input\jobname.tab\closein 1\fi 
%\endgroup\vfill\eject}
%******************************* section ***********************************
\def\section#1\par{\vskip0pt plus.1\vsize\penalty-100\vskip0pt plus-.1
\vsize\bigskip\vskip\parskip
\ifnum\nosection=0\ifappmode\relax\else\writetoc
\fi\fi% ajout
\advance\nosection by 1\global\nosubsection=0\global\neqno=0
\vbox{\noindent\bf{\hyperdef\hypernoname{section}{\prefix\the\nosection}%
{\prefix\the\nosection}\ #1}}
\writetoca{{\string\hyperref{}{section}{\prefix\the\nosection}%
{\prefix\the\nosection}} {#1}}
\message{\the\nosection\ #1}
\mark{\the\nosection}\bigskip\noindent
%\xdef\ecrire{\write\tab{\string\par\string\item{\prefix\the\nosection}
%#1\string\leaderfill {\noexpand\number\pageno}}}\ecrire
}

% appendix
\def\appendix#1#2\par{\bigbreak\nosection=0
\appmodetrue
\notenumber=0
\neqno=0
\def\prefix{A}
\mark{\the\nosection}
\message{APPENDICES}
{\leftline{APPENDICES} \hyperdef\hypernoname{appendix}{app}{ 
\leftline{\uppercase\expandafter{#1}}
\leftline{\uppercase\expandafter{#2}}}}
\bigskip\noindent\nonfrenchspacing
\writetoca{\string\hyperref{}{appendix}{app}{Appendices}.\ #1.\ #2}%
}
% **************************** \subsection *************************
\def\subsection#1\par {\vskip0pt plus.05\vsize\penalty-100\vskip0pt
plus-.05\vsize\bigskip\vskip\parskip\advance\nosubsection by 1
\vbox{\noindent\it{\hyperdef\hypernoname{subsection}{\prefix\the\nosection.%
\the\nosubsection}{\prefix\the\nosection.\the\nosubsection\ #1}}}%
\smallskip\noindent 
\writetoca{{\string\hyperref{}{subsection}{\prefix\the\nosection.%
\the\nosubsection}{\prefix\the\nosection.\the\nosubsection}} {#1}}
\message{\the\nosection.\the\nosubsection\ #1}
} 
\def\note #1{\advance\notenumber by 1
\footnote{$^{\the\notenumber}$}{\sevenrm #1}} 
% ?????

\parindent=1em 
\newinsert\margin
\dimen\margin=\maxdimen
\count\margin=0 \skip\margin=0pt
%*****************************************************************
% IMPORTANT, new version demands chapter be defined before any section,
% section be defined before any subsection
\def\sslbl#1{\DefWarn#1%
\ifdraftmode{\hfill\escapechar-1{\rlap{\hskip-1mm%
\sevenrm\string#1}}}\fi%
\ifnum\nosection=0\xdef#1{}%
\edef\ewrite{\write\eqdf{\string\def\string#1{}}
\write\eqdf{}}\ewrite%
\edef\ewrite{\write\equa{{\string#1}}%
\write\equa{}}\ewrite%
\else%
\ifnum\nosubsection=0%
%\xdef#1{\prefix\the\nosection}%
\xdef#1{\noexpand\hyperref{}{section}{\prefix\the\nosection}{\prefix\the\nosection}}%
\edef\ewrite{\write\eqdf{\string\def\string#1{\prefix%
\the\nosection}}\write\eqdf{}}\ewrite%
\edef\ewrite{\write\equa{{\string#1},\prefix\the\nosection}%
\write\equa{}}\ewrite%
\writedef{#1\leftbracket#1}
\else%
%\xdef#1{\prefix\the\nosection.\the\nosubsection}%
\xdef#1{\noexpand\hyperref{}{subsection}{\prefix\the\nosection.%
\the\nosubsection}{\prefix\the\nosection.\the\nosubsection}}%
\writedef{#1\leftbracket#1}
\edef\ewrite{\write\eqdf{\string\def\string#1{\prefix%
\the\nosection.\the\nosubsection}}\write\eqdf{}}\ewrite%
\edef\ewrite{\write\equa{{\string#1},\prefix\the\nosection%
.\the\nosubsection}\write\equa{}}\ewrite\fi\fi}%
%**********************************************************************
% Autre utilitaire
\def\listcontent{\immediate\closeout\tfile\immediate\openin%
\ch@ckfile=\jobname.tab
\ifeof\ch@ckfile\message{no file \jobname.tab, no table of contents this
pass}%
\else\closein\ch@ckfile\centerline{\bf\iffrancmode Table des
mati\`eres \else Contents\fi}\nobreak\medskip% 
{\baselineskip=12pt\parskip=0pt\catcode`\@=11\input\jobname.tab
\catcode`\@=12\bigbreak\bigskip}\fi}
\newwrite\tfile \def\writetoca#1{}
%       use this to write file with table of contents
\def\writetoc{\immediate\openout\tfile=\jobname.tab
   \def\writetoca##1{{\edef\next{\write\tfile{\noindent ##1
   \string\leaderfill {\string\hyperref{}{page}{\noexpand\number\pageno}%
                       {\noexpand\number\pageno}} \par}}\next}}}

% ********************* references harvmac style ***********************
%     \ref\label{text}
% generates a number, assigns it to \label, generates an entry.
% To list the refs on a separate page,  \listrefs
%
\def\nolabels{\def\wrlabeL##1{}\def\eqlabeL##1{}\def\reflabeL##1{}}
\def\writelabels{\def\wrlabeL##1{\leavevmode\vadjust{\rlap{\smash%
{\line{{\escapechar=` \hfill\rlap{\sevenrm\hskip.03in\string##1}}}}}}}%
\def\eqlabeL##1{{\escapechar-1\rlap{\sevenrm\hskip.05in\string##1}}}%
\def\reflabeL##1{\noexpand\llap{\noexpand\sevenrm\string\string\string##1}}}
\nolabels

\global\newcount\refno \global\refno=1
\newwrite\rfile
\def\ref{[\hyperref{}{reference}{\the\refno}{\the\refno}]\nref}
\def\nref#1{\DefWarn#1%
\xdef#1{[\noexpand\hyperref{}{reference}{\the\refno}{\the\refno}]}%
\writedef{#1\leftbracket#1}%
\ifnum\refno=1\immediate\openout\rfile=\jobname.ref\fi
\chardef\wfile=\rfile\immediate\write\rfile{\noexpand\item{[\noexpand\hyperdef%
\noexpand\hypernoname{reference}{\the\refno}{\the\refno}]\ }%
\reflabeL{#1\hskip.31in}\pctsign}\global\advance\refno by1\findarg}
%       horrible hack to sidestep tex \write limitation
\def\findarg#1#{\begingroup\obeylines\newlinechar=`\^^M\pass@rg}
{\obeylines\gdef\pass@rg#1{\writ@line\relax #1^^M\hbox{}^^M}%
\gdef\writ@line#1^^M{\expandafter\toks0\expandafter{\striprel@x #1}%
\edef\next{\the\toks0}\ifx\next\em@rk\let\next=\endgroup\else\ifx\next\empty%
\else\immediate\write\wfile{\the\toks0}\fi\let\next=\writ@line\fi\next\relax}}
\def\striprel@x#1{} \def\em@rk{\hbox{}}
\def\lref{\begingroup\obeylines\lr@f}
\def\lr@f#1#2{\DefWarn#1\gdef#1{\let#1=\UNd@FiNeD\ref#1{#2}}\endgroup\unskip}

\def\addref#1{\immediate\write\rfile{\noexpand\item{}#1}} %now unnecessary
\def\listrefs{{}\vfill\supereject\immediate\closeout\rfile\writestoppt
\baselineskip=14pt\centerline{{\bf\iffrancmode R\'eferences\else References%
\fi}}
\bigskip{\parindent=20pt%
\frenchspacing\escapechar=` \input \jobname.ref\vfill\eject}\nonfrenchspacing}
\def\startrefs#1{\immediate\openout\rfile=\jobname.ref\refno=#1}
\def\xref{\expandafter\xr@f}\def\xr@f[#1]{#1}
\def\refs#1{\count255=1[\r@fs #1{\hbox{}}]}
\def\r@fs#1{\ifx\UNd@FiNeD#1\message{reflabel \string#1 is undefined.}%
\nref#1{need to supply reference \string#1.}\fi%
\vphantom{\hphantom{#1}}{\let\hyperref=\relax\xdef\next{#1}}%
\ifx\next\em@rk\def\next{}%
\else\ifx\next#1\ifodd\count255\relax\xref#1\count255=0\fi%
\else#1\count255=1\fi\let\next=\r@fs\fi\next}
%************************
%*******
%
\newwrite\lfile
{\escapechar-1\xdef\pctsign{\string\%}\xdef\leftbracket{\string\{}
\xdef\rightbracket{\string\}}\xdef\numbersign{\string\#}}
\def\writedefs{\immediate\openout\lfile=\jobname.def \def\writedef##1{%
{\let\hyperref=\relax\let\hyperdef=\relax\let\hypernoname=\relax
 \immediate\write\lfile{\string\def\string##1\rightbracket}}}}%
\def\writestop{\def\writestoppt{\immediate\write\lfile{\string\pageno%
\the\pageno\string\startrefs\leftbracket\the\refno\rightbracket%
\string\def\string\secsym\leftbracket\secsym\rightbracket%
\string\secno\the\secno\string\meqno\the\meqno}\immediate\closeout\lfile}}
\def\writestoppt{}\def\writedef#1{}
\writedefs
% ******
% bibliography: not very satisfactory
\def\biblio\par{\vskip0pt plus.1\vsize\penalty-100\vskip0pt plus-.1
\vsize\bigskip\vskip\parskip
\message{Bibliographie}
{\leftline{\bf \hyperdef\hypernoname{biblio}{bib}{Bibliographical Notes}}}
\nobreak\medskip\noindent\frenchspacing
\writetoca{\string\hyperref{}{biblio}{bib}{Bibliographical Notes}}}%

%**************** autre version si plusieurs biblio ************************
\def\biblionote{\iffrancmode Notes Bibliographiques\else Bibliographical Notes
\fi}
\def\beginbib\par{\vskip0pt plus.1\vsize\penalty-100\vskip0pt plus-.1
\vsize\bigskip\vskip\parskip
\message{Bibliographie}
{\leftline{\bf \hyperdef\hypernoname{biblio}{\the\nosection}%
{\biblionote}}}
\nobreak\medskip\noindent\frenchspacing
\writetoca{\string\hyperref{}{biblio}{\the\nosection}%
{\biblionote}}}%
\def\endbib{\nonfrenchspacing}

% *************** exercises: same comment
\def\Exercises{\iffrancmode Exercices\else Exercises
\fi}
\def\exerc\par{\vskip0pt plus.1\vsize\penalty-100\vskip0pt plus-.1
\vsize\bigskip\vskip\parskip
\message{Exercises}
{\leftline{\bf \hyperdef\hypernoname{exercise}{\the\nosection}{\Exercises}}}
\nobreak\medskip\noindent\frenchspacing
\writetoca{\string\hyperref{}{exercise}{\the\nosection}{\Exercises}}
}
%*************************************************************************
%Macro de numerotation automatique
%*************************************************************************
% numbering without naming
\def\eqnn{\global\advance\neqno by 1 \ifinner\relax\else%
\eqno\fi(\prefix\the\nosection.\the\neqno)}
%
% numbering and attaching a name: \eqnd{\ename}
\def\eqnd#1{\global\advance\neqno by 1 
{\xdef#1{($\noexpand\hyperref{}{equation}{\prefix\the\nosection.\the\neqno}%
{\prefix\the\nosection.\the\neqno}$)}}%???
\ifinner\relax\else\eqno\fi(\hyperdef\hypernoname{equation}{\prefix\the%
\nosection.\the\neqno}{\prefix\the\nosection.\the\neqno})
\writedef{#1\leftbracket#1}
\ifdraftmode{\escapechar-1{\rlap{\hskip.2mm\sevenrm\string#1}}}\fi
\edef\ewrite{\write\eqdf{\string\def\string#1{($\prefix\the\nosection.%
\the\neqno$)}}%
\write\eqdf{}}\ewrite%
\edef\ewrite{\write\equa{{\string#1},(\prefix\the\nosection.\the\neqno)
{\noexpand\number\pageno}}\write\equa{}}\ewrite}
%
% for eqalignno, allows (1a) (1b)...
\def\checkm@de#1#2{\ifmmode{\def\f@rst##1{##1}\hyperdef\hypernoname{equation}%
{#1}{#2}}\else\hyperref{}{equation}{#1}{#2}\fi}
\def\f@rst#1{\c@t#1a\em@ark}\def\c@t#1#2\em@ark{#1}
\def\eqna#1{\global\advance\neqno by1\ifdraftmode{\hfill%
\escapechar-1{\rlap{\sevenrm\string#1}}}\fi%
\xdef #1##1{(\noexpand\relax\noexpand%
\checkm@de{\prefix\the\nosection.\the\neqno\noexpand\f@rst{##1}1}%
{\hbox{$\prefix\the\nosection.\the\neqno##1$}})}
\writedef{#1\numbersign1\leftbracket#1{\numbersign1}}%
} 
%
% \eqn,\eqnna,eqnnd obsolete pour compatibilite anterieure, 
\def\eqn{\eqnn}

\def\em@rk{\hbox{}} 
\def\xeqn{\expandafter\xe@n}\def\xe@n(#1){#1}
\def\xeqna#1{\expandafter\xe@na#1}\def\xe@na\hbox#1{\xe@nap #1}
\def\xe@nap$(#1)${\hbox{$#1$}}
% \eqns allows to quote several equations, suppressing unnecessary ()
\def\eqns#1{(\e@ns #1{\hbox{}})}
\def\e@ns#1{\ifx\UNd@FiNeD#1\message{eqnlabel \string#1 is undefined.}%
\xdef#1{(?.?)}\fi{\let\hyperref=\relax\xdef\next{#1}}%
\ifx\next\em@rk\def\next{}%
\else\ifx\next#1\xeqn#1\else\def\n@xt{#1}\ifx\n@xt\next#1\else\xeqna#1\fi
\fi\let\next=\e@ns\fi\next}
%**********************************************************************
%*************************** figure macros ****************************
% Pour centrer ajouter 16mm a la taille de la boite
\def\figure#1#2{\global\advance\nofigure by 1 \vglue#1%
{\elevenpoint
\setbox1=\hbox{#2}
\ifdim\wd1=0pt\centerline{Fig.\ \the\nofigure\hskip0.5mm}%
\else\def\caption{Fig.\ \the\nofigure\quad#2\hskip0mm}
\setbox0=\hbox{\caption}
\ifdim\wd0>\hsize\noindent Fig.\ \the\nofigure\quad#2\else
                 \centerline{\caption}\fi\fi}\par}
% le bigskip a la fin a ete enleve!
 % obsolete, for compatibility
%***************
%figure alignee a gauche
\def\lfigure#1#2{\global\advance\nofigure by
1\vglue#1\leftline{\elevenpoint\hskip10truemm  Fig.\
\the\nofigure\quad #2}} 
%***********************************************************************
\catcode`@=12

\input epsf
\input taiwan0.xxx
%\draft
%\writedefs
\francmodefalse
\def\r{{\rm r}}
\def\psib{\psi}
\def\phib{\phi}
\preprint{SPhT/97-018}

\title{Vector models in the large $N$ limit: a few applications} 

\centerline{J.~ZINN-JUSTIN*}
\medskip{\it
CEA-Saclay, Service de Physique Th\'eorique**, F-91191 Gif-sur-Yvette
\goodbreak Cedex, FRANCE} 
\footnote{}{${}^*$email: zinn@spht.saclay.cea.fr}

\footnote{}{${}^{**}$Laboratoire de la Direction des
Sciences de la Mati\`ere du 
Commissariat \`a l'Energie Atomique}

\abstract
In these lecture notes prepared for the\par
{\bf 11$^{\bf th}$ Taiwan Spring School, Taipei 1997},\par
and updated for the Saalburg summer school 1998, we review the solutions of
$O(N)$ or $U(N)$ models in the large $N$ limit and as $1/N$ expansions, in the
case of vector representations. The general idea is 
that invariant composite fields have small fluctuations for $N$ large.
Therefore the method relies on constructing effective field theories for these
composite fields after integration over the initial degrees of freedom.
We illustrate these ideas by showing that the large $N$ expansion
allows to relate the $(\phib^2)^2$ theory and the non-linear $\sigma$-model,
models which are renormalizable in different dimensions.
In the same way large $N$ techniques allow to relate the Gross--Neveu, an
example of a theory with four-fermi self-interaction, with a Yukawa-type
theory renormalizable in four dimensions, a topic relevant for four
dimensional field theory.\par 
Among other issues for which large $N$ methods are also useful we will briefly
discuss finite size effects and finite temperature field theory, because
they involve a crossover between different dimensions.\par
Finally we consider the case of a general scalar $V(\phib^2)$ field 
theory, explain how the large $N$ techniques can be generalized, and discuss
some connected issues like tricritical behaviour and double scaling limit.
Some sections in these notes are directly adapted from the work \par
Zinn-Justin J., 1989, {\it Quantum Field Theory and Critical Phenomena},
Clarendon Press (Oxford third ed. 1996). \par
\medskip
{\it These lecture notes are dedicated to Mrs.~T.D.~Lee, who
recently passed away, as a testimony of gratitude for the long lasting
friendship between our families.} 
\endabstract
\vfill\eject
\listcontent
\vfill\eject
\section Introduction

In these lectures we describe a few applications of large $N$ techniques
to quantum field theories (QFT) with $O(N)$ or $U(N)$ symmetries,
where the fields are in the vector representation. We want to show that large
$N$ results nicely complement results obtained from more conventional
perturbative renormalization group (RG). Indeed the shortcoming of the latter
method is that it mainly applies to gaussian or near gaussian fixed points.
This restricts space dimension to dimensions in which the corresponding
effective QFT is renormalizable, or after dimensional 
continuation, to the neighbourhood of such dimensions. Large $N$ techniques
in some cases allow a study in generic dimensions. They rely on noting that
in the large $N$ limit scalar (in the group sense) composite fields have small
fluctuations (central limit theorem). Therefore if we are able to construct an
effective field theory for the scalars, integrating out the initial degrees of
freedom, we can solve the field theory in a $1/N$ expansion. Note that for
vector representations the number of independent scalars is finite and
independent of $N$, unlike what happens for matrix representations. This
explains why vector models have been solved much more generally than
matrix models.\par
In these lectures we will in particular stress two points: first it is
necessary to always check that the $1/N$ expansion is both IR finite and
renormalizable. Some technical aspects of this question which will be
described in 
section \sssfivNRT. This is essential for the stability of the large $N$
results and the existence of a $1/N$ expansion. Second, the large $N$
expansion is just a technique, with its own (often unknown) limitations.
It should not be discussed in isolation. Instead, as we shall do in the
following examples, it should be combined with other perturbative techniques
and the reliability of the $1/N$ expansion should be inferred from the
general consistency of all results.\par
Second-order phase transitions in classical statistical physics will
provide us with the first illustration of the usefulness of the large $N$
expansion. Due to the divergence of the correlation length at the critical
temperature, systems then have at and near $T_c$ universal properties which
can be described by effective continuum quantum field theories. The
$N$-vector model that we 
discuss below is the simplest example but it has many applications since
it allows to describe the critical properties of systems like
vapour--liquid, binary mixtures, superfluid Helium or ferromagnetic 
transitions as well as the statistical properties of polymers.
Before showing what kind of information can be provided by large $N$
techniques we will first shortly recall what can be learned from 
perturbative renormalization group (RG) methods. Long distance properties can
be described in $d=4-\varepsilon$ dimension by a $(\phib^2)^2$ field
theory. Instead in $d=2+\varepsilon$ the relevant QFT model
is the $O(N)$ non-linear $\sigma$ model. It is somewhat surprising that
the same statistical model can be described by two different theories.
Since the results derived in this  way are valid {\it a priori}\/ only for
$\varepsilon$ small, there is no overlap to test the consistency. The large
$N$ expansion will allow us to discuss generic dimensions and thus to
understand the relation between both field theories.\par
Another domain of application of the large $N$ expansion is finite size
effects 
and finite temperature field theory. In these situations a
dimensional crossover occurs between the large size or zero temperature
situation 
where the infinite volume theory is relevant to a dimensionally reduced
theory in the small volume or high temperature limit. Both effective field
theories being renormalizable in different dimensions, perturbative RG cannot
describe correctly both situations. Again large $N$ techniques will help
us to understand the crossover.\par
Four-fermi interactions have been proposed to generate a composite Higgs
particle in four dimensions, as an alternative to a Yukawa-type theory,
as one finds in the Standard Model. Again, using the specific example of the
Gross--Neveu model, we will use large $N$ techniques to clarify the relations
between these two approaches. We will finally briefly indicate that other
models with chiral properties, like massless QED or the Thirring model, can be
studied by similar techniques.\par 
In the last section we return to scalar boson field theories, and examine
multicritical points (where the large $N$ technique will show some obvious
limitations), and the double scaling limit, a toy model for discussing
problems encountered in matrix models of 2D quantum gravity.
%

%********
% \phi^4 theory
\section{The $N$-vector model near dimension four: Renormalization
Group (RG)}   

The $N$-vector model is a lattice model described in terms of $ N$-vector spin
variables $ {\bf S}_{i} $ of unit length on each lattice site $ i $,
interacting through a short range ferromagnetic $ O(N) $ symmetric  two-body
interaction $V_{ij}$. The partition function of such a model can
be written:\sslbl\scGRG 
$$ Z= \int \prod_{i} \d  {\bf S}_{i}\,\delta \left( {\bf S}^{2}_{i}-1
\right)\exp\left[- {\cal E} \left( {\bf S} \right)/T\right], \eqnd{\eZspin} $$
in which the configuration energy $\cal E$ is:
$$ {\cal E} \left( {\bf S}  \right)=-\sum_{ij}V_{ij}
{\bf S}_{i}\cdot {\bf S}_{j}\,. \eqnd\ehamspin $$
This model has a second order phase transition between a disordered phase
at high temperature, and a low temperature ordered phase where the $O(N)$
symmetry is spontaneously broken, and the order parameter ${\bf S}_i$ has a
non-vanishing expectation value. At a second order phase transition the
correlation length diverges, and therefore a non-trivial long distance physics
can be defined. Scaling and universality properties emerge which we want to
study. \par
To generate correlation functions one can add to $ {\cal E} \left(
{\bf S}  \right)$ a coupling to a space-dependent magnetic field
$$ {\cal E} \left( {\bf S}  \right)=-\sum_{ij}V_{ij}
{\bf S}_{i}\cdot {\bf S}_{j}+\sum_i {\bf H}_i\cdot{\bf S}_i \,.  \eqnn
$$ 
\subsection Mean field theory and the stability of the gaussian fixed point

To derive the critical properties of the $N$-vector model one can proceed
in the following way: one starts from the mean field approximation, valid in
high dimensions. One then shows that the mean field approximation is the
first term in a systematic expansion. One  discovers that for
dimensions $d>4$ the successive terms in the expansion do not modify
the leading mean field behaviour. For $d<4$ instead IR divergences
appear and the mean field approximation is no longer valid. Moreover
a summation of the leading IR divergences to all orders in the
expansion leads to an effective local $\phi^{4}$ field theory. 
The corresponding action is given by the first relevant terms of the
Landau--Ginzburg--Wilson hamiltonian:\sslbl\ssGRGintro
$$ {\cal H} \left(\phib \right) = \int \d^{d}x
\left[ \ud c \left(\nabla \phib \right)^2 + \ud a \phib^2(x)
+b{1 \over 4!}\bigl(\phib^2(x)\bigr)^2 \right] , \eqnd\eLGWham $$
with $ a$, $ b $ and $ c $ being {\it regular}\/ functions of the
temperature for $ T $ close to $ T_c $. \par
Note that the expression \eLGWham~, which in the sense of classical
statistical physics is a configuration energy, is often called
hamiltonian. The reason is that if one starts from a classical
hamiltonian and a functional integral over phase space, the integral
over conjugate momenta is gaussian and thus trivial. From the point of
view of quantum field theory the expression \eLGWham~has the form of an
euclidean action, analytic continuation to imaginary time of the
classical field theory action. We shall thus generally call it the
action. \par  
Alternatively one can imagine starting from the configuration energy 
\ehamspin~and constructing Wilson's renormalization group by integrating out
short distance degrees of freedom. The spin variable ${\bf S}_i$ is then
replaced by a local average, a vector of continuous length of the type of the
field $\phib(x)$.\par 
Mean field theory corresponds to the gaussian fixed point of this
renormalization group. At the critical temperature one finds a massless free
field theory 
$$ {\cal H}_{\rm G} \left(\phib \right) = \int \d^{d}x  \ud c \left(\nabla
\phib \right)^2 .$$
One then performs an analysis of the stability of the gaussian fixed point.
Mean field theory assumes that the order parameter, here the field $\phib(x)$,
is small and varies only on macroscopic scales. Therefore a general 
action can be expanded in powers of the field $\phib(x)$ and derivatives.
$$ {\cal H} \left(\phib \right) = \int \d^{d}x  \ud c \left(\nabla
\phib \right)^2 +\sum_\ell {\cal H}_\ell(\phib),$$
where $\sum_\ell$ means sum over all space integrals $ {\cal H}_\ell(\phib)$
of $O(N)$ symmetric monomials in $\phib$ of degree $n_\ell$ and containing
$m_\ell$ derivatives (often below called operators, a language borrowed from
quantum field theory). \par 
A convenient way to understand the relevance of the  $ {\cal H}_\ell(\phib)$
terms in the large distance (infrared) limit is to rescale
all space or momentum variables, and measure distances in units of the
correlation length, or, at the critical temperature, in some arbitrary unit
much larger than the lattice spacing and corresponding to the typical
distances at which correlations are measured. \par
Let us perform such a rescaling here, and rescale also the field $\phi(x) $ in
such a way that the coefficient of $ \left[ \nabla \phib (x)
\right]^{2} $, to which all contributions will be compared, becomes the
standard 1/2: 
$$ \eqalignno{ x & \mapsto \Lambda x, & \eqnd\exscale \cr  \phib
(x) & \mapsto \zeta \phib (x). & \eqnd\ephiscale \cr} $$
After this rescaling all quantities have a dimension in units of $\Lambda$.
Our choice of normalization for the gradient term implies:
$$ \zeta =c^{-1/2} \Lambda^{1-d/2}, \eqnn $$
which shows that $\phib$ now has in terms of $\Lambda$ its canonical dimension
$d/2-1$. \par
A term $ {\cal H}_\ell(\phib)$ then is multiplied by
$$ {\cal H}_\ell(\phib)\mapsto \Lambda^{d-n_\ell(d-2)/2-m_\ell} {\cal
H}_\ell(\phib).$$ 
For $\Lambda$ large we observe the following:\par
(i) The leading term is the term proportional to $\int\d^d x\,\phib^2(x)$,
which is multiplied by $\Lambda^2$. This is not surprising since it gives a
mass to the field and therefore the theory moves away from the massless
critical theory (the term is called relevant).\par 
(ii) If $d>4$ all other terms are multiplied by negative powers and therefore
become negligible in the long distance limit. They are called irrelevant.
The gaussian fixed point is stable and mean field theory thus correct.
\par
(iii) In four dimensions the $\phi^4$ interaction is independent of $\Lambda$:
it is called marginal while  all other interactions remain irrelevant. The
analysis of the stability of the gaussian fixed point then requires a finer
study which will be based on the field theory perturbative renormalization
group. \par
(iv) Below four dimensions the $\phi^4$ interaction becomes relevant, the
gaussian fixed point is certainly unstable. The question of the existence of
another 
non-trivial fixed point is non-perturbative and cannot be easily answered.
Partial answers are based upon the following assumption: the
dimensions of operators are {\it continuous}\/ functions of the space
dimension. This means that we are going to look for a fixed point which,
when $d$ approached four, coalesces with the gaussian fixed point. Moreover
even at this new fixed point, at least in some
neighbourhood of dimension four, all operators except $(\phib^2)^2$ should
remain irrelevant. The action \eLGWham~should contain all relevant
operators and therefore enough information about the non-trivial fixed
point.\par    
After the rescaling  \eqns{\exscale,\ephiscale} the action $ {\cal H}
\left(\phib \right) $ then becomes: 
$$ {\cal H} \left(\phib \right)= \int \d ^{d}x \left\lbrace\ud
\left[ \nabla \phib (x) \right]^{2}+\ud r\phib
^{2} (x)+{1 \over 4!}g\Lambda^{4-d}\bigl(\phib^2(x)\bigr)^2 
 \right\rbrace , \eqnd{\eLGWphi} $$
with $ r  = a\Lambda^{2}/c$, $  g = b / c^2$.
The action \eLGWphi~generates a perturbative expansion of field theory
type which can be described in terms of Feynman diagrams. These have to be
calculated with a momentum cut-off of order $\Lambda$, reflection of
the initial microscopic structure. The corresponding theory is thus
analogous to regularized quantum field theory. The precise cut-off
procedure can be shown to be irrelevant except that it should satisfy
some general regularity conditions. For example the propagator can be
modified (as in Pauli--Villars's regularization) but the inverse
propagator in momentum space must remain a regular function of
momentum (the forces are short range). \par 
Let us call $ r_{c} $ the value of the parameter $r$ which corresponds, at $g$
fixed, to the 
critical temperature $ T_{c} $ at which the correlation length $\xi$ diverges.
In terms of the scale $ \Lambda $ the critical domain is then defined by:
$$ \left. \eqalign{&{\rm physical\ mass}\ =\xi^{-1}\ll \Lambda\ \Rightarrow\
\left\vert r-r_{c} \right\vert  \ll \Lambda^{2} 
\cr & {\rm  distances}  \gg 1/\Lambda \quad  {\rm or\ momenta} \ll \Lambda ,
\cr & {\rm magnetization}\ M\equiv |\left< \phib(x)\right>|\ll 
\zeta^{-1}\sim \Lambda^{(d/2)-1}. \cr} \right. \eqnd{\ecritdom} $$
Note that these conditions are met if $ \Lambda $ is identified with
the cut-off of a usual field theory. However an inspection of the action
\eLGWphi~also shows that, in contrast with conventional quantum field theory,
the $ \phi^{4} $ coupling constant has a dependence in $ 
\Lambda $ given {\it a priori}. For $ d<4 $ the $ \phi^{4} $ coupling
is very large in terms of the scale relevant for the critical domain. In
the usual formulation of quantum field theory instead the {\it bare}\/
coupling constant also is an adjustable parameter. This implies for
instance that for $d<4$ (super-renormalizable theory) the coupling
constant varies when the correlation length changes. This is a somewhat
artificial situation if one believes that that the initial bare or
microscopic theory has a physical meaning. \par
The critical properties of the field theory (like the long distance
behaviour of correlation functions) can then be analyzed by RG methods in
$4-\varepsilon$ dimension, i.e.~near the so-called upper-critical
dimension (and with some additional assumptions in $d<4$).
\medskip
{\it Dimensions of fields.} Because we deal with translation invariant
theories,  we will generally discuss the scaling behaviour of
correlation functions in momentum variables. Let us relate the scaling
behaviour of connected correlation functions expressed in terms of
space and momentum variables. When functions have a scaling behaviour,
one defines  
$$\left<\prod_{i=1}^n {\cal O}_i(x_i/\lambda)\right>_{\rm c}=\lambda^D
\left<\prod_i {\cal O}_i(x_i)\right>_{\rm c}\ {\rm with}\ D={\sum_i
d_{{\cal O}_i}}\,, \eqnn $$ 
where $ {\cal O}_i$, sometimes called operator, is a local polynomial in the
basic fields (associated with the order parameter), and the quantity
$d_{{\cal O}_i}$, which we sometimes also denote $[{\cal O}_i]$, is called
the dimension of the field (operator) ${\cal O}_i$.\par 
After Fourier transformation and factorization of the
$\delta$-function of momentum conservation, one then finds, in $d$ 
space dimension, 
$$ \left<\prod_{i=1}^n {\cal O}_i(\lambda p_i\right>_{\rm c}=\lambda^{D'}
\left<\prod_i {\cal O}_i(p_i)\right>_{\rm c}\ {\rm with\ now}\ D'={d+\sum_i
(d_{{\cal O}_i}-d)}\,. \eqnd\escalspamom  $$ 
Finally it is convenient to introduce the Legendre transform
$\Gamma(\phi)$ of the generating functional $W(H)=T\ln Z$ of $\phi$-field
connected correlation functions. We denote by $W^{(n)}$ and
$\Gamma^{(n)}$ the corresponding connected and 1PI functions.
One verifies that if one performs a Legendre transformation on
the source associated with the field (operator) ${\cal O}_i$, the
quantity $d_{{\cal O}_i}-d$ in equation \escalspamom~is replaced by
$-d_{{\cal O}_i}$. 
\subsection RG equations for the critical (massless)
theory

The field theory with the action \eLGWphi~can now be studied by field
theoretical methods. From simple power counting arguments one
concludes that the critical (or massless) theory
does not exist in perturbation theory for any dimension smaller than 4. If we
define, by dimensional continuation, a critical theory 
in $ d=4-\varepsilon $ dimensions, even for arbitrarily small $ \varepsilon$
there always exists an order in perturbation $ (\sim 2 \left/\varepsilon ) 
\right. $ at which IR (infrared) divergences appear. Therefore the idea,
originally due to Wilson and Fisher, is to perform a
double series expansion in powers of the coupling constant $ g $ and $
\varepsilon $. Order by order in this  expansion, the critical behaviour
differs from the mean field behaviour only by powers of logarithm, and we can
construct a perturbative critical theory by adjusting $ r $ to its critical
value $ r_{c} \left(T=T_{c} \right)$.\sslbl\ssGRGeps \par
To study the large cut-off limit we then use methods developed for the
construction of the renormalized massless $ \phi^{4} $ field theory. We 
introduce  rescaled (renormalized) correlation functions,
defined by renormalization conditions at a new scale $\mu\ll \Lambda$, and
functions of a renormalized coupling constant $ g_{\r}$. We write here 
equations for Ising-like systems, the field $ \phi $ having only one component.
The generalization to the $N$-vector model with $ O(N) $ symmetry, is
straightforward except in the low temperature phase or in a symmetry breaking
field, a situation which will be examined in section \ssCDeqst.  
Then:
$$\left\lbrace \eqalign{ \Gamma^{(2)}_{\r} \left. \left(p;g_{\r},\mu ,\Lambda
\right) \right\vert_{p^{2}=0} & = 0\,,  \cr {
\partial \over \partial p^{2}}\Gamma^{(2)}_{\r} \left. \left(p;g_{\r},\mu
,\Lambda \right) \right\vert_{p^{2}=\mu^{2}} & = 
1\,,  \cr  \Gamma^{(4)}_{\r} \left(p_{i}=\mu\theta_i;g_{\r},\mu
,\Lambda \right)  & = \mu^{\varepsilon}g_{\r}\,,  \cr}\right. \eqnd\erencond
$$  
in which $\theta_i$ is a numerical vector.
These correlation functions are related to the original ones by the
equations: 
$$ \Gamma^{(n)}_{\r} \left(p_{i};g_{\r},\mu ,\Lambda \right)=Z
^{n/2} \left(g,\Lambda / \mu \right)\Gamma^{(n)} \left(p
_{i};g,\Lambda \right). \eqnd{\erelBren} $$
Renormalization theory tells us that the functions $
\Gamma^{(n)}_\r \left(p_{i};g_{\r},\mu ,\Lambda \right) $ of equation
\erelBren\ have at $ p_{i}$, $ g_{\r} $ and $ \mu $ fixed, a large
cut-off limit which are the renormalized correlation functions $ \Gamma
^{(n)}_{\r} \left(p_{i};g_{\r},\mu \right)$. A detailed analysis actually
shows that at any finite order in perturbation theory:
$$ \Gamma^{(n)}_{\r} \left(p_{i};g_{\r},\mu ,\Lambda \right)=\Gamma
^{(n)}_{\r} \left(p_{i};g_{\r},\mu
\right)+O\left(\Lambda^{-2}(\ln\Lambda)^L\right) , \eqnd{\ecorrect} $$
in which the power $L$ of $ \ln\Lambda $ increases with the order in $
g $ and $ \varepsilon $. Furthermore the renormalized
functions $ \Gamma^{(n)}_{\r} $ do not depend on the specific cut-off
procedure and, given the normalization conditions \erencond, are therefore
universal. Since the renormalized functions $ \Gamma^{(n)}_{\r} $ and
the initial ones $ \Gamma^{(n)} $ are asymptotically proportional, both
functions have the same small momentum or large distance behaviour. To
determine the universal critical behaviour it is thus sufficient to
study the renormalized field theory. And indeed most peturbative
calculations of universal quantities have been performed in this
framework. However, it is interesting to determine not only the
asymptotic critical behaviour, but also the corrections to the 
asymptotic theory. Furthermore, renormalized quantities are not directly
obtained in non-perturbative calculations. For these reasons it is also
useful to express the implications of equation \erelBren\ directly on the
initial theory. 
\medskip
{\it Bare RG equations.} Let us differentiate equation \erelBren~with respect
to $\Lambda $ at $ g _{\r} $ and $ \mu $ fixed,  
taking into account \ecorrect:
$$\left.\Lambda{ \partial \over \partial \Lambda}\right\vert_{g_{\r},\mu
\ {\rm fixed}}Z^{n/2} \left(g,{\Lambda / \mu} \right)\Gamma^{(n)}
\left(p_{i};g,\Lambda \right)  =O\left(\Lambda^{-2}(\ln\Lambda)^L\right) .
\eqnd{\eLambdLa} $$
We now neglect corrections subleading (in perturbation theory) by powers of
$ \Lambda $. \par
Then, using chain rule, we can rewrite equation \eLambdLa~as:
$$ \left[ \Lambda{ \partial \over \partial \Lambda} +\beta
\left(g,{\Lambda / \mu} \right){\partial \over \partial g}-{n \over
2}\eta \left(g,{\Lambda / \mu} \right) \right] \Gamma^{(n)}
\left(p_{i};g,\Lambda \right)=0\,. \eqnd\eRGbare $$
The functions $ \beta $ and $ \eta $, which are dimensionless and
may thus  depend only on the dimensionless quantities $ g $ and $ \Lambda /\mu
$, are defined by:
\eqna\ebeta
$$ \eqalignno{\beta \left(g,{\Lambda / \mu} \right) & = \left.\Lambda{
\partial \over \partial \Lambda} \right\vert_{g_{\r},\mu} g\,, &
\ebeta{a} \cr  \eta \left(g,{\Lambda / \mu} \right) & =
\left.-\Lambda{ \partial \over \partial \Lambda} \right\vert_{g
_{\r},\mu}\ln Z \left(g,{\Lambda / \mu} \right). & \ebeta{b} \cr} $$
However,  the functions $ \beta$ and $ \eta $ can also be directly calculated
from equation \eRGbare~in terms of functions $ \Gamma^{(n)} $ which do not 
depend on $ \mu $. Therefore the functions $ \beta$ and $ \eta $ cannot
depend on the ratio $ \Lambda /\mu $ (in the definitions \ebeta{} 
consistency requires that contributions which goes to zero like some
power of $ \mu \left/\Lambda \right. $, should be  neglected, as in equation
\eLambdLa). Then equation \eRGbare~can be rewritten:  
$$ \left[ \Lambda{ \partial \over \partial \Lambda} +\beta \left(g
\right){\partial \over \partial g}-{n \over 2}\eta (g)
\right] \Gamma^{(n)} \left(p_{i};g,\Lambda \right)=0\,.
\eqnd\eRGBas  $$
Equation \eRGBas~is an equation satisfied when the cut-off is large by
the physical correlation functions of statistical mechanics which are also
the bare correlation functions of quantum field theory. It expresses 
the existence of a renormalized theory.
\subsection RG equations and large distance behaviour: the $\varepsilon
$-expansion 

Equation \eRGBas~can be solved by the method of
characteristics: one introduces a dilatation parameter $ \lambda $, together
with a running coupling constant $ g(\lambda) $ and a scale
dependent field renormalization $ Z (\lambda) $ satisfying the flow equations
\sslbl\sscGRGeps
\eqna\eRGgZphi
$$\deqalignno{ \lambda{ \d  \over \d  \lambda} g \left(\lambda
\right)&=\beta \left(g \left(\lambda \right) \right) ,& \ g \left(1
\right) & =g\,; & \eRGgZphi{a} \cr  \lambda{ \d  \over \d 
\lambda} \ln Z \left(\lambda \right)& =\eta \left(g \left(\lambda \right)
\right), &  \ Z \left(1 \right)& =1\,. & \eRGgZphi{b} \cr} $$
The behaviour of correlation functions for $|p_i|\ll\Lambda $ ($\lambda\to0$)
is then governed by IR fixed points, zeros of the RG $\beta$-function
with a positive slope. \par
The RG functions $\beta $ and $ \eta $ can be calculated in perturbation
theory. From the relation between bare and renormalized coupling constant
and the definition \ebeta{a} it follows that ($\varepsilon=4-d)$:
$$ \beta \left(g,\varepsilon \right)=-\varepsilon g+{N+8 \over 48\pi
^{2}}g^2+O \left(g^{3},g^{2}\varepsilon \right). \eqnd\ebetaone $$
Let us now assume that $g$ initially is sufficiently small, so that
perturbation theory is applicable.  
We see that above or at four dimensions, i.e.~$\varepsilon \le 0$, the function
$\beta$ is positive and  $g(\lambda)$ decreases approaching the origin
$g=0$. We recover that the gaussian fixed point is IR stable for
$d>4$, and find that it ia also stable at $d=4$.\par
Below four dimensions, instead, the gaussian fixed point $g=0$ is IR repulsive.
However, expression \ebetaone~ shows that, for $\varepsilon$ small, $
\beta(g) $ now has a non-trivial zero $g^*$:    
$$\beta(g^*)=0,\quad g^*  = {48\pi^{2} \over N+8}\varepsilon +O
\left(\varepsilon^{2} \right), \quad {\rm with}\
\beta'(g^*)\equiv\omega   = \varepsilon +O \left(\varepsilon^{2} \right).
\eqnd\egstar  $$  
The slope $ \omega $ at the zero is positive. 
This non-gaussian fixed point thus is IR stable, at least in the sense
of an $\varepsilon$-expansion. In four dimensions it merges with the
gaussian fixed point and the eigenvalue $ \omega $ vanishes,
indicating the appearance of a marginal operator.  
\par
The solution of the RG equation then determines the behaviour of $
\Gamma^{(n)} \left(p_{i};g,\Lambda \right) $ for  $|p_i|\ll\Lambda $:
$$  \Gamma^{(n)} \left(\lambda p_{i};g,\Lambda \right) 
\mathop{\sim}_{\lambda \rightarrow 0 } \lambda^{d- \left(n/2 \right)
\left(d-2+\eta \right)}\Gamma^{(n)} \left(p_{i};g^{\ast},\Lambda
\right), \eqnd{\escaleb} $$
where $\eta=\eta(g^*)$. Critical correlation functions have a power law
behaviour for small momenta, independent of the initial value of the 
$ \phi^{4} $ coupling constant $ g$. \par
In particular the small momentum behaviour of the inverse two-point function
is obtained for $n=2$. For the two-point function $ W^{(2)} \left(p \right) $
this yields:  
$$  W^{(2)}(p)=\left[\Gamma^{(2)}(p)\right]^{-1}\mathop{\sim}_{\left\vert p
\right\vert \rightarrow 0 }  1 \left/p^{2-\eta}. \right. \eqnn $$
The spectral representation of the two-point function implies
$\eta>0$. A short calculation yields:
$$ \eta =\frac{N+2}{2(N+8)^2}\varepsilon^{2}+O \left(\varepsilon^{3} \right).
\eqnn $$
The scaling in equation \escaleb~indicates that the field $ \phi (x)$, which
had at the gaussian fixed point a canonical dimension $(d-2)/2$, has now
acquired an ``anomalous" dimension $ d_{\phi}$ (see the discussion of the end
of section \ssGRGintro): 
$$ d_{\phi}=\ud \left(d-2+\eta \right).  $$
These results call for a few comments. Within the framework of
the $ \varepsilon $-expansion, one thus proves that all correlation functions
have, for $d<4$, a long distance behaviour different from the one predicted by
mean field theory. In addition the critical behaviour does not depend on the
initial value of the $ \phi^{4} $ coupling constant $ g$. At least for
$\varepsilon$ small one may hope that the analysis of leading IR singularities
remains valid and thus it does not depend on any other
coupling either. Therefore the critical behaviour is {\it universal}, although
less universal than in mean field theory, in the sense that it depends only on
some small number of qualitative features of the system under consideration. 
\subsection Critical correlation functions with $ \phi\sp{2}(x) $ insertions

RG equations for critical correlation functions with
$\int\d^dx\,\phi^{2}(x) $ insertions can also be derived. The operator
$\phi^2(x)$ has a direct physical interpretation. It is the most singular part
(i.e.~the most relevant) of the energy density \eLGWham. Long distance scaling
properties follow. Moreover these RG equations can be used to derive RG
equations for correlation functions in the whole critical domain. 
\sslbl\ssGRGfii \par
We denote by $ \Gamma^{(l ,n)} \left(q_{1},\ldots
,q_{l };p_{1},\ldots ,p_{n};g,\Lambda \right) $ the mixed 1PI correlation
functions of the order parameter $\phi(x)$ and the  energy density $\ud
\phi^2(x)$ ($ n\,\phi $ fields and $ l\, { 1 \over 2}\phi^{2} $ operators,
with $(l+n)\geq 2 $). Renormalization theory tells us that we can define
renormalized correlation functions $ \Gamma^{(l  ,n)}_{\r}
\left(q_{i};p_{j};g_{\r},\mu \right) $ which, in addition to conditions
\erencond, satisfy:  
$$ \left.\eqalign{ \left.\Gamma^{(1,2)}_{\r} \left(q;p_{1},p_{2};g
_{\r},\mu \right) \right\vert_{{p^{2}_{1}=p^{2}
_{2}=\mu^{2},\ p_{1}\cdot p_{2}=-{1 \over 3}\mu^{2}}} & = 1\,, \cr
\left.\Gamma^{(2,0)}_{\r} \left(q,-q;g_{\r},\mu \right) \right\vert_{{q^{2}={4
\over 3}\mu^{2}}} & = 0\,,  \cr}\right. \eqnn $$ 
and are related to the original ones by:
$$\lim_{\Lambda \rightarrow \infty} Z^{n/2} \left({Z_2/ Z}
\right)^{l }  \left[ \Gamma^{(l  ,n)} \left(q_{i};p
_{j};g,\Lambda \right) -\delta_{n0}\delta_{l  2}\Lambda
^{-\varepsilon}A \right] =\Gamma^{(l  ,n)}_{\r} \left(q_{i};p
_{j};g_{\r},\mu \right). \eqnn $$
$ Z_{2} \left(g,{\Lambda / \mu} \right) $ and $ A
\left(g,{\Lambda / \mu} \right) $ are two new renormalization constants. \par
Differentiating with respect to $ \Lambda $ at $ g_{\r} $ and $
\mu $ fixed, as has been done in section \ssGRGeps, and using chain rule one
obtains a set of RG equations:
$$ \left[ \Lambda{ \partial \over \partial
\Lambda} +\beta \left(g \right){\partial \over \partial g}-{n \over 2}\eta
\left(g \right)-l  \eta_{2} \left(g \right) \right]
\Gamma^{(l  ,n)} =\delta_{n0}\delta_{l  2}\Lambda
^{-\varepsilon}B \left(g \right). \eqnd{\eRGgamln} $$
In addition to  $ \beta $ and $ \eta $ (equations \ebeta{}) two
new RG functions, $ \eta_{2}(g) $ and $ B(g) $, appear: 
$$ \eqalignno{ \eta_{2} \left(g \right) & = -\Lambda \left.{\partial \over
\partial \Lambda} \right\vert_{g_{\r},\mu}\ln \left[ Z_{2}
\left(g, \displaystyle{ \Lambda /\mu} \right) \left/  Z \left(g,
\displaystyle{ \Lambda / \mu} \right) \right.\right] , &
\eqnd{\etadeg} $$
\cr  B \left(g \right) & = \left[ \Lambda \left.{\partial \over 
\partial \Lambda} \right\vert_{g_{\r},\mu}-2\eta_{2} \left(g
\right)-\varepsilon \right] A \left(g,{\Lambda / \mu} \right). &
\eqnn \cr} $$
Note that for $ n=0$, $ l  =2$, the RG
equation \eRGgamln\ is not homogeneous. This is a consequence of the
non-multiplicative character of renormalization in this case. 
\par 
In the homogeneous case, equation \eRGgamln~can be solved exactly in the same
way as equation \eRGbare. A new function $\zeta_2(\lambda)$ has to be
introduced, associated with the RG function $\eta_2(g)$. 
Again the solution of equation \eRGgamln~combined with simple dimensional
analysis leads to the scaling behaviour
$$ \Gamma^{(l  ,n)} \left(\lambda q_{i};\lambda p_{j};g,\Lambda
\right)\mathop{\propto}_{\lambda \rightarrow 0 } \lambda^{d-n 
(d-2+\eta )/2 -l / \nu}, \eqnd{\egalnsca} $$
where  the correlation length exponent $ \nu$ is related to $\eta_2(g^*)$ by:
$$ \nu =  \left[ \eta_{2}(g^*)+2\right]^{-1}. \eqnd{\enugdef} $$
The dimension of the field $\phib^2$ follows (see section \ssGRGintro)
$$d_{\phi^2}=d-1/\nu\,. \eqnn $$
Using equations \eqns{\etadeg,\enugdef} it is easy to calculate $\eta_2(g) $
at one-loop order. At the fixed point $ g = g^{\ast}$ (equation \egstar) one
then obtains the exponent $\nu$:  
$$ 2\nu =1+\frac{(N+2)}{2(N+8)}\varepsilon+O \left(\varepsilon^{2}\right). $$
\medskip
{\it The $\left<\phi^2\phi^2\right>$ correlation function.} The $ \phi^{2} $
(energy density) two-point function $ \Gamma^{(2,0)} $ satisfies an
inhomogeneous  RG equation. To solve it one first looks for a particular
solution, which can be chosen of the form $\Lambda^{-\varepsilon}C_{2}(g)$:
$$ \beta \left(g \right)C'_{2} \left(g \right)-
\left[ \varepsilon +2\eta_{2} \left(g \right) \right] C_{2}
\left(g \right)=B \left(g \right). \eqnd{\einhoCde} $$
The solution is uniquely determined  by imposing its {\it regularity}\/ at
$ g=g^{\ast}$. \par
The general solution of equation \eRGgamln\ is then the sum of this
particular solution and of the general solution of the homogeneous equation
which has a behaviour given by equation \egalnsca:
$$ \Gamma^{(2,0)} \left(\lambda q;g,\Lambda
\right)-\Lambda^{-\varepsilon}C_{2} \left(g \right)\mathop{\sim}_{
\lambda \rightarrow 0 } \lambda^{d-2/\nu}. \eqnn $$
\medskip
{\it Remarks}. \par
(i) The physics we intend to describe corresponds to integer values
of $\varepsilon$, $\varepsilon =1,2 $. Although we can only prove the validity
of all RG results within the framework  of the $ \varepsilon $-expansion,  we
shall eventually assume that their validity extends beyond an infinitesimal
neighbourhood of dimension 4. The large $N$-expansion provides a test of
the plausibility of this assumption. The decisive test comes, of course,
from the comparison with experimental or numerical data.\par
(ii) In four dimensions the $\phi^4$ interaction is marginally irrelevant;
the renormalized coupling constant of the $\phi^4$ field theory goes
to zero only logarithmically when the cut-off becomes infinite. This
induces logarithmic corrections to mean field theory. Moreover, since
no other fixed point seems to exist, this leads to the so-called {\it
triviality property} (see section \ssGRGivD) of the $\phi^4$ quantum
field theory.  
\subsection{Scaling behaviour in the critical domain}

We have described the scaling behaviour 
of correlation functions at criticality, $ T=T_{c}$. We now consider the
critical domain which is defined by the property that the correlation length
is large with respect to the microscopic scale, but finite. \sslbl\scCD \par 
\medskip
{\it Remark.} The temperature is coupled to the total configuration
energy. Therefore a variation of the temperature generates a
variation of all terms contributing to the effective action. However
the most relevant contribution (the most IR
singular) corresponds to the $\phi^2(x)$ operator. 
We can therefore take the difference $t=r-r_c\propto T-T_c$ between the
coefficient of $\phi^2$ in \eLGWphi\ and its critical value as a linear
measure of the deviation from the critical temperature.
Dimensional analysis then yields the relation
$$\Gamma^{(n)} \left(p_{i};t,g,\Lambda \right)=\Lambda^{d-n(d-2)/2}
\Gamma^{(n)} \left(p_{i}\Lambda^{-1};t\Lambda^{-2},g, 1 \right)
.\eqnd\edimnonTc$$
With this parametrization the critical domain corresponds to
$|t|\ll \Lambda^2$.
\medskip
{\it Expansion around the critical theory.} One thus adds to the critical
action a term of the form $\ud t\int\d^d x\, \phi^{2} (x) $.
To derive RG equations in the
critical domain one expands correlation functions in formal power series of
$ t$. The coefficients are critical correlation functions involving $
\phi^{2}(x) $, for which RG equations have been derived in section
\ssGRGfii, inserted at zero momentum. Some care has to be taken to avoid
obvious IR problems. Summing the expansion, one obtains RG equations valid for
$ T\not= T_{c}$, $|T-T_c|\ll 1$.\par 
After summation of the $t$-expansion one finally obtains the RG equation:
$$ \left[ \Lambda{ \partial \over \partial \Lambda} +\beta \left(g
\right){\partial \over \partial g}-{n \over 2}\eta \left(g \right)-\eta_{2}
\left(g \right)t{\partial \over \partial t} \right] \Gamma^{(n)}
\left(p_{i};t,g,\Lambda \right)=0\,. \eqnd{\eganTRG} $$
\medskip
{\it Scaling laws above $T_{c} $.} 
As for previous RG equations, equation \eganTRG~can be integrated by using the
method of characteristics. In addition to the functions $ 
g(\lambda) $ and $ Z(\lambda) $, one needs a running temperature
$ t(\lambda) $. Taking the large $ \Lambda $, or the small $
\lambda $ limit one finally obtains: \sslbl\ssCDscal
$$ \Gamma^{(n)} \left(p_{i};t,g,\Lambda =1 \right)\mathop{\sim}_{\scriptstyle
t\ll 1\atop\scriptstyle \left\vert p_{i} \right\vert \ll 1   }m^{ \left( d-n
\left(d-2+\eta \right)/2 \right)}F^{(n)}_{+} \left(p_{i} \left/m \right.
\right), \eqnd{\egastrsc} $$
with:
$$ m \left(\Lambda =1 \right)=\xi^{-1}\sim t^{\nu}. \eqnd{\eCsiscal} $$
From equation \egastrsc\ we infer
that the quantity $m$ is proportional to the physical mass or inverse
correlation length. Equation \eCsiscal\ then shows that the divergence of the
correlation length $ \xi =m^{-1} $  at $T_c$ is characterized by the exponent
$\nu$.\par
For $ t\not= 0$, the correlation functions are finite at zero momentum and
behave as: 
$$ \Gamma^{(n)} \left(0;t,g,\Lambda \right)\propto t^{\nu \left( d-n
 \left(d-2+\eta \right)/2 \right)}. \eqnd{\esczemom} $$
In particular for $ n=2 $ we obtain the inverse magnetic
susceptibility:
$$ \chi^{-1}=\Gamma^{(2)} \left(p=0;t,g,\Lambda \right)\propto
t^{\nu(2-\eta)}. \eqnn $$ 
The exponent which characterizes the divergence of $ \chi $ is usually called
$ \gamma $. The equation \esczemom\ establishes the relation between
exponents:
$$ \gamma =\nu \left(2-\eta \right). \eqnn $$
\subsection Scaling laws in a magnetic field and below $T_c$

In order to pass continuously from the disordered $ \left(T>T_{c} 
\right) $ to the ordered phase $ \left(T<T_{c} \right) $,  avoiding the
critical singularities at $T_c$,
it is necessary to add to the action an interaction which explicitly
breaks its symmetry. One thus add a small magnetic field to the spin
interactions. One then derives RG equations in a field, or at fixed
magnetization. In this way correlation functions above and below $ T_{c}$
can be continuously connected, and  scaling laws established in the whole
critical domain. The first example is provided by the relation between field
and magnetization, i.e.~the equation of state.\sslbl\ssCDeqst 
\medskip
{\it The equation of state.} 
Let us call $ M $ the expectation value of $\phi (x) $ in a constant field $ H$
(for $N>1$ the quantities $M$ and $H$ should be regarded as the length
of the corresponding vectors).
The thermodynamic potential per unit volume, as a function of $ M $, is by
definition: 
$$\Omega^{-1} \Gamma \left(M,t,g,\Lambda \right)= \sum^{\infty}_{n=0}{M^{n}
\over n!}\Gamma^{(n)} \left(p_{i}=0;t,g,\Lambda \right). \eqnn $$
The magnetic field $ H $ is given by:
$$ H=\Omega^{-1}{\partial \Gamma \over \partial M}= \sum^{\infty}_{n=1}{M^{n}
\over n!}\Gamma^{(n+1)} \left(p_{i}=0;t,g,\Lambda \right). \eqnd{\emagfld} $$
Noting that $ n\equiv M \left(\partial /\partial M \right)$, we
immediately derive from the RG equation \eganTRG, the RG equation
satisfied by $ H \left(M,t,g,\Lambda \right)$:
$$ \left[ \Lambda{ \partial \over \partial \Lambda} +\beta \left(g
\right){\partial \over \partial g}-{1 \over 2}\eta \left(g \right)
\left(1+M{\partial \over \partial M} \right)-\eta_{2} \left(g
\right)t{\partial \over \partial t} \right] H \left(M,t,g,\Lambda
\right)=0\,. \eqnd{\eRGmgfld} $$
To integrate equation \eRGmgfld\ by the method of characteristics we
have to introduce, in addition to the functions $ g (\lambda)$, $
t (\lambda) $ and $ Z (\lambda) $, a new function $ M (\lambda) $.
However one verifies that $M(\lambda)$ is given by $M(\lambda)=M
Z^{-1/2}(\lambda)$.\par 
Then from the arguments outlined in previous sections one derives the scaling
form
$$ H \left(M,t,g ,1 \right)\sim M^{\delta} f\left( tM^{-1/\beta} \right) ,
\eqnd{\escmgfld} $$
with:
$$  \beta =\ud \nu (d-2+\eta)=\nu d_\phi\,,\quad 
\delta  = {d+2-\eta \over d-2+\eta}={d\over d_\phi}-1\, . \eqnd\edeltbet $$ 
Equation \escmgfld~exhibits the scaling properties of the equation of
state. Moreover equations \edeltbet~relate the traditional
critical exponents which characterize the 
vanishing of the spontaneous magnetization and the singular relation between
magnetic field and magnetization at $T_c$ respectively to the exponents $
\eta $ and $ \nu $ introduced previously.
\par
The universal function $ f(x)$  is infinitely differentiable at $ x=0$. 
because when $ M $ is different from zero the theory remains massive
even at $ t=0 $. The magnetic field $ H $ has a
regular expansion in odd powers of $ M $ for $ t>0$. This implies that when
the variable $ x $ becomes large and positive, $ f (x) $ has the
expansion (Griffith's analyticity): 
$$ f (x)= \sum^{\infty}_{p=0}a_{p}x^{\gamma -2p\beta}. \eqnn $$
\par
The appearance of a spontaneous magnetization, below $
T_{c}$, implies that the function $ f (x) $ has a negative
zero $ x_{0}$. Then equation \escmgfld~leads to the relation:
$$ M= \left\vert x_{0} \right\vert^{-\beta} \left(-t \right)^{\beta}\
\ \ {\rm for} \ \ \ H=0\,,\ \ t<0\,. \eqnd{\espontMg} $$
Equation \espontMg\ gives the behaviour of the spontaneous magnetization
when the temperature approaches the critical temperature from below. 
\medskip
{\it Correlation functions in a field.} 
We now examine the behaviour of correlation functions in a field. 
We write expressions for Ising-like systems. In the ordered phase some
qualitative differences appear between systems which have a discrete and a
continuous symmetry. We illustrate these differences with an example at
the end of the section. \par
The correlation functions at fixed magnetization $ M $ are
obtained by expanding the generating functional $ \Gamma \left(M \left(x
\right) \right) $ of 1PI correlation functions, around $ M \left(x 
\right)=M $.  
From the RG equations satisfied by the correlation functions in
zero magnetization (equations \eganTRG) it is then easy to derive:
$$ \left[ \Lambda{ \partial \over \partial
\Lambda} +\beta \left(g \right){\partial \over \partial g}-{1 \over 2}\eta
\left(g \right) \left(n+M{\partial \over \partial M} \right)-\eta_{2}
\left(g \right)t{\partial \over \partial t} \right]
\Gamma^{(n)} \left(p_{i};t,M,g,\Lambda \right)=0\,.\eqnd\eRGbelTc $$
This equation can be solved by exactly the same method as equation
\eRGmgfld. One finds
$$  \Gamma^{(n)} \left(p_{i};t,M,g,\Lambda =1
\right) \sim m^{ \left[ d- \left(d-2+\eta \right)/2
\right]}F^{(n)} \left({p_{i} / m},tm^{-1/\nu} \right),
\eqnd{\egaMTsc} $$ 
for $\left\vert p_{i} \right\vert \ll 1$, $\left\vert t \right\vert \ll 1$,
$M\ll 1 $ and with the definition:
$$ m=M^{\nu /\beta}. \eqnn $$
The r.h.s.\ of equation \egaMTsc\ now depends on two different mass
scales: $ m=M^{\nu /\beta} $ and $ t^{\nu}$. 
\medskip
{\it Correlation functions below $ T_{c} $.}
We have argued above that correlation functions are regular functions of $ t $
for small $ t $, provided $ M $ does not vanish. It is therefore possible to
cross the critical point and to then take the zero external magnetic field
limit. In the limit $ M $ becomes the 
spontaneous magnetization which is given, as a function of $t$, by equation
\espontMg. After elimination of $ M $ in favour of $ t $ in equation \egaMTsc,
one finds the critical behaviour below $ T_{c}$:
$$ \Gamma^{(n)} \left(p_{i};t,M \left(t,H=0 \right),g,1 \right)\sim m
^{d-n\left(d-2+\eta \right)/2}F^{(n)}_{-} \left(p_{i} \left/m \right.
\right), \eqnn $$ 
with:
$$ m= \left\vert x_{0} \right\vert^{-\nu} \left(-t \right)^{\nu},\qquad H=0\,
,\qquad t<0\,. \eqnn $$ 
We conclude that the correlation functions have exactly the same
scaling behaviour above and below $ T_{c}$. \par 
The extension of these considerations to the functions with $ \phi^{2} $
insertions, $ \Gamma^{(l,n)} $ is straightforward. In particular the
same method yields the behaviour of the specific heat below $ T_{c}$:
$$ \Gamma^{(2,0)} \left(q=0,M \left(H=0,t \right)
\right)-\Lambda^{-\varepsilon}C_{2} \left(g \right)
\mathop{\sim}_{\displaystyle{{\rm for} \ \ t<0} } A^{-}\left(-t \right)
^{-\alpha} , \eqnd{\esphemin} $$
which similarly proves that the exponent above and below $T_c$ are the same.
\par
Note that the constant term $ \Lambda^{-\varepsilon}C_{2}\left(g \right) $
which depends explicitly on $ g $ is the same above and below $ T_{c}$, in
contrast with the coefficient of the singular part. \par
The derivation of the equality of exponents above and below $ T_{c}$, relies
on the existence of a path which avoids the critical point, along which the
correlation functions are regular, and the RG equations everywhere satisfied.
\medskip
{\it Remark.} The universal functions characterizing the behaviour of
correlation functions in the critical domain still depend on the
normalization of physical parameters $t$, $H$, $M$, distances or momenta.
Quantities which are independent of these normalizations are
universal pure numbers. Simple examples are provided by the ratios of
the amplitudes of the singularities above and below $T_c$ like
$A^+/A^-$ for the specific heat. 
\medskip
{\it The $O(N) $-symmetric $N$-vector model.} 
We now indicate a few specific properties of models in which the action 
has a continuous $O(N)$ symmetry. \par
The differences concern correlation functions in a field or below $ T_{c}$.
The addition of a magnetic field term in an $ O (N) $ symmetric
action has various effects. \par
First, the magnetization and the magnetic field are now vectors. The
RG equations have exactly the same form as the Ising-like  $N=1$ case
but the scaling forms derived previously apply to the modulus of these
vectors. \par
Second, since magnetic field or magnetization distinguish a direction in
vector space, there now exist $ 2^n $ $n$-point functions, each spin
being either along the magnetization or orthogonal to it. 
When the continuous $O(N)$ symmetry of the action is broken linearly
in the dynamical variables (as in the case of a magnetic field)
these different correlation functions are related by a set
of identities, called WT identities. The simplest one involves the 2-point
function $ \Gamma_{\rm T}^{-1} $, at zero momentum, of the components
orthogonal to $\bf M$, i.e.~the transverse susceptibility $ \chi^{}_{\rm T}$
$$ \Gamma_{\rm T} \left(p=0 \right)= \chi^{-1}_{\rm T}=H / M\,. \eqnn $$
It follows that if
$ H $ goes to zero below $ T_{c}$, $ H/M $ and therefore $ \Gamma_{\rm T} $ at
zero momentum vanish. The latter property implies the existence of $N-1$
(massless) Goldstone modes corresponding to the spontaneous breaking of
the $ O(N) $ symmetry. \par
Note finally that the inverse longitudinal 2-point function $
\Gamma_{\rm L}(p) $ has IR singularities at zero momentum in zero field
below $ T_{c} $ generated by the
Goldstone modes. This is characteristic of continuous symmetries, and will
play an essential role in next section.
\subsection Four dimensions: logarithmic corrections and triviality

Let us just briefly comment about the situation in four dimensions.
If we solve the RG equation \sslbl\ssGRGivD
$$\lambda{ \d  \over \d  \lambda} g (\lambda)=\beta \bigl(g(\lambda)
\bigr),$$  
for the running coupling constant, assuming that $\beta(g)$ remains positive
for all $g>0$ (no non-trivial fixed point), we find that $g(\lambda)$ goes
to zero logarithmically; the operator $\phi^4$ is marginally irrelevant.
Writing generally
$$\beta(g)=\beta_2 g^2+\beta_3 g^3+O(g^4),\quad \beta_2>0\,,$$
we find for $\lambda\to 0$:
$$\ln\lambda=-{1\over\beta_2 g(\lambda)}-{\beta_3\over\beta_2^2} \ln
g(\lambda) +K(g), \eqnn $$
with:
$$K(g)={1\over\beta_2 g}-{\beta_3\over\beta_2^2} \ln g -\int_0^g \d g'
\left({1\over\beta(g')}-{1\over\beta_2 g'{}^2}+{\beta_3\over\beta_2^2 g'}
\right) .$$
Since the running coupling constant goes to zero in the long distance
limit, quantities can be calculated from perturbation theory. From the point
of view of critical phenomena logarithmic corrections to mean field theory
are generated.\par
Finally let us note that empirical evidence coming from lattice calculations 
strongly suggests the absence of any other fixed point.\par 
From the point of view of particle physics one faces the {\it triviality}\/
problem: for any initial bare coupling constant $g$ the renormalized coupling
$g(\mu/\Lambda)$ at scale $\mu$ much smaller than the cut-off $\Lambda$
behaves like
$$g(\mu/\Lambda)\sim{1\over\beta_2 \ln (\Lambda/\mu)} ,\eqnn $$
Therefore if one insists in sending the cut-off to infinity one finds a
free (trivial) field theory. However in the modern point of view of
{\it effective}\/ field theories, one accepts the idea that quantum
field theories may not be consistent on all scales but only in a
limited range. Then the larger is the range the smaller is the low
energy effective coupling constant. In the standard model these
comments may apply to the weak-electromagnetic sector which contains a
$\phi^4$ interaction and trivial QED.
\beginbib

The modern formulation of the RG ideas is due to\rf 
K.G. Wilson, {\it Phys. Rev.} B4 (1971) 3174, 3184;
\nrf and presented in an expanded form in\rf
K.G. Wilson and J. Kogut, {\it Phys. Rep.} 12C (1974) 75.
\nrf The idea of RG transformations was earlier proposed in a simplified form
in\rf 
L.P. Kadanoff, {\it Physics} 2 (1966) 263.
\nrf The systematic classification of operators can be found in\rf
F.J. Wegner, {\it Phys. Rev.} B5 (1972) 4529, B6 (1972) 1891, and in Wegner's
contribution to {\it Phase Transitions and Critical Phenomena,} vol. 6, C.
Domb and M.S. Green eds. (Academic Press, London 1976).
\nrf The idea of the $\varepsilon$-expansion is due to\rf
K.G. Wilson and M.E. Fisher, {\it Phys. Rev.  Lett.} 28 (1972) 240.
\nrf After Wilson's original articles, several authors have realized that the
RG equations derived in renormalized quantum field theories, could be applied
to Critical Phenomena\rf
C. Di Castro, {\it Lett. Nuovo Cimento}. 5 (1972) 69;
G. Mack, {\it Kaiserslautern 1972}, Lecture Notes in Physics  vol. 17, W.
Ruhl and A. Vancura eds. (Springer Verlag, Berlin 1972);
E. Br\'ezin, J.C. Le Guillou and J. Zinn-Justin, {\it Phys. Rev.} D8 (1973)
434, 2418; 
P.K. Mitter, {\it Phys. Rev.} D7 (1973) 2927;
G. Parisi, {\it Carg\`ese Lectures 1973}, published in {\it J. Stat. Phys.} 23
(1980) 49;  
B. Schroer, {\it Phys. Rev.} B8 (1973) 4200;
C. Di Castro, G. Jona-Lasinio and L. Peliti, {\it Ann. Phys. (NY)} 87 (1974)
327;  
F. Jegerlehner and B. Schroer, {\it Acta Phys. Austr. Suppl.} XI (1973) 389
(Springer-Verlag, Berlin).
\nrf The RG equations for the bare theory were first derived in\rf
J. Zinn-Justin, {\it Carg\`ese Lectures 1973}, unpublished, incorporated in
the review\rf
E. Br\'ezin, J.C. Le Guillou and J. Zinn-Justin in {\it
Phase Transitions and Critical Phenomena} vol. 6, C. Domb and M.S. Green
eds. (Academic Press, London 1976). \nrf
Recent work in this area is mainly devoted to a more accurate calculations of
universal quantities, using both the $\varepsilon$-expansion and the
perturbative expansion of the massive $\phi^4$ theory in three dimensions. See
for example\rf
R. Guida and J. Zinn-Justin, {\it Nucl.Phys.} B489 (1997) 626, hep-th/9610223;
{\it Critical Exponents of the N-vector model}, to appear in {\it J. Phys.} A,
cond-mat/9803240; 

S. A. Larin, M. Moennigmann, M. Stroesser, V. Dohm, {\it Phys.
Rev.} B58 (1998), cond-mat/9711069 and cond-mat/9805028;

M. Caselle, M. Hasenbusch, {\it J.Phys.} A31 (1998) 4603, cond-mat/9711080;
{\it Nucl.Phys.Proc.Suppl.} 63 (1998) 613, hep-lat/9709089; 
{\it J.Phys.} A30 (1997) 4963, hep-lat/9701007;
J. Engels, T. Scheideler, (Bielefeld U. preprint BI-TP-98-23, hep-lat/9808057;
A. I. Sokolov, E. V. Orlov, V. A. Ul'kov, S. S. Kashtanov,
{\it Universal effective action for $O(n)$-symmetric $\lambda \phi^4$ model
from renormalization group}, hep-th/9808011;
B. N. Shalaev, S. A. Antonenko, A. I. Sokolov,
{\it Phys. Lett. A} 230 (1997) 105-110, cond-mat/9803388;

G. M\"unster and J. Heitger, {\it Nucl. Phys.} B424 (1994)
582; C. Gutsfeld, J. K\"uster and G. M\"unster, {\it Nucl. Phys.} B479
(1996) 654, cond-mat/9606091;

P. Butera and M. Comi, hep-lat/9703017;  {\it Phys. Rev.} B56 (1997) 8212,
hep-lat/9703018

A.K. Rajantie, {\it Nucl. Phys.} B480 (1996) 729, hep-ph/9606216;

A. Pelissetto, E. Vicari, cond-mat/9805317; {\it Nucl.Phys.} B522 (1998)
605, cond-mat/9801098;

S. Caracciolo, M.S. Causo, A. Pelissetto
{\it Nucl.Phys.Proc.Suppl.} 63 (1998) 652, hep-lat/9711051;
{\it Phys. Rev.} E57 (1998) 1215, cond-mat/9703250

H. Kleinert, S. Thoms, V. Schulte-Frohlinde, quant-ph/9611050;
H. Kleinert, V. Schulte-Frohlinde, {\it Phys. Lett.} B342 (1995) 284,
cond-mat/9503038;

J. Rudnick, W. Lay and D. Jasnow, preprint cond-mat/9803026.
\nrf
Finally let us mention that exact bare RG equations have been proven by using
the method of integration over short distance modes. The problem has a long
story. Earlier work includes\rf
F.J. Wegner and A. Houghton, {\it Phys. Rev.} A8 (1973) 401; J.F. Nicoll {\it
et al}, {\it Phys. Rev. Lett.} 33 (1974) 540,\nrf
but more recent developments have been induced by the exact continuum RG
equations derived by\rf
J. Polchinski, {\it Nucl. Phys.} B231 (1984) 269.
\nrf
For more recent applications of the method see for example the review \par
T.R. Morris, {\it Elements of the Continuous Renormalization Group},
hep-th/9802039,
\nrf and references therein. A few additional references are\rf
J.F. Nicoll and T.S. Chang, {\it Phys. Lett.} 62A (1977) 287;
A. Hasenfratz and P. Hasenfratz, {\it Nucl. Phys.} B270 (1986) 687;
C. Wetterich, {\it Phys. Lett.} B301 (1993) 90; M. Bonini, M. D'Attanasio and
G. Marchesini, {\it Nucl. Phys.} B409 (1993) 441, B444 (1995) 602;
S. Seide, C. Wetterich, Heidelberg U. preprint HD-THEP-98-20, 
cond-mat/9806372.  

\endbib 
\vfill\eject
% n.l.$\sigma$-model
\def\pib{\pi}
\def\omegab{\omega}
\section{The $O(N)$ Spin Model at Low Temperature: the Non-Linear
$\sigma$-Model}   

Let us again consider the lattice model \eqns{\eZspin,\ehamspin}
of section \scGRG~with partition function:\sslbl\scLTs 
$$ Z= \int \prod_{i} \d  {\bf S}_{i}\,\delta \left( {\bf S}^{2}_{i}-1
\right)\exp \sum_{ij}V_{ij}{\bf S}_{i}\cdot {\bf S}_{j}/T \,.  $$
We will now discuss this model from the point of view of a low
temperature expansion. The methods we employ, however, apply only to
continuous symmetries, here to $N\ge 2$. They rely on the property that models
with continuous symmetries, in contrast to models with discrete symmetries,
have a non-trivial long distance physics at any temperature below $T_c$, due
to the massless Goldstone modes. \par
We first prove universal properties of the low temperature, ordered, phase at
fixed temperature. Then, in the non-abelian case, $N>2$, we show that
additional information about critical properties can be obtained, by analyzing
the instability of the ordered phase at low temperature and near two
dimensions, due to Goldstone mode interactions. 
\par 
The analysis is based on the following observation: The  $N$-vector model
\eqns{\eZspin,\ehamspin} can be considered as a lattice regularization of the
non-linear $ \sigma $-model (note $2{\bf S}_i\cdot {\bf S}_j=2-({\bf S}_i-
{\bf S}_j)^2$). The low temperature expansion of the lattice model is the
perturbative expansion of the regularized field theory. The field theory is
renormalizable in dimension two. RG equations, valid in two and 
more generally $2+\varepsilon$ dimension follow. Their solutions will help us
to understand the long distance behaviour of correlation functions.
\par
It is somewhat surprising that two different continuum field theories, the
$(\phib^2)^2$ and the non-linear $\sigma$-model describe the long distance
physics of the same lattice model. This point will be clarified by
an analysis of the $1/N$-expansion of both field theories. This property,
totally mysterious at the classical level, emphasizes the essential nature of
quantum (or statistical) fluctuations.  
\subsection The non-linear $\sigma$-model

We now study the non-linear $\sigma$-model from the point of view of 
renormalization and renormalization group. In continuum notation the field
${\bf S}(x)$ has unit length and the action is \sslbl\ssLTsnls
$${\cal S}({\bf S})={1\over 2t}\int\d^d x\,\partial_\mu{\bf S}(x)\cdot
\partial_\mu{\bf S}(x) ,$$
where $t$ is proportional to the temperature $T$.
To generate perturbation theory we
parametrize the field ${\bf S}(x)$:
$${\bf S}(x) = \{ \sigma(x) , \pib(x) \}, $$
and eliminate locally the field $\sigma(x)$ by:
$$\sigma(x) = \left(1-\pib^2(x)\right)^{1/2}.$$
This parametrization is singular but this does not show up in perturbation
theory which assumes $\pi(x)$ small.
\medskip
{\it The $O(N)$ symmetry.}
The $O(N-1)$ subgroup which leaves the component $\sigma$ invariant 
acts linearly on the $N-1$ component vector $\pi$. However a general $O(N)$
transformation will transform $\pib$ into a linear combination of $\pib$ and
$\sqrt{1-\pib^2}$. The $O(N)$ symmetry is realized non-linearly.
An infinitesimal transformation corresponding to the generators of $O(N)$ not
belonging to $O(N-1)$ takes the form
$$\delta \pib=\omegab \sqrt{1-\pib^2}, $$
where $\omegab$ is a $N-1$ component vector of parameters corresponding to
these generators. 
\smallskip
As we have done for the $ \left( \phib^{2} \right)^{2} $
model, we  scale all distances in order to measure momenta in units of the
inverse lattice spacing $\Lambda $. We thus write the partition function:
$$ Z = \int \left[\bigl(1- \pib^2(x)\bigr)^{-1/2}\d\pib(x)
\right] \exp\left[-S(\pib)\right] ,\eqnn $$
with
$${\cal S}(\pib)={\Lambda^{d-2} \over 2t} \int \d^{d}x \,
g_{ij}(\pib)\partial_\mu \pib_i (x)\partial_\mu \pib_j (x),\eqnn $$ 
where $g_{ij}$ is the metric on the sphere
$$g_{ij}=\delta_{ij}+{\pib_i\pib_j\over 1-\pib^2}.\eqnn $$
Moreover, as expected, the functional measure is related to the
metric by
$$\sqrt{\det(g_{ij})}={1\over \sqrt{1-\pib^2}}\,.$$
\medskip
{\it Propagator, perturbation theory and power counting.}
Unlike the $\phi^4$ field theory, the action is non-polynomial in the fields.
An expansion of the action in powers of $\pib$ generates an infinite number of
interactions. However we note that the power of $t$ in front of a diagram
counts the number of loops. Therefore at a finite loop order, only a finite
number of interactions contribute.\par
The $\pib$ propagator is proportional to: 
$$\Delta_\pib(k)={t \Lambda^{2-d}\over k^2}, $$
The $\pib$ thus has the usual canonical dimension $(d-2)/2$. Since we have
interactions with arbitrary powers of $\pib$, the model is renormalizable 
in two dimensions, where all interactions have dimension two.
\medskip
{\it The role of the functional measure.}
If we try to write the functional measure as an additional interaction
we find
$$\prod_x {1\over\sqrt{1-\pib^2(x)}}=\exp\bigl(-\ud \sum_x
\ln(1-\pib^2(x)\bigr). $$
This quantity is well-defined on the lattice but not in the continuum. This
problem, which already appears in quantum mechanics ($d=1$) reflects the
necessity of a lattice regularization to precisely define the quantum
hamiltonian in the presence of interactions with derivatives. A perturbative
solution is provided by dimensional regularization, where this term can simply
be omitted. In lattice regularization it cancels quadratic divergences.
\medskip
{\it IR divergences, spontaneous symmetry breaking and the role of
dimension two.} We see
that the perturbative phase of the non-linear $\sigma$ model is automatically
a phase in which the $ O(N) $ symmetry is  spontaneously
broken, and the $(N-1) $ components of $ {\bf S}(x) $, $\pib(x)$,
are massless Goldstone modes. 
\smallskip
(i) For $ d\leq 2 $ we know from the  Mermin--Wagner theorem that SSB with
ordering $ \left( \left<{\bf S} \right> \not= 0 \right) $ is impossible in a
model with continuous symmetry and short range forces. Correspondingly
IR divergences appear in the perturbative expansion of the non-linear $\sigma$
model for $d\le2$, for example $\left<\sigma\right>$ diverges at order $t$ as
$\int \d^d p/p^2$. For $ d\leq 2 $ the
critical temperature $ T_{c} $ vanishes and perturbation theory makes sense
only in presence of an IR cut-off which breaks explicitly the symmetry and 
orders the spins (thus selecting a classical minimum of the action). Therefore
nothing can be said about the long distance 
properties of the unbroken theory directly from perturbation theory. 
\par
(ii) For $d> 2$ instead, perturbation theory which predicts spontaneous
symmetry breaking (SSB), is not IR divergent. This is consistent with the
property that in the $N$-vector model, for $ d>2 $, the $O(N)$  symmetry is
spontaneously broken at low temperature. At $ T<T_{c} $ fixed, the large
distance behaviour of the theory is  dominated by the massless or spin wave
excitations. On the other hand nothing can be said, in perturbation theory, of
a possible critical region $ T\sim T_{c} $.
\smallskip
To go somewhat beyond perturbation theory we shall use field theory RG
methods. It is therefore necessary to first define the model in two dimensions
where it is renormalizable. There IR divergences have to be dealt
with. We thus introduce an IR cut-off in the form of a magnetic
field in the $\sigma$ direction (a constant source for the $\sigma$ field)
$${\cal S}(\pib,h)={\Lambda^{d-2} \over t} \int \d^{d}x\left\{{1 \over 2}
\left[ \left(\partial_{\mu} \pib (x )\right)^{2}+{ \left( \pib \cdot
\partial_{\mu} \pib (x) \right)^{2} \over 1- \pib^2(x)} \right]
-h \sqrt{ 1- \pib^{2}(x)} \right\}. \eqnd\eactpih $$ 
Expanding the additional term in powers of $\pib$ we see that it generates a
mass term
$$\Delta_\pib(k)={t \Lambda^{2-d}\over k^2+h}, $$
and additional interactions of dimension 0 in $d=2$.\par
We then proceed in formal analogy with the case of the $ \left( \phib^{2}
\right)^{2} $ field theory, i.e.\ study the theory in $ 2+\varepsilon $
dimension as a double series expansion in the temperature $ t$ and $
\varepsilon $. In this way the perturbative expansion is renormalizable
and RG equations follow.
\subsection RG equations

Using power counting and some non-trivial WT identies (quadratic in the 1PI 
functional) one can show that the renormalized action takes the form:     
$$ {\cal S}_\r (\pib_\r, h_\r) = {\mu^{d-2}Z \over 2t_\r Z_{t}} \int
\d^{d}x \left[ \left(\partial_{\mu} \pib_\r  \right)^{2}+ \left(\partial_\mu
\sigma_\r  \right)^{2} \right]  -{\mu^{d-2} \over t_\r }h_\r  \int 
\sigma_\r  (x) \d ^{d}x\,, \eqnd{\eactsigr} $$
in which $ \mu $ is the renormalization scale and:
$$ \sigma_\r  (x)= \left[ Z^{-1}- \pib^{2}_\r  \right]^{1/2}. \eqnn $$ 
Note that the renormalization constants can and thus will be chosen $h$
independent. This is automatically realized in the minimal subtraction
scheme.\par 
The relation:
$$ \pib_\r  (x)=Z^{-1/2} \pib (x),\eqnn $$
implies
$$\mu^{d-2}{ h_\r  \over t_\r }=\Lambda^{d-2}Z^{1/2} {h \over t}.
\eqnd{\ehrenor} $$
With our conventions the coupling constant, which is proportional to
the temperature, is dimensionless. The relation between the cut-off dependent
and the renormalized correlation functions is:
$$ Z^{n/2} \left(\Lambda/ \mu ,t \right)\Gamma^{(n)} \left(p
_{i};t,h,\Lambda \right)=\Gamma^{(n)}_\r  \left(p_{i};t_\r ,h
_\r ,\mu \right). \eqnn $$
Differentiating with respect to $ \Lambda $ at renormalized
parameters fixed, we obtain the RG equations:
$$\left[ \Lambda{ \partial \over \partial
\Lambda} +\beta (t){\partial \over \partial t}-{n \over 2}\zeta
(t)+ \rho(t) h{\partial \over \partial h}
\right] \Gamma^{(n)} \left(p_{i};t,h,\Lambda \right)=0\, ,
\eqnd{\eRGsigma}  $$ 
where the RG functions are defined by: 
$$\left.\eqalign{ \beta(t)&=\left.\Lambda{ \partial \over \partial \Lambda}
\right\vert _{\rm ren.\, fixed}t \,,  \cr
 \zeta (t)&=\left.\Lambda{ \partial \over \partial \Lambda} \right\vert_{ {\rm
ren.\, fixed}} \left(-\ln Z \right) \,,  \cr
\rho(t)&=\left.\Lambda{ \partial \over \partial \Lambda} \right\vert
_{\rm ren.\, fixed}\ln h \,.  \cr}\right. \eqnn $$
The coefficient of $ \partial /\partial h $ can be
derived from equation \ehrenor\ which implies (taking the logarithm of both
members): 
$$ 0=h^{-1}\Lambda{ \partial \over \partial \Lambda} h+d-2-{1 \over
2}\zeta (t)-{\beta (t) \over t}, \eqnn $$
and therefore:
$$\rho(t)=2-d+{1 \over 2}\zeta (t)+{\beta (t) \over
t}. \eqnn $$
To be able to  discuss correlation functions involving the
$ \sigma $-field, we also need the RG equations satisfied by connected
correlation functions $ W^{(n)} $:
$$ \left[ \Lambda{ \partial \over \partial
\Lambda} +\beta (t){\partial \over \partial t}+{n \over 2}\zeta
(t) + \left({1 \over 2}\zeta (t)+{\beta
(t) \over t}-\varepsilon \right)h{\partial \over \partial h}
\right] W^{(n)}=0\,, \eqnd{\eRGsigWn} $$
in which we now have set:
$$ d=2+\varepsilon\, . \eqnn $$
The two RG functions can be obtained
at one-loop order from a calculation of the 2-point function $ \Gamma^{(2)}$: 
$$ \Gamma^{(2)} \left(p \right)={\Lambda^{\varepsilon} \over t} \left(p
^{2}+h \right)+ \left[ p^{2}+\ud \left(N-1 \right) h
\right]{ 1 \over \left(2\pi \right)^{d}} \int^{\Lambda}{ \d^{d}q \over
q^{2}+h}+O (t)\,. \eqnn $$ 
Applying the RG equation \eRGsigma\ to $ \Gamma^{(2)} $ and
identifying the coefficients of $ p^{2} $ and $ h $, we derive two
equations which determine $ \beta (t) $ and $ \zeta \left(t\right) $ at
one-loop order 
\eqna\eRGfusig
$$ \eqalignno{ \beta (t) & = \varepsilon t-{ \left(N-2 \right)
\over 2\pi} t^{2}+O \left(t^{3},t^{2}\varepsilon \right), & \eRGfusig{a}
%\eqnnd{\ebetasig}
\cr \zeta (t) & = { \left(N-1 \right) \over 2\pi} t+O
\left(t^{2},t\varepsilon \right). & \eRGfusig{b}  \cr} $$  
\subsection Discussion of the RG flow

From the expression of $ \beta (t) $ in equation \eRGfusig{a}
we immediately conclude:\sslbl\ssLTsRG \par
For $ d\leq 2 \left(\varepsilon \leq 0 \right) $, $ t=0 $ is an
unstable IR fixed point, the IR instability being induced by the vanishing
mass of the would-be Goldstone bosons. The
spectrum of the theory thus is not given by perturbation theory and the
perturbative assumption of spontaneous symmetry breaking at low temperature is
inconsistent. As mentioned before, this result agrees with rigorous arguments.
Note that since the model depends only on one coupling constant, $t=0$ is also
a UV stable fixed point (the property of large momentum asymptotic freedom).
Section \ssLTsiiD\ contains a short discussion of the physics in two
dimensions for $N>2$. The abelian case $ N=2$ is special and has to be
discussed separately. \par 
For $ d>2$, i.e.\ $\varepsilon >0$, $ t=0 $ is a stable IR
fixed point, the $ O(N) $ symmetry is spontaneously broken at low
temperature in zero field. The effective coupling constant, which
determines the large distance behaviour, approaches the origin for all
temperatures  $ t < t_{c} $, $t_{c}$ being the first non-trivial zero of
$\beta(t)$. Therefore the large distance properties of the
model can be obtained from low temperature expansion and renormalization
group, replacing the perturbative parameters by effective parameters obtained
by solving the RG equations. 
\medskip
{\it The critical temperature.} Finally we observe that, at least for $
\varepsilon $ positive and small, and $N>2$, the RG function $ \beta (t) $
has a non-trivial zero $t_{c}$:  
$$  t_{c}={2 \pi \varepsilon \over N-2}+O \left(\varepsilon
^{2} \right)\Rightarrow \beta \left(t_{c} \right)=0\,,{\rm\ and}\quad
\beta' \left(t_{c} \right) =-\varepsilon +O \left(\varepsilon^{2}
\right).\eqnn $$
Since $ t_{c} $ is an unstable IR fixed point, it is by
definition a critical temperature. Consequences of this property 
are studied below. Let us only immediately note that $ t_{c} $ is also a UV
fixed point, i.e.\ it governs the large momentum behaviour of the renormalized
theory. The large momentum behaviour of correlation functions is not given by
perturbation theory but by the fixed point. As a consequence the perturbative
result that the theory cannot be rendered
finite for $ d>2 $ with a finite number of renormalization constants, cannot
be trusted. \par
We now discuss more precisely the solutions of the RG equations.
\subsection Integration of the RG equations

We first examine the implications of the RG equations for the
large distance behaviour of correlation functions for $d>2$ where $t=0$ is an
IR fixed point. Equation \eRGsigma~can be solved as usual by 
the method of characteristics, i.e.~by introducing a scaling parameter
$\lambda$ and running parameters. It is here convenient to proceed somewhat
differently by looking for a solution of the form
$$ \Gamma^{(n)} \left(p_{i};t,h,\Lambda \right)=\xi^{-d}(t) M_{0}^{-n}
(t)F^{(n)} \bigl(p_{i}\xi(t),h/h_0(t) \bigr). \eqnd\eRGsolve $$
The ansatz \eRGsolve~solves the RG equations provided the unknown functions
$M_0(t)$, $\xi(t)$ and $h_0(t)$ are chosen to be
$$ \eqalignno{ M_{0} (t) & = \exp\left[-{1 \over 2} \int^{t}
_{0}{\zeta \left(t' \right) \over \beta \left(t'
\right)} \d  t'\right] , & \eqnd{\eMzero} \cr \xi (t) & =
\Lambda^{-1}t^{1/\varepsilon}\exp\left[\int^{t}_{0} \left({1 \over
\beta \left(t' \right)}-{1 \over \varepsilon t'} \right)
\d  t'\right] , & \eqnd{\ecsidef} \cr }
$$ 
with then
$$h_0(t)=t M_{0}^{-1}(t)\xi^{-d} (t)\Lambda^{2-d}.\eqnn $$
Note that the function $\xi(t)$ has in zero field the
nature of a correlation length. \par
For the connected correlation functions the same analysis leads to:
$$ W^{(n)}\left(p_{i};t,H,\Lambda \right)=\xi^{d(n-1)}(t)
M_{0}^{n}(t)G^{(n)} \bigl(p_{i}\xi(t),h/h_0(t)\bigr). \eqnd{\eRGWnsol} $$
\par
It is  convenient to also introduce the function $K(t)$
$$ K(t)=M_0(t) \left[\Lambda\xi(t)\right]^{d-2}/t=1+O(t). \eqnd\edefK $$
Combining equation \eRGsolve~with dimensional analysis we can rewrite the
scaling relation in an equivalent form 
$$\eqalignno{\Gamma^{(n)}(p_i,t,h,\Lambda)&\sim M_0^{-n}(t)[K(t)h]^{d/2} &\cr
&\quad \times \Gamma^{(n)}\left({p_i \over[K(t)h]^{1/2}},{t 
\left[K(t)\right]^{d/2} \over M_0(t)}
\left({h\over \Lambda^2}\right)^{(d-2)/2}, 1,1\right).&\eqnd\eRGsigsolam
\cr}$$  
Let us apply this result to the determination of the singularities near the
coexistence curve, i.e.\ at $t$ fixed below the critical temperature when the
magnetic field $h$ goes to zero. 
\medskip
{\it The coexistence curve.} The magnetization is given by
$$M(t,h,\Lambda)\equiv\left<\sigma(x)\right>=
\Lambda^{-\varepsilon}t{\partial \Gamma^{(0)} \over  \partial h}
,\eqnd{\eRGsigM} $$ 
($ \Gamma^{(0)} $ is the magnetic field dependent free energy). 
At one-loop order in a field one finds
$$M=1-{N-1 \over 2}\Lambda^{-\varepsilon}  t { 1 \over \left(2\pi \right)^{d}}
\int^{\Lambda}{ \d^{d}q \over q^{2}+h}+O\left(t^2\right).$$ 
Thus from relation \eRGsigsolam~follows
$$M(t,h,\Lambda=1)=M_0(t)-{N-1 \over 2}t\left[K(t)\right]^{d/2}
h^{(d-2)/2} {\Gamma(1-d/2) \over(4\pi)^{d/2}}+O\left(h,h^{d-2}\right).$$   
This result shows that $M_0(t)$ is the spontaneous magnetization and 
displays the singularity of the scaling equation of state (section \ssCDeqst) 
on the coexistence curve ($H=0$, $T<T_c$) for $N>1$, and in all dimensions
$d>2$.  
\medskip
{\it The equation of state in the critical domain.}
Let now instead use the scaling form \eRGsolve
$$ M=\Lambda^{2-d}t{\partial \Gamma^{(0)} \over \partial h}=M_{0} (t)F^{(0)}
\bigl(h/h_0(t)\bigr). \eqnn $$ 
Inversion of this relation yields the scaling form of the equation of state:
$$ h= h_0(t) f\left({M\over M_{0}(t)} \right), \eqnd{\estatsig} $$
and the 1PI correlation functions can thus be written in terms of the
magnetization as:
$$ \Gamma^{(n)} \left(p_{i},t,M,\Lambda \right)=\xi^{-d} \left(t
\right)M_{0}^{-n} (t)F^{(n)} \left(p_{i}\xi \left(t
\right),{M \over M_{0} (t)} \right). \eqnd{\eRGmgsol} $$
The equations \eqns{\estatsig,\eRGmgsol} are consistent with
the equations  \eqns{\escmgfld,\egaMTsc}: the appearance of two different
functions $\xi(t)$ and $M_0(t)$ corresponds to the existence of two
independent critical exponents $\nu,\beta$ in the $(\phib^2)^2$ field theory.
They extend, in the large distance limit, the  scaling form  of correlation
functions, valid in the critical region, to all temperatures below $t_{c}$.
There is however one important difference between the RG equations of the
$\left(\phib^{2}\right)^{2}$ theory and of the $\sigma$-model: the
$\left(\phib^{2}\right)^{2}$ theory depends on two coupling constants, the
coefficient of $\phib^{2}$ which plays the role of the temperature, and the
coefficient of $\left(\phib^{2}\right)^{2}$ which has no equivalent here. The
correlation functions of the continuum  $\left(\phib^{2}\right)^{2}$ theory
have the exact scaling form \eRGmgsol\ only at the IR fixed point. In
contrast, in the case of the $\sigma$-model, it has been possible to eliminate
all corrections to scaling corresponding to irrelevant operators order by
order in perturbation theory. We are therefore led to a remarkable
conclusion: the correlation functions of the $O(N)$ non-linear model are
identical to the correlation functions of the  $\left(\phib^{2}\right)^{2}$
field theory  at the IR fixed point. This conclusion is supported by the
analysis of the scaling behaviour performed within the $1/N$ expansion (see
equation \epartsig). 
\medskip
{\it Critical exponents.} Let us now study more precisely
what happens when $t$ approaches $ t_{c} $ (for $N>2$). The function $\xi(t)$
diverges as:   
$$ \xi (t)\sim \Lambda^{-1} \left(t_{c}-t \right)^{1
\left/\beta' \left(t_{c} \right) \right.}. \eqnn $$
We conclude that the correlation
length exponent $\nu$ is given by
$$ \nu =-{1 \over \beta' \left(t_{c} \right)}. \eqnn $$
For $d$ close to 2 the exponent $ \nu $ thus behaves like:
$$ \nu \sim 1/\varepsilon\,. \eqnn $$
The function $ M_{0} (t) $ vanishes at $ t_{c} $:
$$ \ln M_{0} (t)=-{1 \over 2}{\zeta \left(t_{c} \right) \over
\beta' \left(t_{c} \right)}\ln \left(t_{c}-t \right)+\,
\hbox{  const.}\ .\eqnn $$
This yields the exponent $\beta$ and thus also $\eta$ through the scaling
relation $\beta=\ud \nu(d-2+\eta)$: 
$$ \eta =\zeta \left(t_{c} \right)-\varepsilon\, . \eqnn $$
A leading order we find:
$$ \eta ={\varepsilon \over N-2}+O \left(\varepsilon^{2} \right).
\eqnd{\etasig} $$
We finally note that the singularity of $ \Gamma^{(n)} $
coming from the prefactor $ \xi^{-d}M_{0}^{-n} $ indeed agrees near $ t_{c}
$ with the result of equation \egastrsc. \par
Consideration of operators with four derivatives allows also to calculate the
exponent $\omega$ which characterizes leading corrections to scaling. One
finds 
$$\omega=4-d-2\varepsilon/(N-2)+O(\varepsilon^2)\,.$$
\medskip
{\it The nature of the correlation length $\xi(t)$.} The length scale $\xi(t)
$ is a cross-over scale between two different behaviours of correlation
functions. For distances large compared to $ \xi (t) $, the behaviour of
correlation functions is governed by the Goldstone modes (spin wave
excitations) and can thus be deduced from the perturbative low temperature
expansion. However when $ t $ approaches $ t_{c}$, $\xi(t) $ becomes large.
There then exist distances large with respect to the microscopic scale but
small with respect to $ \xi(t) $ in which correlation functions have a
critical behaviour. In this situation we can construct continuum 
correlation functions consistent on all scales, the critical behaviour being
also the large momentum behaviour of the renormalized field theory.
\medskip
{\it General comment.} From the consideration of the low temperature
expansion we have been able to describe, for theories with a continuous
symmetry, not only the complete structure of the low temperature phase, and
this was expected, but also, in the non-abelian case, the critical behaviour
near two dimensions . \par 
This result is somewhat surprising: Indeed perturbation theory
is only sensitive to the local structure of the sphere $ {\bf S}^{2}=1
$ while the restoration of symmetry involves the sphere globally. 
This explains the peculiarity of the abelian case $ N=2 $ because locally a
circle cannot be distinguished from a non-compact straight line. For $ N>2 $
the sphere has instead a local characteristic curvature. Still different
regular compact manifolds may have the same local metric, and therefore the
same perturbation theory. They all have the same low temperature physics.
However the previous results concerning the critical behaviour are 
physically relevant only if they are still valid when $\varepsilon$ is not
infinitesimal and $t$ approaches $t_{c}$, a condition which cannot be checked
directly. In particular the low temperature expansion misses in  general
terms decreasing like $\exp\left({\rm const.}/t\right)$ which may in some
cases be essential for the physics. Therefore in section \sssfivNRT~we
will establish a direct relation between the $O(N)$ $\sigma$
model and the $\left(\phib^{2}\right)^{2}$ field theory to all orders
in the large $N$ expansion.
This gives us some confidence that the previous considerations are  valid for
the $N$-vector model at least for $N$ sufficiently large. On the other hand
the physics of $N=2$ is not well reproduced. Cardy and
Hamber have speculated about the RG flow for $N$ close to 2 and dimension $d$
close to 2, incorporating phenomenologically the Kosterlitz--Thouless
transition in their analysis.
\subsection The dimension two

Dimension two is of special interest from the particle physics point of view.
The RG function $ \beta (t) $ is then:\sslbl\ssLTsiiD
$$ \beta (t)=-{(N-2) \over 2\pi} t^{2}+O \left(t^{3} \right). \eqnn $$
The non-linear $ \sigma $-model for $ N>2 $ is the simplest example
of a so-called asymptotically free field theory (UV free) since the first
coefficient of the $ \beta $-function is negative, in contrast with the $
\phi^{4} $ field theory. Therefore the large momentum behaviour of correlation
functions is entirely calculable from perturbation theory and RG arguments.
There is, however a counterpart, the theory is IR unstable and thus, in zero
field $h$, the spectrum of the theory is not perturbative. Contrary to
perturbative indications, it consists of $ N $ massive degenerate states since
the $ O(N) $ symmetry is not broken. Asymptotic freedom and the
non-perturbative character of the spectrum are also properties of QCD, the
theory of strong interactions, in four dimensions, . 
\par  
If we now define a function $ \xi(t) $ by:
$$ \xi (t)=\mu^{-1}\exp\left[\int^{t}{ \d  t'\over \beta \left(t'
\right)}\right], \eqnn $$ 
we can again integrate the RG equations and we find that
$ \xi (t) $ is the correlation length in zero field. In addition
we can use the explicit expression of the $\beta$-function
to calculate the correlation length or the physical mass for small
$ t $: 
$$ \xi^{-1}(t)= m (t)= K \mu t^{-1/(N-2)} \e^{-2\pi /[(N-2)t]}\left(1 +
O(t)\right). \eqnn $$ 
However the exact value of the integration constant $K$, which 
gives the physical mass in the RG scale, can only be calculated by
non-perturbative techniques. \par  
Finally the scaling forms \eqns{\eRGsolve,\eRGWnsol} imply that the
perturbative 
expansion at fixed magnetic field is valid, at low momenta or large
distances, and for  $h/h_0(t)$ large.

\beginbib

Early references on the $\sigma$-model include:\rf
M. Gell-Mann and M. L\'evy, {\it Nuovo Cimento} 16 (1960) 705;
S. Weinberg, {\it Phys. Rev.} 166 (1968) 1568;
J. Schwinger, {\it Phys. Rev.} 167 (1968) 1432;
W.A.~Bardeen and B.W. Lee, {\it Nuclear and Particle Physics}, Montreal 1967,
B. Margolis and C.S.~Lam  eds. (Benjamin, New York 1969);
K. Meetz, {\it J. Math. Phys.} 10 (1969) 589.
\nrf The quantization was discussed in:\rf
I.S. Gerstein, R. Jackiw, B.W. Lee and S. Weinberg, {\it Phys. Rev.} D3 (1971)
2486;
J. Honerkamp and K. Meetz, {\it Phys. Rev.} D3 (1971) 1996;
J. Honerkamp, {\it Nucl. Phys.} B36 (1972) 130;
A.A. Slavnov and L.D. Faddeev, {\it Theor. Math. Phys.} 8 (1971) 843.
\nrf Pauli--Villars's regularization for chiral models has been proposed
by\rf 
A.A. Slavnov, {\it Nucl. Phys.} B31 (1971) 301.
\nrf The model was studied as a limit of the linear $\sigma$-model in\rf
D. Bessis and J. Zinn-Justin, {\it Phys. Rev.} D5 (1972) 1313.
\nrf The renormalization group properties were discussed in\rf
A.M. Polyakov, {\it Phys. Lett.} 59B (1975) 79;
E. Br\'ezin and J. Zinn-Justin, {\it Phys. Rev. Lett.} 36 (1976) 691; {\it
Phys. Rev.} B14 (1976) 3110;
W.A. Bardeen, B.W. Lee and R.E. Shrock, {\it Phys. Rev.} D14 (1976) 985.
\nrf The renormalizability in two dimensions was proven in\rf
E. Br\'ezin, J.C. Le Guillou and J. Zinn-Justin, {\it Phys. Rev.} D14 (1976)
2615; B14 (1976) 4976.
\nrf Higher order calculations of critical exponents are due to\rf
S. Hikami, {\it Nucl. Phys.} B215[FS7] (1983) 555;
W. Bernreuther and F.J. Wegner, {\it Phys. Rev. Lett.} 57 (1986) 1383;
F. Wegner, {\it Nucl. Phys.} B316 (1989) 663.
\nrf Composite operators are discussed in \rf
F.J. Wegner, {\it Z. Phys.} B78 (1990) 33; G.E. Castilla and S. Chakravarty,
{\it Phys. Rev. Lett.} 71 (1993) 384, cond-mat/960588.
\nrf Finally the 2D non-linear $\sigma$-model has recently been the subject of
extensive analytic \rf
P. Hasenfratz, M. Maggiore and F. Niedermayer, {\it Phys. Lett.} B245
(1990) 522; P. Hasenfratz and F. Niedermayer, {\it Phys. Lett.} B245
(1990) 529; J. Balog, M. Niedermaier,{\it Phys. Lett.} B386 (1996) 224,
hep-th/9604161; {\it Phys.Rev.Lett.} 78 (1997) 4151, hep-th/9701156;
{\it Nucl.Phys.} B500 (1997) 421, hep-th/9612039;
D.-S. Shin, {\it Nucl. Phys.} B496 (1997) 408, hep-lat/9611006; 
\nrf and numerical investigations, see for example\rf
A. Cucchieri, T. Mendes, A. Pelissetto, Alan D. Sokal, 
{\it J. Stat. Phys.} 86 (1997) 581, hep-lat/9509021; M. Hasenbusch,
{\it Phys. Rev.} D53 (1996) 3445; M. Hasenbusch, R.R. Horgan, {\it
Phys. Rev.} D53 (1996) 5075, hep-lat/9511004; T. Neuhaus, preprint WUB-96-30
hep-lat/9608043; N. Schultka,  cond-mat/9611043; 
M. Campostrini, A. Pelissetto, P. Rossi, E. Vicari,
{\it  Phys. Rev.} D54 (1996) 1782, hep-lat/9602011;
{\it  Phys. Lett.} B402 (1997) 141, hep-lat/9702010;       
S. Caracciolo, R.G. Edwards, A. Pelissetto, A.D.
Sokal, {\it Phys. Rev .Lett.} 75 (1995) 1891, hep-lat/9411009; {\it
Phys. Rev. Lett.} 76 (1996) 1179, hep-lat/9602013;
S. Caracciolo, R.G. Edwards, T. Mendes, A. Pelissetto, A.D. Sokal,
{\it Nucl. Phys. Proc. Suppl.} 47 (1996) 763, hep-lat/9509033;
G. Mana, A. Pelissetto, A.D. Sokal,
{\it Phys. Rev.} D54 (1996) 1252, hep-lat/9602015; 
B. Alles, A. Buonanno, G. Cella, {\it Nucl. Phys.} B500 (1997) 513, 
hep-lat/9701001;
B. Alles, G. Cella, M. Dilaver, Y. Gunduc, hep-lat/9808003.
\endbib

\vfill\eject
%large $N$ techniques
% add finite size effects
\def\vv{v}
\def\tu{\tau}
\def\ro{\rho}
\def\efigi{1}
\section{$\left(\phib \sp{2} \right) \sp{2} $ Field Theory and Non-Linear
$\sigma$ Model in the Large $N$ Limit}  
 
In the preceding sections
we have derived universal properties of critical systems within the
frameworks of the formal $ \varepsilon =4-d$ 
and $\varepsilon=d-2$ expansions (at least for $N>2$). It is therefore 
reassuring to verify, at least in some limiting case, that the results 
obtained in this way remain valid even when $ \varepsilon$ 
is no longer infinitesimal. We show in this section that, in the case 
of the
$O(N)$ symmetric $\left(\phib^{2} \right)^{2} $ field theory,  the
same universal properties can also be derived at fixed dimension 
in the large $N$ limit, and more generally order by order in the
large $ N $-expansion. We then examine the non-linear
$\sigma$-model in the same limit.
\sslbl\scfivN  
\subsection Introduction
 
We again consider the partition function:\sslbl\ssfivNi 
$$ Z= \int \left[ \d \phib (x) \right] \exp \left[-S(\phib)\right] ,
\eqnd\eONpart $$  
where $S(\phib)$ is the $ O(N) $ symmetric action
\eLGWphi~($u=\Lambda^{4-d}g$): 
$$ S \left( \phib \right)= \int \left\lbrace{ 1 \over 2} \left[
\partial_{\mu} \phib (x) \right]^{2}+{1 \over 2}r
\phib^{2} (x)+{u \over 4!} \left[ \phib^{2}
(x) \right]^{2} \right\rbrace \d ^{d}x\,. \eqnd{\eactON} $$ A
cut-off $ \Lambda $, consistent with the symmetry, is implied. \par
The solution of the model in the large $N$ limit is based on a idea of
mean field theory type: it can be expected that for $N$ large the
$O(N)$ invariant quantities self-average and therefore have small
fluctuations. Thus for example 
$$\left\langle \phib^{2}(x)\phib^{2}(y)\right\rangle
\mathop{\sim}_{N
\rightarrow \infty}\left\langle \phib^{2}(x)\right\rangle  \left\langle
\phib^{2}(y)\right\rangle .$$ 
This suggests to take $\phib^2(x)$ as a dynamical variable.
Technically, in the case of the $\left(\phib^{2} \right)^{2} $
theory, this can be achieved by using an identity similar to the
Hubbard transformation: 
$$\exp\left[{1 \over 2}r \phib^2+{u \over 4!} \left( \phib^2
\right)^2\right] \propto \int\d \lambda \exp \left(
{3 \over 2u}\lambda^2-{3r \over u}\lambda -{1 \over 2}\lambda
\phib^2\right) , \eqnd\eONHubb $$ 
where the integration contour is parallel to the imaginary axis. By
introducing a field $\lambda(x)$ the identity can be used for each
point $x$ inside the functional integral \eONpart. The new
functional integral is then gaussian in $\phib $ and the integral
over the field $ \phib $ can be performed. The dependence on $N$ of
the partition function becomes explicit. Actually it is convenient
to separate the components of $ \phib $ into one component $\sigma
$, and $ N-1 $ components $ \pib $, and integrate only over $ \pib $
(for $T<T_c$ it may even be convenient to integrate over only $N-2$
components). For $N$ large the difference is negligible. To generate
$\sigma$ correlation functions we also add a source $ H(x)$ to the
action
$$ Z (H)= \int \left[ \d \lambda (x)
\right] \left[ \d  \sigma (x) \right] \exp \left[ -S_N
\left(\lambda ,\sigma \right)+ \int \d^{d}x\, H (x)\sigma (x) \right] ,
\eqnd{\eZeff} $$
with:
$$ \vbox{\eqalignno{ S_N \left(\lambda ,\sigma \right)= &
\int \left[{ 1 \over 2} \left(\partial_{\mu}\sigma \right)^{2}
-{3 \over 2u}\lambda^2 (x) +{3r\over u}\lambda(x)+{1 \over 2}\lambda
(x) \sigma^{2} (x) \right] \d^{d}x & \cr &\quad +{ \left(N-1
\right) \over 2}\tr\ln \left[ -\Delta +\lambda (\cdot) \right] . &
\eqnd\eactONef \cr}} $$ 
\medskip
{\it $\lambda$-field correlation functions.} In this formalism it is
natural to also calculate correlation functions involving the
$\lambda$-field. These have a simple interpretation in the initial
$\phib$-field formalism. Indeed let us add a source $j_{\lambda}$
for $\lambda$ in the action
\eactONef. Then reintroducing the $\phib$-field and integrating over $\lambda$
we recover instead of action \eactON,
$$S(\phib)-(u/6)\phib^2 j_{\lambda} +(u/6)j_{\lambda}^2
-rj_\lambda\,.
\eqnd\ejlam $$    
Therefore $j_{\lambda}$ generates the $\phib^2$ correlation functions,
up to a multiplicative factor and a translation of the connected
2-point function.
\subsection Large $ N $ limit: the critical domain
  
We now take the large $ N $ limit at $ Nu $ fixed. With this 
condition $S_N$ is of order $N$ and the functional integral can be
calculated for $N$ large by steepest descent.   We expect
$\sigma=O(N^{1/2})$, $\lambda=O(1)$. We look for a uniform saddle
point ($\sigma(x),\lambda(x)$ space-independent), 
$$ \sigma (x)=\sigma\, ,\qquad \lambda (x)=\lambda\, . $$
Differentiating then action \eactONef\ with respect to $\sigma$ and
$\lambda$ we obtain the saddle point equations:
\eqna\esaddleN
$$ \eqalignno{ \lambda\sigma & = 0\,, & \esaddleN{a} \cr {\sigma
^2\over N}-{6 \over Nu} \left(\lambda-r \right)+{1 \over \left(2\pi
\right) ^{d}} \int^\Lambda { \d ^{d}p \over p^{2}+\lambda} & =
0\,. &
\esaddleN{b}  \cr} $$
\medskip
{\it Remark.} In the large $N$ limit the leading perturbative
contributions come from chains of ``bubble'' diagrams of the
form displayed in figure \efigi. These diagrams form asymptotically
a geometrical series which is summed by the algebraic techniques
explained above.
\midinsert
\epsfxsize=100.mm
\epsfysize=11.mm
\centerline{\epsfbox{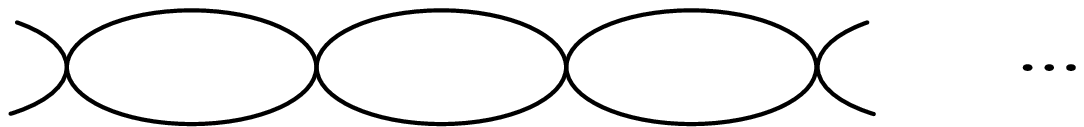}}
\figure{3.mm}{Leading diagrams in the limit $N\to \infty$.}
\endinsert
{\it The low temperature phase.} Equation \esaddleN{a} implies either
$
\sigma=0 $ or $\lambda=0$. In the low temperature phase $ \sigma $, the
average value of the field, does not vanish. Equation \esaddleN{b}
then yields: 
$${\sigma^{2}\over N}=-{6 \over Nu}r-{1 \over (2\pi)^{d}} \int
^\Lambda { \d ^{d}p \over p^{2}}. \eqnd{\esponmag} $$ Note that this
equation has solutions only for $ d>2$. This is a manifestation of
the Mermin--Wagner--Coleman theorem: in a system with only short
range forces a continuous symmetry cannot be broken for $ d\leq 2$,
in the sense that the average $\sigma$ of the order parameter
necessarily vanishes. Physically the would-be Goldstone modes are
responsible for this property: being massless, as we know from
general arguments and as the propagator in the r.h.s.\ of \esponmag\
confirms, they induce an IR instability for $d\le 2$.
\par
Setting
$$\eqalignno{ r_{c}&=-{Nu \over 6}{1 \over \left(2\pi \right)^{d}}
\int ^\Lambda{\d^{d}p \over p^2}\,,& \eqnd\ercritic \cr
r&=r_c+(u/6)\,\tu\,, &\eqnd\eTmTc \cr} $$ we can rewrite equation
\esponmag:
$$ \sigma^{2}=- \tu= (-\tu)^{2\beta}\quad {\rm with} \
\beta=\ud\, \cdot \eqnd{\expbeta} $$
%$$ \sigma^{2}=-6t/u\propto (-t)^{2\beta}\ \Rightarrow \
%\beta=\ud\,. \eqnd{\expbeta} $$
%Equation \esponmag\ has a solution $ \sigma $ provided:
%$$ { 6 \over Nu}r+{1 \over(2\pi)^{d}} \int^\Lambda { \d^{d}p \over
%p^{2}}<0\,. \eqnn $$ 
The expectation value of the field vanishes for $r=r_c$, which
therefore corresponds to the critical temperature. Moreover we find
that for $ N $ large the exponent $\beta $ remains classical, i.e.\
mean-field like, in all dimensions.  
\medskip
{\it The high temperature phase.} Above $ T_{c} $, $ \sigma $
vanishes. In expression \eactONef\ we see that the
$\sigma$-propagator then becomes
$$\Delta_\sigma={1\over p^2+\lambda}\,.\eqnd\eDeltasigN$$ Therefore
$\lambda^{1/2} $ is at this order the physical mass, i.e.\ the
inverse correlation length $\xi^{-1}$ of the field $\sigma $ 
$$m=\xi^{-1}=\lambda^{1/2} \, . \eqnd{\emass} $$ From equation
\esaddleN{b} we can verify that $\partial r/\partial\lambda$ is
positive. The minimum value of $r$, obtained for $\lambda=0$, is
$r_c$. Using equations \eqns{\ercritic,\eTmTc} in equation
\esaddleN{b} we then find:
$$ {6\over u}+{N \over \left(2\pi \right)^{d}}
\int^\Lambda { \d ^{d}p \over p^{2} \left(p^{2}+m^{2}
\right)} ={\tu\over m^2}\,. \eqnd{\ecorleng} $$
\smallskip
(i) For $ d>4 $ the integral in \ecorleng~has a limit for $ m=0 $
and therefore at leading order:
$$ m^{2}=\xi^{-2}\sim \tu \quad {\rm and\ thus} \quad \nu =\ud\, ,
\eqnn $$ which is the mean field result.\par 
(ii) For $ 2<d<4 $ instead, the integral
behaves for $m$ small like  (setting $d=4-\varepsilon$):
$$D_1(m^2)\equiv{ 1 \over \left(2\pi \right)^{d}} \int^\Lambda {\d^{d}p \over
p^{2} \left(p^{2}+m^{2} \right)}  =C(d) m^{-\varepsilon}
-a(d)\Lambda^{-\varepsilon} +O\left({m^{2-\varepsilon} \Lambda^{-2}}
\right), \eqnd\eintasym $$ 
with
\eqna\eintNcor
$$ \eqalignno{N_d &={2 \over(4\pi)^{d/2} \Gamma(d/2) } & \eintNcor{a}
\cr  C(d)&=- { \pi \over2\sin(\pi d/2)}N_d \, , & \eintNcor{b} \cr}
$$  where we have introduced for convenience the usual loop factor
$N_d$.  The constant $a(d)$ which characterizes the leading
correction in equation 
\eintasym, depends explicitly on the regularization, i.e.~the way large
momenta are cut.\par The leading contribution, for $m\to 0$, to the
l.h.s.~of equation
\ecorleng~now comes from the integral. Keeping only the leading term in 
\eintasym~we obtain: 
$$ m=\xi^{-1}\sim \tu^{1/(2-\varepsilon)}, \eqnd{\ecorlenb} $$ which
shows that the exponent $ \nu $ is not classical:
$$ \nu ={1 \over 2-\varepsilon} ={1 \over d-2}\cdot \eqnd{\enuNlim}
$$
\par
(iii) For $d=4$ the l.h.s.\ is still dominated by the integral:
$$D_1(m^2)={ 1 \over(2\pi)^4} \int^\Lambda { \d^{4}p \over p^{2}
\left(p^{2}+m^{2} \right)} \mathop{\sim}_{\displaystyle m\to 0}
{1\over8\pi^2}\ln(\Lambda/m).$$ The correlation length no longer has
a power law behaviour but instead a mean-field behaviour modified by
a logarithm. This is typical of a situation where the gaussian fixed
point is stable, in the presence of a marginal operator. \par (iv)
Examining equation \esaddleN{b} for $\sigma=0$ and $d=2$  we find
that the correlation length becomes large only for $r\to-\infty$.
This peculiar situation will be discussed in the framework of the
non-linear  $\sigma$-model.
\smallskip
Finally, in the critical limit $\tu=0$, $\lambda$ vanishes and thus 
from the form \eDeltasigN\ of the $\sigma$-propagator we find that
the critical exponent $\eta$ remains classical for all $d$
$$\eta=0\ \Rightarrow \ d_\phi=\ud(d-2)\,.\eqnd\eetaNg$$ We verify
that the exponents $\beta,\nu,\eta$ satisfy the scaling relation
proven within the framework of the $\varepsilon$-expansion
$$\beta=\nu d_\phi\,.$$
\medskip
{\it Singular free energy and scaling equation of state.} In a
constant magnetic field $H$ in the $\sigma$ direction, the free
energy $W(H)/\Omega$ per unit volume is given by 
$$W(H)/\Omega =\ln Z/\Omega  ={3\over 2u}\lambda^2-{3r\over
u}\lambda-{1\over2}
\lambda\sigma^2+H\sigma-{N\over2}\tr(-\Delta+\lambda), $$
where $\Omega$ is the total space volume and $\lambda,\sigma$ the
saddle point values are given by equation \esaddleN{b} and the
modified saddle point  equation \esaddleN{a}:
$$\lambda \sigma = H\,. \eqnd{\esaddfld} $$ The thermodynamical
potential $\Gamma(M)$ is the Legendre transform of $W(H)$. First
$$M=\Omega^{-1}{\partial W\over\partial H}=\sigma\,,$$ because
partial derivatives of $W$ with respect to $\lambda,\sigma$ vanish
as a consequence of the saddle point equations. It follows
$$V(M)\equiv\Gamma(M)/\Omega=HM-W(H)/\Omega =-{3\over 2u}\lambda^2+{3r\over
u}\lambda+{1\over2} \lambda M^2 +{N\over2}\tr(-\Delta+\lambda).$$ 
As a property of the Legendre transform, the saddle point equation for
$\lambda$ is now obtained by writing that the derivative of $\Gamma$
vanishes.
\par
The term $\tr\ln$ can be evaluated for large $\Lambda$ in terms of
$r_c$ and  the quantities defined in \eintasym. One finds
$$\eqalign{
\tr\ln[(\Delta-\lambda)\Delta^{-1}]&={1\over(2\pi)^d}\int\d^d
p\,\ln[(p^2+\lambda)/p^2] 
\cr &=-2{C(d)\over d}\lambda^{d/2}-{6 r_c\over
Nu}\lambda+{a(d)\over2}\lambda^2\Lambda^{4-d}+O(\lambda^{1+d/2}\Lambda^{-2}).
\cr} $$ 
The thermodynamical potential becomes
$$V(M)={3\over 2}\left({1\over u^*}-{1\over u}\right)\lambda^2
+{3(r-r_c)\over u}\lambda+{1\over2}\lambda M^2-{NC(d)\over
d}\lambda^{d/2} ,
\eqnn $$
where we have defined
$$u^*={6\over Na(d)}\Lambda^\varepsilon.\eqnd\eustari $$ 
Note that for $\lambda$ small the term proportional to $\lambda^2$ is
negligible with respect to the singular term $\lambda^{d/2}$ for $d<4$. At
leading order in the critical domain
$$V(M)= {1\over2}\tu\lambda+{1\over2}\lambda
M^2-{NC(d)\over d}\lambda^{d/2} , \eqnn $$ where $\tu$ has been
defined in \eTmTc.\par The saddle point equation for $\lambda$ takes
the simple form
$$\tu + M ^2-NC(d)\lambda^{d/2-1}=0 , $$ and thus
$$\lambda=\left[{1\over NC(d)}\left(\tu
+M^2\right)\right]^{2/(d-2)}. $$ It follows that the leading
contribution, in the critical domain, to the thermodynamical
potential is given by 
$$ V(M) \sim {(d-2)\over 2d}{1\over\bigl(NC(d)
\bigr)^{2/(d-2)}} (\tu +M^2)^{d/(d-2)}.\eqnd\ethermscN $$ Various
quantities can be derived from $V(M)$, for example the equation
of state by differentiating with respect to $M$. The resulting
scaling equation of state is
$$H={\partial V\over \partial M}= h _0
M^{\delta}f\left(\tu/M^2 \right), \eqnd{\estatNli} $$  in which
$h_0$ is a normalization constant, The exponent $\delta$ is given by:
$$\delta= {d+2 \over d-2}\,, \eqnn $$ in agreement with the general
scaling relation relation $\delta=d/ d_\phi-1$, and the function
$f(x)$ by: 
$$f(x)= (1+ x)^{2/(d-2)}. \eqnd{\estatNgb} $$ The asymptotic form of
$f(x)$ for $x$ large implies $\gamma=2/(d-2)$ again in agreement
with the scaling relation $\gamma=\nu(2-\eta)$. Taking into account
the values of the critical exponents $\gamma$ and $\beta$ it is then
easy to verify that the function $f$ satisfies all required 
properties like for example Griffith's analyticity (see section
\ssCDeqst). In particular the equation of state can be cast into the
parametric form:
$$\eqalign{\sigma& =R^{1/2}\theta\,,\cr
\tu & =3R\left(1-\theta^2\right),\cr
H& =h_0 R^{\delta/2}\theta\left(3-2\theta^2\right)^{2/(d-2)}.\cr}$$
\medskip
\midinsert
\epsfxsize=50.mm
\epsfysize=16.mm
\centerline{\epsfbox{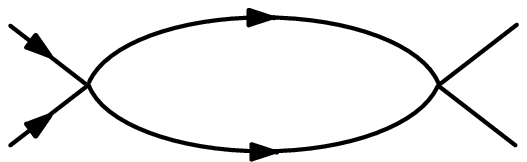}}
\vskip-16.7mm
\centerline{$q$}
\vskip5mm
\centerline{$p-q$}
\figure{3.mm}{The ``bubble" diagram $B_\Lambda(p,m)$.}
\endinsert
\def\efigii{2}
\medskip
{\it Leading corrections to scaling.} The $\lambda^2$ term yields the
leading corrections to scaling. It is subleading by a power of
$\tu$
$$\lambda^2/\lambda^{d/2}=O(\tu^{(4-d)/(d-2)}).$$ We conclude
$$\omega\nu=(4-d)/(d-2)\ \Rightarrow\ \omega=4-d\,.\eqnd\eNomega $$ 
We have identified the exponent $\omega$ which governs the leading
corrections to scaling. Note that for the special value $u=u^*$ this
correction vanishes.
\medskip
{\it Specific heat exponent. Amplitude ratios.}  Differentiating
twice $V(M)$ with respect to $\tu$ we obtain the specific heat
at fixed magnetization
$$C_H= {1\over (d-2)}{1\over\bigl(NC(d) \bigr)^{2/(d-2)}} (\tu
+M^2)^{(4-d)/(d-2)}.\eqnd\esphCH $$ For $M=0$ we identify the
specific exponent $\alpha$
$$\alpha={4-d\over d-2}, \eqnn $$ which indeed is equal to $2-d\nu$,
as predicted by scaling laws. Among the ratio of amplitudes one can
calculate for example $R^+_\xi$ and $R_c$ (for definitions see
chapter 28 of main reference)
$$(R^+_\xi)^d={4N\over(d-2)^3}{\Gamma(3-d/2)\over(4\pi)^{d/2}},\quad
R_c={4-d\over(d-2)^2}. \eqnn $$
\medskip
{\it The $\lambda$ and $(\phib)^2$ two-point functions.}
Differentiating twice the action \eactONef\ with respect to
$\lambda(x)$, then replacing the field $\lambda(x)$ by its
expectation value $m^2$, we find the {$ \lambda $-propagator} $
\Delta_{\lambda}(p)$ above $T_c$ 
$$ \Delta_\lambda(p)=-{2 \over N} \left[ {6\over
Nu}+B_\Lambda(p,m)\right]^{-1} \,, \eqnn $$ where $B_\Lambda(p,m)$
is the bubble diagram of figure \efigii:
$$B_\Lambda(p,m)={1 \over \left(2\pi \right)^{d}} \int^\Lambda { \d
^{d}q
\over \left(q^{2}+m^{2} \right) \left[ \left(p-q \right)^{2}+m^{2}
\right]} . \eqnd\ediagbul $$ 
The $\lambda$-propagator is negative because the $\lambda$-field is
imaginary. As noted in \ssfivNi, it is simply related to the
$\phib^{2}$ 2-point function
$$\left<\phib^2 \phib^2\right>={B_\Lambda(p,m)\over
1+(Nu/6)B_\Lambda(p,m)}.
\eqnn $$
At zero momentum we recover the specific heat. The small $m$ expansion of
$B_\Lambda(0,m)$ can be derived from the expansion \eintasym. 
One finds
$$\eqalignno{B_\Lambda(0,m)&={1\over(2\pi)^d} \int^\Lambda {\d^{d}q \over
\left(q^{2}+m^{2} \right)^2} \cr &={\partial \over \partial m^2}
\bigl(m^2 D_1(m^2)\bigr) 
\mathop{=}_{m\ll\Lambda}
 (d/2-1)C(d) m^{-\varepsilon}-a(d)\Lambda^{-\varepsilon}
+\cdots\ .\hskip8mm &\eqnd\eBLamze \cr}$$
The singular part of the specific heat thus
vanishes as $m^{\varepsilon}$, in agreement with equation
\esphCH~for $M=0$.
\par 
In the critical theory ($ m=0 $ at this order) for $2\le d\le 4$ the
denominator is also dominated at low momentum by the integral
$$B_\Lambda(p,0)= {1 \over \left(2\pi \right)^{d}} \int^\Lambda {
\d^{d}q \over q^{2}(p-q)^{2}} \mathop{=}_{2<d<4} b(\varepsilon)p^{-\varepsilon}
-a(d)\Lambda^{-\varepsilon}
+O\left(\Lambda^{-2}p^{2-\varepsilon}\right),\eqnd\ebullecrit  $$
where
$$ b \left(\varepsilon \right)=-{\pi \over\sin(\pi d/2)} 
{\Gamma^2 (d/ 2) \over \Gamma (d-1)}N_d \,, \eqnd\econstb $$
and thus:
$$ \Delta_{\lambda}(p)\mathop{\sim}_{p\to 0} -{2 \over N
b(\varepsilon)} p^{\varepsilon}. \eqnd{\eprocrit} $$  We again
verify consistency with scaling relations. In particular we note 
that in the large $N$ limit the {\it dimension $[\lambda]$ of the
field $\lambda$}\/ is
$$[\lambda]=\ud(d+\varepsilon)=2\,, \eqnn $$ 
a result important for the $1/N$ perturbation theory.
\smallskip
{\it Remarks.} \par (i) For $d=4$ the behaviour of the propagator is
still dominated by the integral which has a logarithmic behaviour
$\Delta_\lambda\propto 1/\ln(\Lambda/p)$.\par (ii) Note therefore
that for $d\le 4$ the contributions generated by the term
proportional to $\lambda^2(x)$ in \eactONef\ always are negligible
in the critical domain.
\subsection  RG functions and leading corrections to scaling

 {\it The RG functions.} For a more detailed verification of the
consistency of the large $N$ results with the RG framework, we now
calculate RG functions at leading order. One first easily verifies
that, at leading order for $\Lambda$ large, $m$ solution of equation
\ecorleng~satisfies
\sslbl\sssEGRN
$$\Lambda{\partial m\over\partial\Lambda}+N\varepsilon a(d)
\Lambda^{-\varepsilon}{u^2\over6}{\partial m\over\partial u}=0\,,$$
where the constant $ a (\varepsilon) $ has been defined in
\eintasym. It depends on the cut-off procedure but for
$\varepsilon=4-d$ small satisfies
$$ a(\varepsilon)\sim 1/ (8\pi^{2}\varepsilon). \eqnn
$$ We then set (equation \eustari):
$$ u=g\Lambda^{\varepsilon},\quad g^*=u^*\Lambda^{-\varepsilon}=
6/(Na)\,.\eqnd\eustar $$ In the new variables $\Lambda,g,\tu$ we
obtain an equation which expresses that $m$ is RG invariant
$$ \left( \Lambda{ \partial \over \partial \Lambda} +\beta \left(g
\right){\partial \over \partial g}-\eta_{2}(g)\tu{\partial
\over \partial \tu} \right) m(\tu,g,\Lambda)=0\,, \eqnn $$
with
$$ \eqalignno{ \beta (g) & = -\varepsilon g(1- g/g^*) , &
\eqnd{\ebetaN} \cr
\nu^{-1}(g)= 2+\eta_{2}(g)& = 2-\varepsilon g/g^*. & \eqnn \cr} $$ 
When $a(d)$ is positive (but this not true for all regularizations,
see the discussion below), one finds an IR fixed point $g^*$, as
well as exponents  $\omega=\varepsilon$, and $\nu^{-1}=d-2$, in
agreement with equations \eqns{\eNomega,\enuNlim}. In the framework of the
$\varepsilon$-expansion $\omega$ is associated with the leading
corrections to scaling. In the large $N$ limit $\omega$ remains
smaller than two for $\varepsilon<2$, and this extends the property to
all dimensions $2\le d\le 4$.\par Finally, applying the RG
equations to the propagator \eDeltasigN, we find
$$\eta(g)=0\ , \eqnn $$
a result consistent with the value \eetaNg~found for $\eta$.
\medskip
{\it Leading corrections to scaling.} From the general RG analysis we
expect the leading corrections to scaling to vanish for $u=u^*$.
This property has already been verified for the free energy. Let us now
consider the correlation length or mass $m$ 
given by equation \ecorleng. If we keep the leading correction to
the integral for $m$ small (equation \eintasym) we find
$$ {6\over u}- {6\over u^*} +N C(d) m^{-\varepsilon}
+O\left({m^{2-\varepsilon} \Lambda^{-2}}
\right) ={\tu\over m^2}\,, \eqnd\ecorlengb $$
where equation \eustar~has been used.  We see that the leading
correction again vanishes for $u=u^*$. Actually all correction terms
suppressed by powers of order $\varepsilon$ for $d\to 4$ vanish
simultaneously as expected from the RG analysis of the $\phi^4$
field theory. Moreover one verifies that the leading correction is
proportional to $(u-u^*)\tu^{\varepsilon/(2-\varepsilon)}$, which
leads to $\omega\nu=\varepsilon/(2-\varepsilon)$, in agreement with
equations \eqns{\eNomega,\enuNlim}.
\par
In the same way if we keep the leading correction to the
$\lambda$-propagator in the critical theory (equation
\ebullecrit) we find: 
$$ \Delta_{\lambda} \left(p \right)=-{2 \over N} \left[{6\over Nu}-
{6\over Nu^*}+ b(\varepsilon)p^{-\varepsilon}
\right]^{-1}, \eqnd{\epropNli}  $$
where terms of order $ \Lambda^{-2} $ and $1/N$ have been neglected.
The leading corrections to scaling again exactly cancel for $
u=u^{\ast} $ as expected.
\smallskip
{\it Discussion.} \par (i) One can show that a perturbation due to
irrelevant operators is equivalent, at leading order in the critical
region, to a modification of the $(\phib^2)^2$ coupling.  This can
be explicitly  verified here. The amplitude of the leading
correction to scaling has been found to be proportional to
$6/Nu-a(d)\Lambda^{-\varepsilon}$ where the value of $a(d)$ depends
on the cut-off procedure and thus of contributions of irrelevant
operators. Let us call $u'$ the $(\phib^2)^2$ coupling constant in
another scheme where $a$ is replaced by $a'$. Identifying the
leading correction to scaling we find the  relation:
$${6\Lambda^{\varepsilon} \over Nu}-a(d)={6\Lambda^{\varepsilon}
\over Nu'}-a'(d),$$ 
homographic relation which is consistent with the special form
\ebetaN\ of the $\beta$-function.\par (ii) {\it The sign of
$a(d)$.} It is generally assumed that $a(d)>0$.  This is indeed what
one finds in the simplest regularization schemes, like the simplest
Pauli--Villars's regularization where $a(d)$ is positive in all
dimensions $2 < d < 4$. Moreover $a(d)$ is always positive near four
dimensions where it diverges like
$$a(d)\mathop{\sim}_{d\to 4} {1\over 8\pi^2\varepsilon}.$$ Then
there exists an IR fixed point, non-trivial zero of the
$\beta$-function. For this value $u^*$ the leading corrections to
scaling vanish.\par However for $d$ fixed, $d<4$, this is not a
universal feature. For example in the case of simple lattice
regularizations it has been shown that in $d=3$ the sign is
arbitrary. \par However, if $a(d)$ is negative, the RG method for
large $N$ (at least in the perturbative framework) is confronted
with a serious difficulty.  Indeed the coupling flows in the IR
limit to large values where the large $N$ expansion is no longer
reliable. It is not known whether this signals a real physical
problem, or is just an artifact of the large $N$ limit. \par
Another way of stating the problem is to examine directly the
relation between bare and renormalized coupling constant. Calling
$g_{\rm r} m^{4-d}$ the renormalized 4-point function at zero
momentum, we find
$$m^{4-d}g_{\rm r}={\Lambda^{4-d}g\over 1+\Lambda^{4-d}g N
B_\Lambda(0,m)/6} . \eqnd\egrenor$$ In the limit $m\ll\Lambda$ the
relation can be written
$${1\over g_{\rm r}}={(d-2) N C(d)\over
12}+\left(m\over\Lambda\right)^{4-d}
\left({1\over g}-{N a(d)\over6}\right).\eqnd\egrenor $$
We see that when $a(d)<0$ the renormalized IR fixed point value cannot
be reached by varying $g>0$ for any finite value of $m/\Lambda$. In
the same way leading corrections to scaling can no longer be
cancelled.
\def\efigiii{3}
\subsection Small coupling constant and large momentum
expansions for $d<4$

Section \sssfivNRT~is devoted to a systematic discussion of the $1/N$ 
expansion. However the $1/N$ correction to
the two-point function will help us to immediately understand the
problem of the massless field theory for $d<4$.\par 
We have seen that, in the framework at the $1/N$ expansion, we can
calculate at fixed dimension $d<4$ in the critical limit ($T=T_c, m^2=0$). This
implies that the terms of the $1/N$ expansion cannot be expanded in
a power series of the coupling constant, at least with integer
powers. Note that since the gaussian fixed point is an UV fixed
point, the small coupling expansion is also a large momentum
expansion. To understand the phenomenon we consider the 
$\left<\sigma\sigma\right>$
correlation function at order $1/N$.  At this order only one diagram
contributes (figure \efigiii), containing two $\lambda^{2}\sigma$ vertices. 
After mass renormalization and in the  large cut-off limit we find: 
$$\Gamma^{(2)}_{\sigma\sigma}(p)= p^{2} +{2\over N (2\pi)^{d}}\int
{\d^{d}q \over(6/Nu)+b(\varepsilon)q^{-\varepsilon}}\left({1 \over
(p+q)^2} -{1 \over q^2}\right) +O\left({1 \over N^{2}}\right)
. \eqnd{\eONpropi} $$ 
An analytic study of the integral reveals that it has an expansion of the form
$$\sum_{k\ge 1} \alpha_k u^k p^{2-k\varepsilon}+\beta_k
u^{(2+2k)/\varepsilon} p^{-2k} .\eqnn $$ 
The coefficients $\alpha_k,\beta_k$ can be obtained by performing a Mellin
transformation over $u$ on the integral. Indeed if a function $f(u)$
behaves like $u^t$ for $u$ small, then the Mellin transform $M(s)$
$$M(s)=\int_0^\infty\d u\,u^{-1-s}f(u), $$
has a pole at $s=t$. Applying the transformation to the
integral, and inverting $q$ and $u$ integrations we have to calculate
the integral
$$\int_0^\infty\d
u\,{u^{-1-s}\over(6/Nu)+b(\varepsilon)q^{-\varepsilon}}
={N\over 6}\left(Nb(\varepsilon)q^{-\varepsilon}\over
6\right)^{1-s}{\pi
\over \sin\pi s}\,\cdot $$
Then the value of the remaining $q$ integral follows from the generic
result \eintmunu.\par
The terms with integer powers of $u$ correspond to the formal
perturbative expansion where each integral is calculated for
$\varepsilon$ small enough. $\alpha_k$ 
has poles at $\varepsilon=(2l+2)/k$ for which  the corresponding power of 
$p^2$ is $-l$, i.e.\ an integer. One verifies that $\beta_l$ has a pole at
the same value of $\varepsilon$ and that the singular contributions cancel in
the sum. For these dimensions logarithms of $u$ appear in the expansion.
\midinsert
\epsfxsize=50.mm
\epsfysize=16.mm
\centerline{\epsfbox{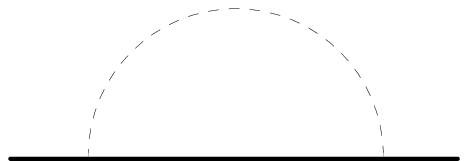}}
\vskip-17.mm
\centerline{$\lambda$}
\vskip7.mm
\centerline{$\sigma$}
\figure{3.mm}{The diagram contributing to $\Gamma^{(2)}_{\sigma\sigma}$ at
order  $1/N$.} 
\endinsert
\subsection The non-linear $\sigma$-model in the large $N$ limit

We have noticed that the term proportional to $\int\d^d x\,\lambda^2(x)$,
which has dimension $4-d$ for large $N$ in all dimensions, is irrelevant in
the critical domain for $d<4$ and can thus be omitted at leading order (this
also applies to $d=4$ where it is marginal but yields only logarithmic
corrections).   
Actually the constant part in the inverse propagator as written in equation
\epropNli~plays the role of a large momentum cut-off. Let us thus consider
the action \eacteffb~without the $\lambda^2$ term. If
we then work backwards, reintroduce the initial field $\phib$ and 
integrate over $ \lambda (x) $ we find \sslbl\ssLTsN 
$$ Z= \int \left[ \d  \phib (x)\right]\delta
\left[ \phib^{2} (x)-{6 \over u} \left(m^{2}-r \right)
\right] \exp\left[- \int{ 1 \over 2}\left(\partial_{\mu} \phib(x)\right)^{2}
\d ^{d}x\right]. \eqnd{\epartsig} $$ 
Under this form we recognize the partition function of the $ O(N) $ symmetric
non-linear $ \sigma $-model in an unconventional normalization. We have 
therefore discovered a remarkable correspondence: to all orders in an $ 1/N $
expansion the renormalized non-linear $ \sigma $-model is identical to the
renormalized $ \left( \phib^{2} \right)^{2} $ field theory at the
IR fixed point. 
\medskip
{\it The large $N$ limit.} In
order to more explicitly show the correspondence between the set of
parameters used in the two models, let us directly solve the $\sigma$-model in
the large $N$ limit. We rewrite the partition function:
$$Z= \int \left[\d\phi(x)\d\lambda(x)\right]
\exp\left[-S(\phib,\lambda)\right] ,\eqnn  $$ 
with:
$$S(\phib,\lambda) = {1 \over 2t}\int \d^{d}x \left[ \left(
\partial_{\mu}\phib \right)^{2} + \lambda \left(\phib^{2} -1 \right)\right].
\eqnd\eactsigla $$
Integrating, as we did in section \ssfivNi, over $N-1$ components of $\phib$
and calling $\sigma$ the remaining component, we obtain:
$$Z= \int \left[\d\sigma(x)\d\lambda(x)\right] 
\exp\left[-S_N(\sigma,\lambda)\right] , \eqnn $$ 
with:
$$S_N \left(\sigma,\lambda  \right)= 
 {1 \over 2t}\int \left[ \left(\partial_{\mu}\sigma \right)^{2}+
\left(\sigma^{2} (x)-1\right) \lambda (x)
\right] \d^{d}x +{1 \over 2} (N-1 ) \tr\ln \left[
-\Delta +\lambda (\cdot) \right] .  \eqnn $$
The large $N$ limit is here taken at $tN$ fixed. The saddle point equations,
analogous to equations \esaddleN{}, are:
\eqna\emgNsig
$$\eqalignno{m^2\sigma &=0\, ,& \emgNsig{a} \cr
\sigma^{2}& = 1 - {(N-1)t \over (2\pi)^{d}} \int^{\Lambda}{\d^{d}p \over p^{2}
+ m^2} \,,& \emgNsig{b}\cr} $$
where we have set $\left<\lambda(x)\right>=m^2$. At low temperature $\sigma $
is different from zero and thus $m$, which is the mass of the $\pi$-field,
vanishes. Equation \emgNsig{b} gives the spontaneous magnetization: 
$$\sigma^{2} = 1 - {(N-1)t \over (2\pi)^{d}} \int^{\Lambda}{\d^{d}p \over
p^{2}} .\eqnd\emagNsig $$
Setting
$${1 \over t_{c}} = {(N-1) \over (2\pi)^{d}} \int^{\Lambda}{\d^{d}p \over
p^{2}} ,\eqnn $$
we can write equation \emagNsig:
$$\sigma^{2} = 1 - t/t_{c}\, . \eqnd\emgTNsig $$
Thus $t_c$ is the critical temperature where $\sigma$ vanishes.\par
Above $t_{c}$, $\sigma$ instead vanishes and $m$, which is now the
common mass of the $\pi$- and $\sigma$-field, is for $2<d<4$
given by: 
$$ {1 \over t_{c}}- {1 \over t} = m^{d-2} {(N-1) \over (2\pi)^{d}}
\int{\d^{d}p \over p^{2}\left(p^{2}+1 \right)}+O\left(m^2\Lambda^{d-4}
\right) .\eqnd{\emasNsig} $$
We recover the scaling form of the correlation length $\xi =1/ m$.
From the equations \eqns{\emgTNsig,\emasNsig}, we can also derive the RG
functions at leading order for $N$ large:
$$ \beta (t) = \varepsilon t-{ N \over 2\pi} t^{2}\, ,\qquad
 \zeta (t) = {N \over 2\pi} t\,. \eqnn  $$
It is also easy to calculate the thermodynamical potential, Legendre
transform of $W(H)=t\ln Z(H)$:
$$V(M)=\Gamma(M)/\Omega={d-2\over2d}{1\over\bigl(NC(d)\bigr)^{2/(d-2)}}
(M^2-1+t/t_c)^{d/(d-2)},\eqnn $$
a result which extends equation \ethermscN~to all temperatures below
$t_c$. The calculation of other physical quantities and the
expansion in $1/N$ follow from the considerations of previous sections
and section \sssfivNRT.
\medskip
{\it Two dimensions and the question of Borel summability.} For $d=2$ the
critical  temperature vanishes and the parameter $m$ has the form:
$$m \sim \Lambda \e^{-2\pi /(Nt)},\eqnn $$
in agreement with the RG predictions. Note that the field 2-point
function takes in the large $N$-limit the form:
$$\Gamma^{(2)}_{\sigma\sigma}(p)=p^2 + m^2\,. \eqnn $$
The mass term vanishes to all orders in the expansion in powers of
the coupling constant $t$, preventing any perturbative calculation of
the mass of the field. The perturbation series is
trivially not Borel summable. Most likely this property is also true for the
model at finite $N$. On the other hand if we break the $O(N)$ symmetry by
a magnetic field, adding a term $h\sigma$ to the action, the physical mass
becomes calculable in perturbation theory.
\medskip
{\it Corrections to scaling and the dimension four.} In equation \emasNsig\
we have neglected corrections to scaling. If we take into account the leading
correction we get instead:
$$m^2\left(C(d)m^{d-4}-a(d)\Lambda^{d-4}\right)\propto t-t_c\,,$$
where $a(d)$, as we have already explained, is a constant which explicitly
depends on the cut-off procedure and can thus be varied by changing
contributions of irrelevant operators. 
By comparing with the results of section \sssEGRN, we discover that,
although the non-linear $\sigma$-model superficially depends on one parameter
less than the corresponding $\phib^4$ field theory, actually this parameter is
hidden in the cut-off function. This remark becomes important in the four
dimensional limit where most leading contributions come from the leading
corrections to scaling. For example for $d=4$ equation \emasNsig\ takes a
different form, the dominant term in the r.h.s.\ is proportional to
$m^2\ln m$. We recognize in the factor $\ln m$ the effective
$\phi^4$ coupling at mass scale $m$. Beyond the $1/N$ expansion,
to describe with perturbation theory and renormalization  group the physics
of the non-linear $\sigma$ model it is necessary to introduce
the operator $\int\d^d x\,\lambda^{2}(x)$, which irrelevant for $d<4$,
becomes marginal, and to return to the $\phi^4$ field theory. 

\def\efigiv{4}
\subsection The $1/N$-expansion: an alternative field theory

{\it Preliminary remarks. Power counting.}
Higher order terms in the steepest descent calculation of the
functional integral \eZeff\ generate  a systematic $ 1/N $ expansion.
Let us first slightly rewrite  action \eactONef. We shift the field
$\lambda(x)$ by its expectation value $m^2$  (equation \emass),
$\lambda(x) \mapsto m^2+\lambda(x)$:\sslbl\sssfivNRT %\sslbl\ssfivNe
$$\eqalignno{ S_N \left( \sigma ,\lambda
\right) &  = {1 \over 2}\int \d ^{d}x \left[
\left(\partial_{\mu}\sigma \right)^{2}+ m^{2}\sigma
^{2}+ \lambda (x)\sigma^{2} (x) - {3\over u}\lambda^{2} (x) 
 -{6 \over u} \left(m^{2}-r\right) \lambda(x) \right] \cr &\quad +{ \left(N-1
\right) \over 2}  \tr\ln \left[ -\Delta +m^2+\lambda (\cdot) \right]
. &\eqnd{\eacteffb} \cr}$$ 
We now analyze the terms in the action \eacteffb~from the
point of view of large $N$ power counting. The
dimension of the field $ \sigma (x) $ is $ (d-2)/2 $. From the critical
behaviour \eprocrit\ of the $\lambda$-propagator we have deduced the
canonical dimension  $[\lambda]$ of the field $ \lambda (x) $:  
$$ 2 \left[ \lambda \right] -\varepsilon =d\, \qquad {\rm i.e.} \quad \left[
\lambda \right] =2\,. $$ 
As noted above, $\lambda^2$ has dimension $4>d$ and is thus irrelevant.
The interaction term $ \int \lambda(x)\sigma^{2} (x)\d^d x $ has
dimension zero. It is easy to verify that the non-local interactions
involving the {$ \lambda$-field},  coming from the expansion of the $ \tr\ln $,
have all also the canonical dimension zero:
$$ \left[ \tr \left[ \lambda (x) \left(-\Delta +m^{2} \right)^{-1}
\right]^{k} \right] =k \left[ \lambda \right] -2k=0\,. $$
This power counting property has the following implication: In contrast with
usual perturbation theory,  the $ 1/N $ expansion generates only logarithmic
corrections to the leading long distance behaviour for any fixed dimension
$d$, $ 2<d\leq 4$. The situation is thus similar to the situation one
encounters for the $ \varepsilon $-expansion (at the IR fixed point)
and one expects to be able to 
calculate universal quantities like critical exponents for example as power
series in $ 1/N$. However, because the interactions are non-local, the results
of renormalization theory do not immediately apply. We now construct 
an alternative quasi-local field theory, for which the standard RG
analysis is valid, and which reduces to the large $N$ field theory in
some limit. 
\medskip
{\it An alternative field theory.}
To be able to use the standard results of renormalization theory we
reformulate the critical theory to deal with the non-local
interactions. Neglecting corrections to scaling we start from the
non-linear $\sigma$-model in the form \eactsigla:   
$$ \eqalignno{ Z & = \int \left[ \d  \lambda (x)
\right] \left[ \d  \phib (x) \right]
\exp\left[-S \left(\phib ,\lambda \right)\right] , & \eqnn
\cr S ( \phib ,\lambda) & = {1 \over 2t}\int \d^{d}x \left[ \left(
\partial_{\mu}\phib \right)^{2} + \lambda \left(\phib^{2} -1 \right)\right].
  & \eqnd{\eactphla}   \cr} $$
The difficulty arises from the $\lambda$-propagator, absent in the
perturbative formulation, and generated by the large $N$ summation.
We thus add to the action \eactphla~a term quadratic in $\lambda$
which at tree level of standard perturbation theory generates a
$\lambda$-propagator of the form \eprocrit.
The modified action $S_{\vv}$ then is
$$ S_{\vv} \left( \phib ,\lambda \right) = {1 \over 2}\int \d^{d}x \left\{
{1\over t}\left[\left( \partial_{\mu}\phib \right)^{2} + \lambda
\left(\phib^{2} -1 \right)\right]-
{1\over\vv^2}\lambda(-\partial^2)^{-\varepsilon/2}\lambda\right\} 
. \eqnd{\eactSg}$$
In the limit where the parameter $\vv$ goes to infinity the coefficient of
the additional term vanishes, and the initial action is recovered. \par
We below consider only the critical theory. This means that the
couplings of all relevant interactions will be set to their critical values.
These interactions contain a term linear in $\lambda$ and a polynomial in
$\phib^2$ of degree depending on the dimension. Note that in some discrete
dimensions some monomials become just renormalizable. We therefore work
in generic dimensions. The quantities we shall calculate are regular in the
dimension. 
The field theory with the  action \eactSg~ can be studied with 
standard field theory methods. The peculiar form of the $\lambda$ quadratic
term, which is not strictly local, does not create a problem. Similar terms
are encountered in statistical systems with long range forces. The simple
consequence is 
that the $\lambda$-field is not be renormalized because counter-terms are
always local.\par
It is convenient to rescale $\phib\mapsto \phib\sqrt{t}$, $\lambda \mapsto
\vv\lambda$: 
$$ S_{\vv} \left( \phib ,\lambda \right) = {1 \over 2}\int \d^{d}x \left[
\left( \partial_{\mu}\phib \right)^{2} + \vv\lambda\phib^{2}
- \lambda(-\partial^2)^{-\varepsilon/2}\lambda +{\rm relevant\
terms}\right].$$  
The renormalized critical action then reads:
$$[S_\vv]_{\rm ren} =  {1 \over 2}\int \d^{d}x \left[
Z_\phi \left( \partial_{\mu}\phib \right)^{2} +\vv_\r Z_{\vv}\lambda\phib^{2}
- \lambda(-\partial^2)^{-\varepsilon/2}\lambda +{\rm relevant\
terms}\right] . \eqnd\eactSren $$
It follows that the RG equations for 1PI correlation functions of $l$
$\lambda$ fields and $n$ $\phib$ fields in the critical theory take the
form: 
$$\left[\Lambda {\partial \over \partial \Lambda}+
\beta_{\vv^2}(\vv){\partial \over \partial \vv^2}-{n\over 2}\eta(\vv)\right]
\Gamma^{(l,n)}=0\,.\eqnd \eqRGm $$ 
We can then calculate the RG functions as power series in $1/N$. It
is easy to verify that $\vv^2$ has to be taken of order $1/N$. Therefore 
to generate a $1/N$ expansion one first has to sum the multiple insertions of
the one-loop $\lambda $ two-point function, contributions which form a
geometrical series. The $\lambda$ propagator then becomes
$$\Delta_\lambda(p)=-{2 p^{4-d}\over b(\varepsilon) D(\vv)}\,, \eqnn $$
where we have defined
$$D(\vv)=2/b(\varepsilon)+ N\vv^2.$$
The solution to the RG equations \eqRGm\ can be written:
$$\Gamma^{(l,n)}(\tau p, \vv,\Lambda)=Z^{-n/2}(\tau)
\tau^{d-2l-n(d-2)/2} \Gamma^{(l,n)}(p, \vv(\tau),\Lambda)  ,
\eqnd{\esolRG} $$ 
with the usual definitions
$$\tau{\d \vv^2\over \d \tau}=\beta(\vv(\tau))\,,\quad \tau{\d \ln Z
\over \d \tau}=\eta(\vv(\tau))\, .$$
We are interested in the neighbourhood of the fixed point $\vv^2=\infty$.
One verifies that the RG function $\eta(\vv)$ approaches the exponent 
$\eta$ obtained by direct calculation, and the RG $\beta$-function
behaves like  $\vv^2$. The flow equation
for the coupling constant becomes:
$$\tau{\d \vv^2\over \d \tau}=\rho \vv^2 ,\ \Rightarrow\ \vv^2(\tau)\sim
\tau^{\rho}.\eqnn $$ 
We then note that to each power of the $\lambda$ field corresponds a power of
$\vv$. It follows
$$\eqalignno{\Gamma^{(l,n)}(\tau p,\vv,\Lambda)&\propto
v^l(\tau)\tau^{d-2l-n(d-2+\eta)} & \cr
&\propto\tau^{d-(2-\rho/2)l-n(d-2+\eta)} .& \eqnn \cr}
$$ 
To compare with the result \egalnsca~obtained from the perturbative
renormalization group one has still to take into account that the functions
$\Gamma^{(l,n)}$ defined here are obtained by an additional Legendre 
transformation with respect to the source of $\phib^2$. Therefore
$$ 2-\rho/2=d_{\phib^2}=d-1/\nu \,. \eqnn $$
\midinsert
\epsfxsize=37.6mm
\epsfysize=18.2mm
\centerline{\epsfbox{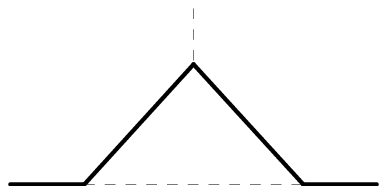}}
\figure{3.mm}{Diagram contributing to $\Gamma^{(3)}_{\sigma\sigma\lambda}$ at 
order  $1/N$.} 
\endinsert
\midinsert
\epsfxsize=34mm
\epsfysize=27.2mm
\centerline{\epsfbox{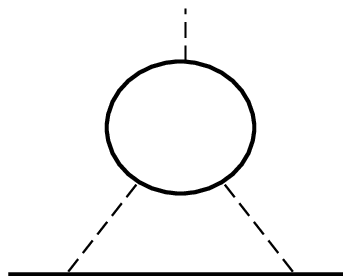}}
\figure{3.mm}{Diagram contributing to $\Gamma^{(3)}_{\sigma\sigma\lambda}$ at
order  $1/N$.} 
\endinsert

\def\efigv{5}
\medskip
{\it RG functions at order $1/N$.}
Most calculations at order $1/N$ rely on the evaluation of the generic
integral 
$${1\over(2\pi)^d}\int{\d^d q \over (p+q)^{2\mu}
q^{2\nu}}=p^{d-2\mu-2\nu}{\Gamma(\mu+\nu-d/2)\Gamma(d/2-\mu)\Gamma(d/2-\nu) 
\over (4\pi)^{d/2}\Gamma(\mu)\Gamma(\nu)\Gamma(d-\mu-\nu)}\,
. \eqnd\eintmunu $$ 
For later purpose it is convenient to set:
$$X_1={2 N_d\over b(\varepsilon)}= {4 \Gamma(d-2)\over
\Gamma(d/2)\Gamma(2-d/2)\Gamma^2(d/2-1)}= %2to4 bug 
{4 \sin(\pi\varepsilon/2) \Gamma(2-\varepsilon)\over
\pi\Gamma(1-\varepsilon/2) \Gamma(2-\varepsilon/2)} .\eqnd{\eXone}$$
To compare with fixed dimension results note $X_1\sim 2(4-d)$ for $d\to 4$
and $X_1\sim(d-2)$ for $d\to 2$.\par
The calculation of the $\left<\phi\phi\right>$ correlation function
at order $1/N$ involves the evaluation of the diagram
of figure \efigiii. We want to determine the coefficient of
$p^2\ln\Lambda/ p$. Since we work at one-loop order we can instead
replace the $\lambda$ propagator $q^{-\varepsilon}$ by $q^{2\nu}$ and send the
cut-off to infinity. We then use the result \eintmunu~with
$\mu=1$. In the limit $2\nu \to-\varepsilon$ the integral has a
pole. The residue of the pole  yields the coefficient of
$p^2\ln\Lambda$ and the finite part contains the $p^2\ln p$ contribution 
$$\Gamma^{(2)}_{\sigma\sigma}(p)= p^{2} +
{\varepsilon\over 4-\varepsilon} {2 N_d\over b(\varepsilon) D(\vv)} \vv^2
p^2\ln(\Lambda/p) .$$
Expressing that the function satisfies the RG equation we obtain the
function $\eta(\vv)$.\par 
The second RG function can be deduced from the divergent parts
of the $\left<\phi\phi\lambda\right>$ function 
$$\Gamma^{(3)}_{\sigma\sigma\lambda}=\vv+A_1 \vv^3 D^{-1}(\vv)\ln\Lambda +A_2
\vv^5 D^{-2}(\vv)\ln \Lambda +\ {\rm finite}\ ,$$ 
with
$$\eqalign{A_1 &=-{2\over b(\varepsilon) } N_d=-X_1 \cr
A_2 &=- {4N\over b^2(\varepsilon)} (d-3)b(\varepsilon) N_d=-2N(d-3)X_1\,,
\cr} $$ 
where  $A_1$ and $A_2$ correspond to the diagrams of figures \efigiv~and
\efigv~respectively. \par
Applying the RG equation one finds the relation at order $1/N$
$$\beta_{\vv^2}(\vv)=2\vv^2\eta(\vv)-2A_1 \vv^4 D^{-1}(\vv)-2A_2
\vv^6 D^{-2}(\vv) . \eqnn $$
We thus obtain
$$\eqalignno{\eta(\vv) &={ \varepsilon\vv^2 \over 4-\varepsilon}X_1
D^{-1}(\vv), & \eqnn \cr
\beta_{\vv^2}(\vv) &={8 \vv^4 \over 4-\varepsilon}X_1 D^{-1}(\vv)
+4N(1-\varepsilon) \vv^6 X_1 D^{-2}(\vv), &\eqnn \cr}$$ 
where the first term in $\beta_{\vv^2}$ comes from $A_1$ and $\eta$
and the second from $A_2$.\par
Extracting the large $\vv^2$ behaviour we find
$$\eqalignno{\eta  & =  { \varepsilon \over N (4-\varepsilon)} X_1 +O(1/N^2).
,& \eqnn \cr 
\rho&={4(3-\varepsilon)(2-\varepsilon)\over N(4-\varepsilon)}X_1 > 0\,,\cr}$$
and thus
$${1\over\nu}=d-2 + {2(3-\varepsilon)(2-\varepsilon)\over N(4-\varepsilon)}X_1
+O(1/N^2) .\eqnn $$
\subsection Additional results
 
The calculations beyond the order $1/N$ are rather technical. The reason
is easy to understand: Because the effective
field theory is  renormalizable in all dimensions $2\leq d \leq 4$, the
dimensional regularization, which is so useful in perturbative calculations,
no longer works. Therefore either one keeps a true cut-off or one
introduces more sophisticated regularization schemes. For 
details the reader is referred to the literature. 
\medskip
{\it Generic dimensions}. The exponents $\gamma$ and  $\eta$ 
are known up to order $1/N^{2}$ and $1/N^3$ respectively in arbitrary
dimensions but the expressions are too complicated to be reproduced here.
The expansion of $\gamma$ up to order
$1/N$ can be directly deduced from the results of the preceding
sections:
$$ \gamma  =  { 1 \over 1- \varepsilon / 2}\left(1-{3 \over 
2N}X_1 \right)+O \left({1 \over N ^{2}} \right) .\eqnn $$
The exponents $ \omega $ and $\theta=\omega\nu$, governing the leading
corrections to scaling, can also be calculated for example from the
$\left<\lambda^2\lambda\lambda\right>$ function:
$$ \eqalignno{\omega & = \varepsilon\left(1-{2(3-\varepsilon)^2 \over
(4-\varepsilon)N}X_1\right) +O \left({1 \over N^{2}}
\right), & \eqnn \cr
\theta=\omega \nu &={\varepsilon \over 2-\varepsilon}
\left(1-{2(3-\varepsilon) \over N}X_1\right) +O \left({1 \over N^{2}}
\right). & \eqnn \cr}$$ 
Note that the exponents are regular functions of $\varepsilon$ up to
$\varepsilon =2$ and free of renormalon singularities at $\varepsilon
=0$.\par
The equation of state and the spin--spin correlation function in zero field
are also known at order $ 1/N $, but since the expressions are
complicated we refer the reader to the literature for details.
\medskip  
{\it Three dimensional results.} Let us give the expansion of $\eta$  in
three dimensions at the order presently available:
$$ \eta  = {\eta_{1} \over N}+{\eta_{2} \over N^{2}}+ {\eta_{3} \over N^{3}}
+O \left({1 \over N^{4}} \right), $$ 
with
$$\eta_1 ={\textstyle{8 \over 3\pi^{2}}}\, ,\quad \eta_2= -{\textstyle {8
\over 3} \eta_1^2}\, ,\quad 
\eta_{3} = \eta_1^3 {\textstyle\left[ -{797 \over 18} - { 61 \over 24}\pi^{2}
+ {27 \over 8}\psi''\left({1 / 2}\right) + {9 \over 2}\pi^{2} \ln 2
\right]}, $$  
$\psi(x)$ being the logarithmic derivative of the $\Gamma$ function.\par
The exponent $\gamma$ is  known only up to order $1/N^{2}$:
$$\gamma  = 2 -{\textstyle {24 \over N \pi^{2}}} + {\textstyle{64 \over
N^{2}\pi^{4}}\left({44 \over 9} - \pi^{2} \right)+O \left({1 \over N^{3}}
\right)}. $$ 
Note that the $1/N$ expansion seems to be rapidly divergent and certainly a
direct summation of these terms does not provide very good 
estimates of critical exponents in 3 dimensions for useful values of $N$.
\subsection Dimension four: triviality, renormalons, Higgs mass
 
A number of issues concerning the physics of the $(\phib^2)^2$  theory
in four dimensions can be addressed within the framework of the large $N$
expansion. For  simplicity reasons we consider here only the critical
(i.e~massless) theory. 
\medskip
{\it Triviality and UV renormalons.} It is
easy to verify that the renormalized coupling constant $g_\r$, 
defined as the  value of the vertex $\left<\sigma\sigma\sigma\sigma\right>$ at
momenta of order $\mu\ll\Lambda$, is given by:
$$g_\r={g\over1+\frac{1}{6}N g B_\Lambda(\mu)}\,,\eqnd\egfivivren $$
where  $B_\Lambda(p)$ corresponds to the bubble diagram  (figure \efigii) 
$$B_\Lambda(p)\mathop{\sim}_{p\ll \Lambda} {1\over 8\pi^2}\ln (\Lambda/p)+\
{\rm const.}\,.  \eqnd\eBivasym $$
We see that when the ratio $\mu/\Lambda$ goes to zero, the renormalized
coupling constant vanishes, for that all $g$. %% (see section \sssLMofiv)
This is the so-called {\it triviality}\/ property. In the standard treatment
of quantum field field, one usually insists in taking the 
infinite cut-off $\Lambda$ limit. Here one then finds only a free field
theory. 
Another way of formulating the problem is the following: it is impossible to
construct in four dimensions a $\phi^4 $ field  theory consistent (in the
sense of satisfying all usual physical requirements) on all scales for non
zero coupling. Of course in the logic of {\it effective} field theories this
is no longer an issue. The triviality property just implies that the
renormalized or effective charge is logarithmically small as indicated by
equations \eqns{\egfivivren,\eBivasym}. Note that if $g$ is generic (not too
small) and $\Lambda/\mu$ large, $g_r$ is essentially independent of the
initial coupling constant. Only if the bare coupling is small is the
renormalized coupling  an adjustable, but bounded, quantity.\par
Let us now imagine that we work formally and, ignoring the problem, we
express the leading contribution to the four-point function in terms of the
renormalized constant:  
$${g\over 1+{N\over 48\pi^2}g\ln (\Lambda/p)}={g_\r\over 1+{N\over
48\pi^2}g_\r\ln (\mu/p)} \,.$$
We then find that the function has a pole for
$$p=\mu\e^{48\pi^2/(Ng_\r)}.$$
This pole corresponds to the Landau ghost for this theory which has
$g=0$ as an IR fixed point. If we calculate contributions of 
higher orders, for example to the two-point function, this pole makes
the loop integrals  diverge. In an expansion in powers of
$g_\r$, each term is instead calculable 
but one finds, after renormalization, UV contributions of the type
$$\int^\infty{\d^4 q\over q^6}\left(-{Ng_\r\over48\pi^2}\ln(\mu/q)\right)^k
\mathop{\propto}_{k\to\infty}\left({Ng_\r\over96\pi^2}\right)^k k!\,.$$
The perturbative manifestation of the Landau  ghost is the appearance of
contributions to the perturbation series which are not Borel summable. 
By contrast the contributions due to the finite momentum region, which can be
evaluated by a semiclassical analysis, are Borel summable, but invisible for
$N$ large. This effect is called  UV renormalon effect. Note 
finally that this UV problem is independent of the mass of the field $\phib$,
that we have taken zero for simplicity  reasons.
\medskip
{\it IR renormalons.}
We now illustrate the problem of IR renormalons with the same example
of the massless  $(\phib^2)^2$ theory (but now zero mass is essential), in
four dimensions, in the large $N$ limit. We calculate the contribution 
of the small momentum region to the mass renormalization, at
cut-off $\Lambda$ fixed. In the large $N$ limit the mass renormalization is
then proportional to (see equation \eONpropi) 
$$\int^\Lambda{\d^4 q\over q^2\bigl(1+\frac{1}{6}NgB_\Lambda(q)\bigr)}
\sim\int{\d^4 q\over q^2\bigl(1+{N\over 48\pi^2}g\ln (\Lambda/q) \bigr)}\,.$$
It is easy to expand this expression in powers of the coupling constant $g$.
The term of order $k$ in the limit $k\to\infty$  behaves as $(-1)^k 
(N/ 96\pi^2)^k k!$. This contribution has the alternating sign of the
semiclassical contribution. Note that more generally for $N$ finite
on finds $(-\beta_2/2)^k k!$. IR singularities are responsible for
additional, Borel summable, contributions to the large order behaviour.\par
In a theory asymptotically free for large momentum, clearly the roles
of IR and UV singularities are interchanged.
\medskip
{\it The mass of the $\sigma$ field in the phase of broken  symmetry}.
The $\phi^4$ theory is a piece of the Standard Model, and the field
$\sigma$ then represents the Higgs field. %%In section \sssmHiggs~we have
%% shown that  
With some reasonable assumptions it is possible to establish for finite $N$
a  semi-quantitative bound on the Higgs mass. Let us examine here what
happens for $N$ large.\par
In the phase of broken symmetry the action, after translation of
average  values, includes a term proportional to $\sigma\lambda$ and thus the
propagators of the fields $\sigma$ and $\lambda$ are elements of a  $2\times2$
matrix $\bf M$: 
$${\bf M^{-1}}(p)=\pmatrix{p^2 & \sigma \cr \sigma & -3/u-\ud N B_\Lambda(p)
\cr} \ ,$$
where $\sigma= \left<\sigma(x)\right>$.
In four dimensions $B_\Lambda$ is given by equation \eBivasym.
It is convenient to introduce a  mass scale $M$, RG invariant, such
that $${48\pi^2\over Nu}+8\pi^2 B_\Lambda(p)\sim \ln(M/p),$$
and thus
$$M\propto \e^{48\pi^2/Nu}\Lambda\,.$$
The mass of the field $\sigma$ at this order is a solution to the equation
$\det{\bf M}=0$. One finds
$$p^2 \ln (M/p)=-(16\pi^2/N)\sigma^2\ \Rightarrow\ m_\sigma^2\ln (i M/
m_\sigma) =(16\pi^2/N)\sigma^2.$$
The mass $m_\sigma$ solution to the equation is complex, because the particle
$\sigma$ can  decay into massless Goldstone bosons. At $\sigma$ fixed, the
mass decreases when the cut-off increases or when the coupling  constant goes
to zero. Expressing  that the mass must be smaller than the cut-off, one
obtains an upper-bound on $m_\sigma$ (but which slightly depends on the chosen
regularization). %% that one can compare with the one obtained in the general
%% discussion of section \sssmHiggs. 
%
\subsection Finite size effects

Another question can be studied in the large  $N$ limit, 
finite size effects. It is difficult to discuss all possible finite size
effects because the results depend both on the geometry of the system and on
the boundary conditions. In particular one must discuss separately boundary
conditions depending whether they break or not translation invariance. In
the first case new effects appear which are surface effects, and that we do
not examine here. 
Note that the periodic conditions are not the only ones which
preserve translation invariance. For systems which have a symmetry one can
glue the boundaries after having made a group transformation. Thus
here one could 
also choose  antiperiodic conditions or more generally fields differing by a
transformation of the $O(N)$ group.\par 
Moreover if we are interested only in qualitative aspects we
can limit ourselves to a simple geometry, in each direction the system having
the same finite size $L$, all other sizes being infinite (but we thus
exclude some questions concerning  crossover regimes).
Even so the number of different possible situations remains large, 
and we limit ourselves here to two examples.
\par
We consider the example of periodic boundary conditions in two cases: 
finite volume (the geometry of the hypercube or rather hypertorus)
in this section, and QFT at finite temperature in next section.
\par
From the point of view of renormalization  group, finite size effects, which
only affect the IR domain, do not change UV divergences. The RG equations
remain the same, only the solutions are modified by the appearance of new
dimensional quantities. Thus if finite sizes are characterized by only one
length $L$, solutions will be functions of an additional  argument $L/\xi$
where $\xi$ is the correlation length. \par
A property characteristic of a system of finite size is the
quantification of momenta in Fourier space.
For periodic conditions, if we call $L$ the
size du system in each direction, we have
$$p_\mu=2\pi n_\mu /L \,,\quad n_\mu\in {\Bbb Z}\,.$$
In particular, in a massless theory the zero mode
${\bf p}=0$  now corresponds to an isolated pole of the propagator.
This automatically leads to IR divergences in all dimensions.
Therefore in equations \esaddleN{} the solution $\sigma\ne 0$ no longer
exists. This is not surprising: there are no phase  transitions in a
finite volume. Neglecting corrections to scaling laws we can then write
equation \emgNsig{b}: 
$$1= (N-1)t L^{-d}\sum_{n_\mu}{1\over m^2+(2\pi {\bf n}/L)^2}\,,
\eqnd\esaddNL $$
where the sums are cut by a cut-off $\Lambda$.\par
To discuss the equation it is convenient to introduce the function
$A(s)$ (related to Jacobi's elliptic functions)
$$A(s)= \sum^{+ \infty}_{n=- \infty} \e^{-sn^{2}}. \eqnd\eJacobi $$
Using Poisson's transformation it is easy to show
$$A(s) = (\pi/s)^{1/2} A\left(\pi^2/s\right).\eqnd \ePoisson $$
Using this definition, and introducing the critical temperature $t_c$,
one can write equation \esaddNL~(for $2<d<4$)
$${1\over t}-{1\over t_c}=(N-1) L^{-d}\int_0^\infty\d s\left(\e^{-s m^2}
A^d(4\pi^2 s/L^2) -L^d(4\pi s)^{-d/2}\right). \eqnd\eFSsigN $$
Setting $s\mapsto L^2 s$ and introducing the function $F$:
$$F(z)=\int_0^\infty\d s\left(\e^{-sz^2} A^d(4\pi^2 s )-(4\pi
s)^{-d/2}\right), \eqnn $$
we can rewrite the relation
$${1\over t}-{1\over t_c}=(N-1) L^{2-d}F(mL) . \eqnn $$
For $|t-t_c|\ll \Lambda^{d-2}$ we find a scaling form 
which is in agreement with the RG result, which predicts ($1/\nu=d-2+O(1/N)$):
$$L m(t,L)=L/\xi(t,L)=f\bigl((t-t_c)L^{1/\nu}\bigr) . $$
Here the length $\xi$ has the meaning of a correlation length only for
$\xi<L$. Since $\eta=0$, the magnetic susceptibility $\chi$ in zero field
instead is always given by $\chi=t/m^2$. \par
One verifies that for $t>t_c$ fixed, $L\to\infty$ and thus $mL\to\infty$ one
recovers the infinite volume limit. On the contrary in the low
temperature phase for $t<t_c$ fixed, $L\to\infty$, $mL$ goes to zero.
Thus the contribution of the zero mode dominates in the r.h.s.~of equation
\esaddNL. Using the relation \ePoisson~one then finds 
$$\eqalign{F(z)&={1\over z^2}+K(d)+O\left(z^2\right), \cr
K(d)&=\int^\infty_0\d s\left[A^d(4\pi^2 s)-1-(4\pi
s)^{-d/2}\right], \cr}$$
and thus
$$\chi(L,t)={t\over m^2}={1\over N-1}(1-t/t_c)L^d-t L^2
K(d)+O\left(L^{4-d}/(t-t_c)\right)   . \eqnn $$
We see that the susceptibility diverges with the volume, an indication of
the existence of a broken symmetry phase.\par
Note finally that it is instructive to make a similar analysis
for different boundary conditions which have no zero mode.
\par
For $d=2$ the regime where finite size effects are observables
corresponds to $t\ln(L\Lambda)=O(1)$, i.e.~to a regime of low
temperature. The zero mode dominates for $t\ln(L\Lambda)\ll 1$, and the
susceptibility is then given by
$$\chi(t,L)\sim \frac{1}{N}L^2\left[1+O(t\ln(L\Lambda))\right] \,.$$
\subsection Field theory at finite temperature

Quantum field theory at finite temperature can be considered as a system
which has a finite size in one direction. Indeed the partition function 
is given by $\tr\e^{-L H}$, where $H$ is the hamiltonian and
$L^{-1}$ the temperature. For a scalar field theory with euclidean lagrangian
density ${\cal L}(\phi)$ this leads to the functional integral
$$Z=\int[\d\phi]\exp\left[-\int_0^L\d\tau\int\d^{d-1}x\,{\cal
L}(\phi)\right],$$ 
where the field $\phi$ satisfies periodic boundary conditions only in one
direction  
$$\phi(\tau=0,x)=\phi(\tau=L,x).$$
Let us again consider, as an example, the non-linear $\sigma$ model.
We find a finite size system, but the interpretation of parameters is
different. The variable $t$ now represents the coupling  constant of the QFT.
Since $L$ is the inverse temperature, the limit $L\to\infty$ corresponds to
the limit of vanishing temperature.\par 
The saddle point equation \emgNsig{b}, in the symmetric phase 
$\sigma=0$, becomes
$$1= (N-1)t {1\over (2\pi)^{d-1}L}\int\d^{d-1} k \sum_{n}{1\over m^2+k^2+(2\pi
n/L)^2}\,. \eqnd\esaddNTf $$
On immediately verifies that the IR problem induced by the zero mode has the
following consequences: since one integrates only over $d-1$ dimensions, a 
phase transition is only possible for $d>3$. Qualitatively at large distance
the condition of finite temperature leads to a property of
{\it dimensional reduction}\/ $d\mapsto d-1$. 
The large $N$ expansion is thus particularly well suited
to the study of this problem which exhibits a crossover between
two different dimensions.\par
Again using Schwinger's  representation of the propagator,
integrating over $k$ and introducing the function \eJacobi~we can
rewrite equation \esaddNTf:
$$\eqalignno{{1\over t}-{1\over t_c}&= {N-1\over (4\pi)^{(d-1)/2}} L^{2-d}
G(mL)&\eqnn \cr G(z)&=\int_0^\infty \d s\,s^{-(d-1)/2}\left[\e^{- z^2
s}A(4\pi^2 s) -(4\pi s)^{-1/2}\right].&\eqnn \cr} $$
Here $\xi_L=m^{-1}$ has really the meaning of a correlation length. \par
This equation has a scaling form for $d<4$. The behaviour of the system then
depends on the ratio between $L$ and the correlation length  
$\xi_\infty$ of the system at zero temperature.
For $t>t_c$ fixed and $L$ large (with respect to $1/\Lambda$) we
recover the zero temperature limit. For $t-t_c$ small we find
a crossover between a regime of small and high temperature.
In the regime $t<t_c$ fixed and $L$ large, we have to examine the  
behaviour of $G(z)$ for $z$ small. \par
At $d=3$:
$$G(z)=-2\ln z +\ {\rm const.}\  .$$
Hence 
$${1\over m^2}\propto \chi(L,t)\propto L^2\exp\left[{4\pi L\over
N}\left({1\over t}-{1\over t_c}\right) \right]. \eqnn $$
One finds that $\xi_L$ remains finite below $t_c$ for all non vanishing
temperatures, and has when the coupling constant $t$  goes to zero or
$L\to\infty$ the
exponential behaviour characteristic of the dimension two. \par
For $d=4$ the situation is different because a transition is possible
in dimension $d-1=3$. This is consistent with the existence of the quantity
$G(0)>0$ which appears in the relation between coupling constant and
temperature at the critical point:
$${1\over t}-{1\over t_c}={(N-1)G(0)\over (4\pi)^{3/2}}{1\over
L^2}\,.\eqnn $$
For a coupling constant $t$ which corresponds to a phase of broken symmetry
at zero temperature ($t<t_c$), one now finds a transition temperature
$L^{-1}\propto \sqrt{t_c-t}$. Studying more generally the saddle point
equations one can derive all other properties of this system. 

\subsection Other methods. General vector field theories

The large $N$ limit can be obtained by several other algebraic methods.
Without being exhaustive, let us list a few. Schwinger--Dyson equations
for $N$ large lead to a self-consistent equation for the two-point function.
From the point of view of stochastic quantization or critical dynamics the
Langevin equation also becomes linear and self-consistent for $N$ large.
One replaces $\phib^2(x,t)$ by $\left<\phib^2(x,t)\right>$ 
($\left<\cdot\right>$ means noise average) at leading order.
Finally a version of the Hartree--Fock approximation also yields the large $N$
result. \sslbl\ssfivNge 
\medskip
{\it General vector field theories.} We now briefly explain how the algebraic
method presented in section \ssfivNi~can be generalized to actions which  have
a more complicated dependence in one or several vector fields. Again in a
general $O(N)$ symmetric field theory the composite fields with small
fluctuations are the scalars constructed from all vectors. The strategy is
then to introduce pairs of fields and Lagrange multipliers for all independent
$O(N)$ invariant scalar products constructed from the many-component fields.
\par  
Let us first take the example of one field $\phib$ and assume that the
interaction is an arbitrary function of the only invariant $\phib^{2}(x)$
$$S(\phib)= \int\d^d x \left\{\ud \left[ \partial_{\mu} \phib (x) \right]^{2}
+V\left(\phib^2\right) \right\} .\eqnd{\eactONg}$$
We then introduce two fields $\ro(x)$ and $\lambda(x)$ and use the identity: 
$$\exp\left[-\int \d^{d}x\,V(\phib^{2})\right] \propto \int
\left[\d\ro(x)\,\d\lambda(x) \right] \exp\left\{-\int
\d^{d}x\left[\ud\lambda\left(\phib^{2}-\ro\right) +   
V(\ro) \right]\right\}.\eqnd\egeniden $$
In the special case in which $V(\ro)$ is a quadratic function, the
integral over $\ro$ can be performed. In all cases, however, the identity
\egeniden\ transforms the action into a quadratic form in $\phib$ 
and therefore the integration over $\phib$ can be performed and the
dependence in $N$ becomes explicit. This method will be applied
in section \ssdblescal~to the study of multi-critical points and double
scaling limit.
\par
If the action is an $O(N)$ invariant function of two fields $\phib_{1}$ and
$\phib_{2}$ the potential depends on the three scalar products
$\phib_{1}\cdot\phib_{2}$, $\phib^{2}_{1}$ and $\phib^{2}_{2}$. Then three
pairs of fields are required. 

\beginbib

As shown by Stanley the large $N$-limit of the classical $N$-vector model
coincides with the spherical model solved by Berlin and Kac\rf
T.H. Berlin and M. Kac, {\it Phys. Rev.} 86 (1952) 821;
H.E. Stanley, {\it Phys. Rev.} 176 (1968) 718.
\nrf Early work on calculating critical properties includes\rf
R. Abe, Prog. Theor. Phys. 48 (1972) 1414; 49 (1973) 113, 1074, 1877;
S.K. Ma, {\it Phys. Rev. Lett.} 29 (1972) 1311; {\it Phys. Rev.} A7 (1973)
2172; 
M. Suzuki, {\it Phys. Lett.} 42A (1972) 5; {\it Prog. Theor. Phys.} 49 (1973)
424, 1106, 1440; 
R.A. Ferrel and D.J. Scalapino, {\it Phys. Rev. Lett.} 29 (1972) 413;
K.G. Wilson, {\it Phys. Rev.} D7 (1973) 2911.
\nrf The contribution of order $1/N$ to the equation of state is given in\rf
E. Br\'ezin and D.J. Wallace, {\it Phys. Rev.} B7 (1973) 1967.
\nrf The spin--spin correlation in zero field is obtained in\rf
M.E. Fisher and A. Aharony, {\it Phys. Rev. Lett.} 31 (1973) 1238;
A. Aharony, {\it Phys. Rev.} B10 (1974) 2834;
R. Abe and S. Hikami, {\it Prog. Theor. Phys.} 51 (1974) 1041.
\nrf The exponent $\omega$ has been calculated to order $1/N$ in\rf
S.K. Ma, {\it Phys. Rev.} A10 (1974) 1818.
\nrf See also the contributions of S.K. Ma and E. Br\'ezin, J.C. Le Guillou
and J. Zinn-Justin to\rf
{\it Phase Transitions and Critical Phenomena} vol. 6, C. Domb and M.S. Green
eds. (Academic Press, London 1976).  
\nrf The consistency of the $1/N$ expansion to all orders has been proven
in\rf 
I. Ya Aref'eva, E.R. Nissimov and S.J. Pacheva, {\it Commun. Math. Phys.} 71
(1980) 213; 
A.N. Vasil'ev and M.Yu. Nalimov, {\it Teor. Mat. Fiz.} 55 (1983) 163.
\nrf At present the longest $1/N$ series for exponents and amplitudes are
found in\rf 
I. Kondor and T. Temesvari, {\it J. Physique Lett. (Paris)} 39 (1978)
L99;  
Y. Okabe and M. Oku, {\it Prog. Theor. Phys.} 60 (1978) 1277, 1287; 61 (1979)
443; 
A.N. Vasil'ev, Yu.M. Pis'mak and Yu.R. Honkonen, {\it Teor. Mat. Fiz.} 46
(1981) 157; 50 (1982) 195. 
\nrf See also\rf
I. Kondor, T. Temesvari and L. Herenyi, {\it Phys. Rev.} B22 (1980) 1451.
\nrf Renormalization of operators is discussed in \rf
K. Lang and W. R\"uhl, {\it Nucl. Phys.} B400 (1993) 597; {\it Z. Phys.} C61
(1994) 459. 
\nrf The case of long range forces has been discussed in\rf
S.K. Ma, {\it Phys. Rev.} A7 (1973) 2172.
\nrf For the Hartree--Fock point of view and QFT at finite temperature see\rf
W.A. Bardeen and M. Moshe,  {\it Phys. Rev.} D28 (1983) 1372.\par
Results concerning the $\beta$-function at order $1/N$ in the massive
theory renormalized at zero momentum have been recently reported in \rf
A. Pelissetto and E. Vicari, {\it Nucl. Phys.} B519 (1998) 626,
cond-mat/9711078.\nrf
A calculation of the dimensions of composite operators to order $1/N^2$ is 
reported in
S. E. Derkachov, A. N. Manashov
{\it Nucl. Phys.} B522 (1998) 301, hep-th/9710015;
{\it Phys. Rev. Lett.} 79 (1997) 1423, hep-th/9705020, 
\nrf and the consequences for the stability of the fixed point of the 
non-linear $\sigma$ model discussed.
\nrf
Some finite size calculations are reported in\rf
S. Caracciolo and A. Pelissetto, preprint hep-lat/9804001. 
\endbib

\vfill\eject
%four fermi interaction
\def\ggn{u}

\section{Gross--Neveu and Gross--Neveu--Yukawa Models}

To illustrate the techniques developed in sections \scLTs, \scfivN,
we now discuss models with fermions exhibiting the phenomenon of chiral phase
transition. Again we consider two different field theory models with
the same symmetries, the Gross--Neveu (GN) and the Gross--Neveu--Yukawa
(GNY) models. The GN model is renormalizable in two dimensions, and 
describes in perturbation theory only one phase, the symmetric phase.
The GNY model is renormalizable in four dimensions and instead allows
a perturbative analysis of the chiral phase transition.
We now show that the physics of these models can indeed be studied by the
same techniques as ferromagnetic systems, that is RG equations near
two and four dimensions, and large $N$ expansion.
\subsection The Gross--Neveu model

The GN model is described in 
terms of a $U(N)$ symmetric action for a set of $N$ massless Dirac fermions
$\{\psi^i, \bar\psi^i \}$:\sslbl\appGNmod 
$$S\left(\bar \psib, \psib\right)= -\int \d^d x \left[ \bar\psib\cdot
\sla{\partial} \psib +\ud G\left(\bar\psib\cdot \psib \right)^2 \right].$$
The GN model has in even dimensions a discrete chiral symmetry:  
$$\psib \mapsto \gamma_S \psib, \quad\bar \psib \mapsto -\bar\psib \gamma_S\,
, \eqnd{\echirald}$$
which prevents the addition of a fermion mass term while in odd dimensions a
mass term breaks space parity. Actually the two
symmetry operations can be written in a form
$${\bf x}=\{x_1,x_2,\ldots, x_d\}\mapsto\tilde {\bf
x}=\{-x_1,x_2,\ldots,x_d\}, 
\quad \cases{\psi(x)\mapsto \gamma_1 \psi(\tilde x), \cr
\bar\psi(x)\mapsto -\bar\psi(\tilde x)\gamma_1 \cr},$$
valid in all dimensions.
\par
This model illustrates the physics of spontaneous fermion mass 
generation and, in even dimensions, chiral symmetry breaking. 
It is renormalizable and asymptotically free in two dimensions.
However, as in the case of the non-linear $\sigma$ model, the perturbative
GN model describes only one phase. The main difference is that the
role of the spontaneously  broken and the explicitly symmetric phase
are interchanged. Indeed it is always the massless phase which is
unstable in low dimensions. \par 
Since the symmetry breaking mechanism is non-perturbative
it will eventually be instructive to compare the GN model
with a different model with the same symmetries: the
Gross--Neveu--Yukawa model. 
\medskip
{\it RG equations near and in two dimensions.}
The GN model is renormalizable in two dimensions, and in perturbation
theory describes only the massless symmetric phase.
Perturbative calculations in two dimensions can be made with an IR cut-off
of the form of a mass term $M\bar\psi\psi$, which breaks softly the chiral
symmetry. It is possible to use dimensional regularization in
practical calculations. Note that in two dimensions the symmetry group
is really $O(2N)$, as one verifies after some relabelling of the
fields. Therefore the $(\bar \psi \psi)^2$ interaction is
multiplicatively renormalized. It is convenient to introduce here a
dimensionless coupling constant 
$$\ggn=G\Lambda^{2-d}. \eqnn $$
As  a function of the cut-off $\Lambda$ the bare correlation functions 
satisfy the RG equations:
$$\left[ \Lambda{ \partial \over \partial
\Lambda} +\beta (\ggn){\partial \over \partial \ggn}-{n \over
2}\eta_{\psi}(\ggn)-\eta_M(\ggn)M{\partial \over \partial M} \right]
 \Gamma^{(n)}\left(p_{i};\ggn,M,\Lambda \right)=0\, .\eqnd{\eRGGN}$$ 
A direct calculation of the $\beta$-function in $d=2+\varepsilon$ 
dimension yields: % constested in 2+\epsilon !!
$$\beta(\ggn)=\varepsilon \ggn-(N'-2){\ggn^2 \over
2\pi}+(N'-2){\ggn^3 \over4\pi^2}+{(N'-2)(N'-7)\over 32\pi^3}\ggn^4
+O\left(\ggn^5\right),\eqnd{\ebetGNii}$$
Note that for $d=2$ $N'=2N$.\par
The special case $N'=2$, for which the $\beta$-function vanishes identically
in two dimensions, corresponds to the Thirring model (because for $N'=2$
$(\bar\psi \gamma_{\mu}\psi)^2=-2 (\bar \psi\psi )^2$). The latter model is to
the equivalent the sine-Gordon or the $O(2)$ vector model.  \par
Finally the field and mass RG functions are 
$$\eqalignno{\eta_{\psi}(\ggn)&={N'-1\over 8 \pi^2}\ggn^2-{(N'-1)(N'-2) \over 
32\pi^3} \ggn^3+{(N'-1)(N'-2)(N'-5)\over 128\pi^4}\ggn^4
, \hskip10mm &\eqnd\egamii \cr
\eta_M(\ggn)&={N'-1\over 2\pi}\ggn-{N'-1\over 8\pi^2}\ggn^2-{(2N'-3)(N'-1)\over
32\pi^3} \ggn^3. & \cr} 
$$
As in the case of the non-linear $\sigma$  model, the solution of the RG
equations \eRGGN~involves a length scale $\xi$ of the type of a correlation
length which is a RG invariant
$$\xi^{-1}(\ggn)\equiv \Lambda(\ggn) \propto \Lambda\exp\left[-\int^\ggn{\d
\ggn'\over \beta(\ggn')}\right]\,. \eqnd\eferMass $$   
\medskip
{\it Two dimensions.}
For $d=2$ the model is asymptotically free. In the chiral theory ($M=0$)
the spectrum, then,
is non-perturbative, and many arguments lead to the conclusion that
the chiral symmetry is always broken and a fermion  mass generated.
From the statistical point of view this corresponds to the existence
of a gap in the spectrum of fermion excitation (as in a superfluid or
superconductor). All masses are proportional to the mass parameter
$\Lambda(\ggn)$ which is a RG invariant. Its dependence in the
coupling constant is given by equation \eferMass: 
$$\Lambda(\ggn)\propto \Lambda
\ggn^{1/(N'-2)}\e^{-2\pi/(N'-2)\ggn}\bigl(1+O(\ggn)\bigr)\,. \eqnn $$    
We see that the continuum limit, which is reached when the masses are
small compared to the cut-off, corresponds to $\ggn\to 0$.\par
$S$-matrix considerations have then led to the conjecture that, for $N$
finite, the spectrum is: 
$$m_n=\Lambda(\ggn) {2(N-1) \over \pi}\sin\left({n\pi \over
2(N-1)}\right),\quad 
n=1,2\ldots <N\,,\ N>2\,, $$ 
To each mass value corresponds a representation of the $O(2N)$ group.
The nature of the representation leads to the conclusion that $n$ odd
corresponds to fermions and $n$ even to bosons.\par
This result is consistent with the spectrum for $N$ large evaluated by 
semiclassical methods. In particular the ratio of the masses of the
fundamental fermion and the lowest lying boson is: 
$${m_{\sigma}\over m_{\psi}}=2\cos\left({\pi \over 2(N-1)}\right)=2+O(1/N^2).
\eqnd{\eGNspec}$$
The large $N$ limit will be recovered in section \sssGNYN.\par
Note that the two first values of $N$ are special, the model $N=2$ is
conjectured to be equivalent to two decoupled sine-Gordon models.
\medskip
{\it Dimension $d=2+\varepsilon$.} 
As in the case of the $\sigma$-model, asymptotic freedom implies the
existence of a non-trivial UV fixed point $\ggn_c$, in $2+\varepsilon$
dimension 
$$\ggn_c={2\pi \over N'-2}\varepsilon\left(1-{\varepsilon \over N'-2}\right)
+O\left(\varepsilon^3\right). $$
$\ggn_c$ is also the critical coupling constant for the transition between a
phase in which the chiral symmetry is spontaneously broken and a massless
small $\ggn$ phase.  \par
At the fixed point one finds the correlation length exponent $\nu$:
$$\nu^{-1}=-\beta'(\ggn_c)=\varepsilon-{\varepsilon^2 \over N'-2}
+O\left(\varepsilon^3\right).\eqnd{\expnuii}$$
The fermion field dimension $[\psi]$ is:
$$2[\psi]=d-1+\eta_{\psi}(\ggn_c)= 1+\varepsilon+ {N'-1 \over
2(N'-2)^2}\varepsilon^2 +O\left(\varepsilon^3\right) .\eqnd{\expetaii}$$
The dimension of the composite field $\sigma=\bar\psib\psib$ is given by
$$[\sigma]=d-1-\eta_M(\ggn_c)=1-{\varepsilon \over N'-2}\,. $$
As for the $\sigma$-model the existence of a non-trivial UV fixed point
implies that large momentum behaviour is not given by perturbation theory
above two dimensions, and this explains why the perturbative result
that the model cannot be defined in higher dimensions cannot be trusted. 
However, to investigate whether the $\varepsilon$ expansion makes
sense beyond an infinitesimal neighbourhood of dimension two other methods are
required, like the $1/N$ expansion which will be considered in section
\sssGNYN. 
\subsection The Gross--Neveu--Yukawa model

The Gross--Neveu--Yukawa (GNY) model has the same chiral and $U(N)$
symmetries as the GN model. The action is ($\varepsilon=4-d$): 
$$S\left(\bar \psib, \psib,\sigma\right)=\int \d^d x \left[- \bar\psib\cdot
\left(\sla{\partial}+g\Lambda^{\varepsilon/2}\sigma \right) \psib
+\ud\left(\partial_{\mu}\sigma\right)^2+\ud m^2 \sigma^2+{\lambda \over 4!}
\Lambda^{\varepsilon}\sigma^4 \right],\eqnd{\eactGNg} $$
where $\sigma$ is an additional scalar field, $\Lambda$ the momentum cut-off,
and $g,\lambda$ dimensionless ``bare" i.e.~effective coupling constants
at large momentum scale $\Lambda$.\par
The action still has a reflection symmetry, $\sigma$ transforming into
$-\sigma$ when the fermions transform by \echirald. 
In contrast with the GN model, however, the chiral transition can here
be discussed by perturbative methods. An analogous situation has already been
encountered when comparing the $(\phib^2)^2$ field theory with the non-linear
$\sigma$ model. Even more, the GN model is renormalizable in dimension two and
the GNY model in dimension four. 
\medskip
{\it The phase transition.} Examining the action \eactGNg~we see that in the
tree approximation when $m^2$ is negative the chiral symmetry is spontaneously
broken. The $\sigma$ expectation value gives a mass to the fermions, a
mechanism reminiscent of the Standard Model of weak-electromagnetic
interactions:  
$$m_{\psi}=g\left<\sigma \right>,\eqnn $$
while the $\sigma$ mass then is:
$$m^2_{\sigma}={\lambda \over 3g^2}m^2_{\psi}\,.\eqnd{\eratiobf}$$
As a result of interactions the transition value $m^2_c$ of the parameter
$m^2$ will be modified. In what follows we set
$$m^2=m^2_c+t\,, \eqnn $$
where the new parameter $t$, in the language of phase transitions, plays 
the role of the deviation from the critical temperature.\par
To study the model beyond the tree approximation we now discuss RG equations
near four dimensions. 
\subsection RG equations near four dimensions

The model \eactGNg~is trivial above four dimensions, renormalizable in four
dimensions and can thus be studied near dimension 4 by RG techniques. 
%Calling $\mu$ the renormalization scale, setting $d=4-\varepsilon$, we can
%write the renormalized action: 
%$$\eqalignno{ S_\r\left(\bar \psib, \psib,\sigma\right)& =\int \d^d x
%\biggl[- Z_{\psi}\bar\psib\cdot\sla{\partial} \psib
%-\Lambda^{\varepsilon/2}gZ_\psi\sigma \bar\psib\cdot \psib & \cr
%&\quad +\ud Z_{\sigma}\left(\partial_{\mu}\sigma\right)^2+\ud (Z_\sigma m^2_c
%+ Z_m t) \sigma^2+\Lambda^{\varepsilon} Z^2_\sigma{\lambda \over 4!} 
%\sigma^4 \biggr],\hskip10truemm&\eqnd{\eactGNr} \cr}$$
%where $m^2_c$ is the critical bare mass squared, the critical temperature in
%statistical language, and $t$ characterizes the deviation from the critical
%temperature. 
Five renormalization constants are required, corresponding to the two
field renormalizations, the $\sigma$ mass, and the two coupling constants.
The RG equations thus involve five RG functions. The 1PI correlation
functions $\Gamma^{(l,n)}$, for $l$ $\psi$ and $n$ $\sigma$ fields, then 
satisfy
$$\left(\Lambda{\partial\over \partial \Lambda}+\beta_{g^2}{\partial \over
\partial 
g^2}+\beta_{\lambda}{\partial \over \partial\lambda} -\ud l\eta_{\psi}-\ud
n\eta_{\sigma}-\eta_m t{\partial \over \partial t} \right)\Gamma^{(l,n)}=0\,.
\eqnd{\eRGiv}$$ 
\midinsert
\epsfxsize=100.mm
\epsfysize=45.mm
\centerline{\epsfbox{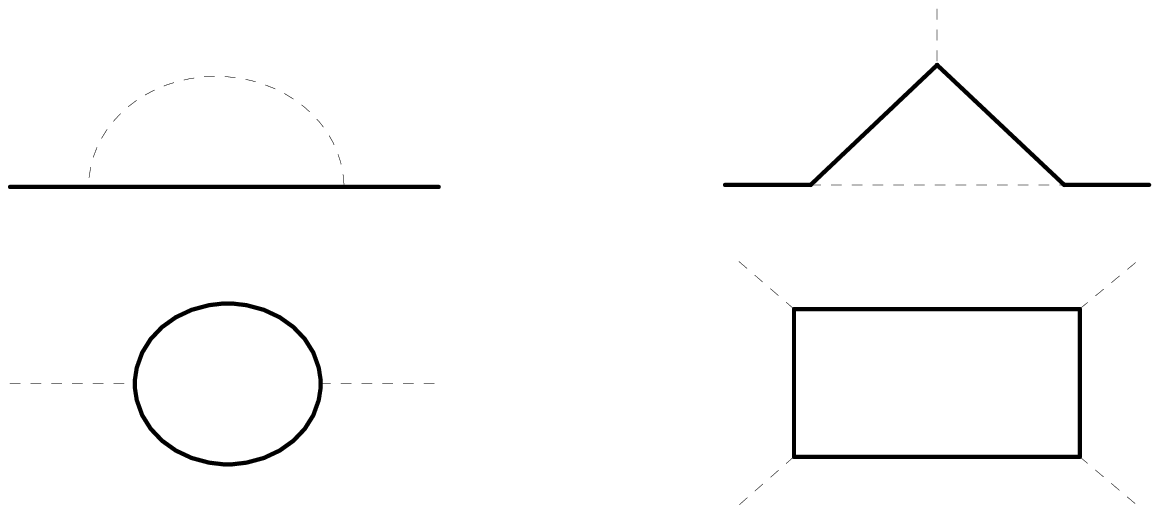}}
\figure{3.mm}{One-loop diagrams: fermions are represented by solid lines.} 
\endinsert
\def\efigvi{6}

\medskip
{\it The RG functions.} The RG functions at one-loop
order involve the calculation of the diagrams of figure \efigvi. One finds:
$$\eqalignno{\beta_{\lambda}&= -\varepsilon \lambda+{1\over 8\pi^2}
\left({3\over 2}\lambda^2+4N\lambda g^2-24N g^4\right), &\eqnd{\ebetl} \cr 
\beta_{g^2}&= -\varepsilon g^2+{2N+3 \over8\pi^2} g^4.  &\eqnd{\ebetgde}
\cr}$$  
Note that in these expressions for convenience we have set in the algebra of
$\gamma$ matrices $\tr {\bf 1}=4$ as in four dimensions. To extrapolate the
results to other dimensions one has to replace everywhere $N$ by $N'/4$,
where $N'=N\tr{\bf1}$ is the total number of fermion degrees of freedom.
\medskip
{\it Dimension four.} 
In four dimensions the origin $\lambda=g^2=0$ is IR stable. 
Indeed the second equation implies that $g$ goes to zero, and the first then
that  $\lambda$ also goes to zero. As a consequence if the bare coupling
constants are generic, i.e.~if the effective couplings at cut-off scale are
of order 1, the effective couplings at scale $\mu\ll\Lambda$ go to zero 
and in a way asymptotically independent from the bare couplings.
One finds
$$g^2(\mu)\sim {8\pi^2 \over (2N+3)\ln(\Lambda/\mu)}\,,\quad
\lambda(\mu)\sim {8\pi^2 \tilde \lambda^*\over \ln(\Lambda/\mu)}\,,$$
where we have defined
$$\tilde \lambda_*={48 N\over (2N+3)\left[(2N-3)+\sqrt{4N^2+132N+9}\right]}.
\eqnn $$ 
This result allows to use renormalized perturbation theory to calculation
physical observables. For example we can evaluate the ratio between the
masses of the scalar and fermion fields. It is then optimal to take for
$\mu$ a value of order $\langle\sigma\rangle$. A remarkable consequence
follows: the ratio \eratiobf~of scalar and fermion masses is fixed
$${m^2_{\sigma}\over m^2_{\psi}}={\lambda_* \over
3g_*^2}={16N\over (2N-3)+\sqrt{4N^2+132N+9} }, \eqnd\emassratio $$
while in the classical limit it seems arbitrary.\par
Of course if the bare couplings are ``unnaturally'' small the same will
apply to the renormalized couplings at scale $\mu$ and the ratio will be
modified. 
\medskip
{\it Dimension $d=4-\varepsilon$.}
One then finds a non-trivial IR fixed point (we recall $N'=N\tr{\bf1}$):
$$g_*^{2}={16\pi^2\varepsilon \over N'+6},\quad \lambda_*=8\pi^2\varepsilon
\tilde \lambda_* \,.\eqnd{\estar}$$
The  matrix of derivatives of the $\beta$-functions has two eigenvalues
$\omega,\omega'$,
$$\omega_1=\varepsilon, \quad \omega_2=\varepsilon
\sqrt{N'{}^2+132N'+36}/(N'+6), 
\eqn$$
and thus the fixed point is IR stable. The first  eigenvalue is always the
smallest.\par 
The field renormalization RG functions are at the same order:
$$\eta_{\sigma}={N'\over 16\pi^2}  g^2,\qquad
\eta_{\psi}= {1 \over 16\pi^2}  g^2 .\eqnn $$
At the fixed point one finds
$$\eta_{\sigma}={N'\varepsilon\over N'+6},\quad \eta_{\psi}={\varepsilon
\over (N'+6)} , \eqnd\expeps $$
and thus the dimensions $d_\psi$ and $d_\sigma$ of the fields
$$d_\psi={3\over2}-{N'+4 \over2(N'+6)}\varepsilon\,,\qquad 
d_\sigma=1-{3\over N'+6}\varepsilon\,. \eqnn $$
The  RG function $\eta_m$ corresponding
to the mass operator is at one-loop order:  
$$\eta_m=-{\lambda \over 16\pi^2}-\eta_{\sigma}\,,$$
and thus the exponent $\nu$:
$${1\over\nu}=2+\eta_m =2-{\varepsilon\over2}\tilde\lambda_*
-  {N'\varepsilon\over N'+6}=  2-\varepsilon{5N'+6+\sqrt{N'{}^2+132N'+36}
\over6(N'+6)}.\eqnn $$
Finally we can evaluate the ratio of masses \eratiobf\ at the fixed point:
$${m^2_{\sigma}\over m^2_{\psi}}={\lambda_* \over
3g_*^2}={8N'\over (N'-6)+\sqrt{N'{}^2+132N'+36}} .$$
In $d=4$ and $d=4-\varepsilon$ the existence of an IR fixed point has the 
same consequence: If we assume that the $\sigma$ expectation value is much
smaller than the cut-off and that the coupling
constants are generic at the cut-off scale, then {\it the ratio  of fermion
and scalar masses is fixed}.
\subsection GNY and GN models in the large $N$ limit

We now show that the GN model plays with respect to the GNY
model \eactGNg~the role the non-linear $\sigma$-model plays with respect to
the $\phi^4$ field theory.
%The large $N$ behaviour of the expressions of section
%. For example
%we find $\lambda_*\sim 48\pi^2/N$ and $d-2+\eta_{\sigma}=2$. This reminds us
%the $(\phib^2)^2$ field theory and suggests a study of the large $N$ limit.
For this purpose we start from the action \eactGNg~of the GNY model
and integrate over $N-1$ fermion fields. We also rescale for
convenience $\Lambda^{(4-d)/2}g\sigma$ into $\sigma$, and then get
the large $N$ action:\sslbl\sssGNYN
$$\left.\eqalign{ S_N\left(\bar \psi, \psi,\sigma\right)&=\int \d^d x \left\{-
\bar\psi \left(\sla{\partial} +\sigma \right) \psi
+\Lambda^{d-4}\left[ {1\over2g^2}\left(\partial_{\mu}\sigma\right)^2+{m^2
\over2g^2}\sigma^2+{\lambda\over 4!g^4} 
\sigma^4 \right]\right\}\cr &\quad -(N-1)\tr \ln\left(\sla{\partial}+\sigma
\right).\cr}\right.\eqnd{\eactefGN} $$
To take the large $N$ limit we assume  $\sigma$ finite and $g^2 ,
\lambda =O(1/N)$.\par
Let us call $V(\sigma)$ the action per unit volume for constant field
$\sigma(x)$ and vanishing fermion fields
$$\eqalignno{V(\sigma)&=\Lambda^{d-4}\left({m^2\over2g^2}\sigma^2+{\lambda\over
4!g^4} \sigma^4 \right) -N\tr \ln\left(\sla{\partial}+\sigma
\right) \cr
&=\Lambda^{d-4}\left({m^2\over2g^2}\sigma^2+{\lambda\over
4!g^4} \sigma^4 \right) -{ N'\over 2}\int^\Lambda{\d^d q\over
(2\pi)^d}\ln(q^2+\sigma^2) . &\eqnn \cr}$$
The expectation value of $\sigma$ for $N$ large is given by a {\it
gap}\/ equation: 
$$V'(\sigma)\Lambda^{4-d}={m^2\over g^2} \sigma+{\lambda \over
6g^4}\sigma^3-N'\Lambda^{4-d}{\sigma \over (2\pi)^d} \int^\Lambda
{\d^d q \over q^2+\sigma^2}=0\,.\eqnd\evac $$ 
It is also useful to calculate the second derivative to check
stability of the extrema
$$V''(\sigma)\Lambda^{4-d}={m^2\over g^2}+{\lambda \over
2g^4}\sigma^2+N' \Lambda^{4-d} \int^\Lambda
{\d^d q\over (2\pi)^d} {\sigma^2-q^2 \over( q^2+\sigma^2)^2}.$$ 
The solution $\sigma=0$ is stable provided
$$V''(0)>0\ \Leftrightarrow\ {m^2\over g^2}>N' \Lambda^{4-d}{1\over
(2\pi)^d} \int^\Lambda {\d^d q\over q^2} .$$
Instead the non-trivial solution to the gap equation exists only
for
$${m^2\over g^2}>N' \Lambda^{4-d}{1\over (2\pi)^d} \int^\Lambda {\d^d
q\over q^2}  ,$$
but then it is stable. We conclude that the critical temperature or
critical bare mass is given by:
$${m^2_c\over g^2}=N'  \Lambda^{4-d}{1\over (2\pi)^d} \int^\Lambda {\d^d q
\over q^2},\eqnd\eTcrit $$ 
which shows that the fermions favour the
chiral transition. In particular when $d$ approaches 2 we observe that
$m^2_c\to +\infty$ which implies that the chiral symmetry is always broken in
2 dimensions. Using equation \eTcrit~and setting 
$$t=\Lambda^{d-4}(m^2-m^2_c)/g^2 , \eqnn $$
we can write the equation for the non-trivial solution
$$t +\Lambda^{d-4}{\lambda \over 6g^4}\sigma^2+N'{\sigma^2 \over
(2\pi)^d} \int^\Lambda {\d^d q \over q^2( q^2+\sigma^2)}=0\,.$$
We now expand the integral for $\sigma$ small (equation \eintasym)
$$D_1(\sigma^2)={ 1 \over \left(2\pi \right)^{d}} \int^{\Lambda}{ \d 
^{d}q \over q^{2} \left(q^{2}+\sigma^{2} \right)}  =C(d)\sigma^{-\varepsilon}
-a(d)\Lambda^{-\varepsilon} + O \left({\sigma^{2-\varepsilon} \over
\Lambda^{2}} \right). \eqnn $$ 
Keeping only the leading terms for $t\to 0$ we obtain for $d<4$ the
scaling behaviour 
$$  \sigma\sim (-t/N'C)^{1/(d-2)}.\eqnd{\expbet} $$
Since, at leading order, the fermion mass $m_{\psi}=\sigma$, it immediately 
follows that the exponent $\nu$ is also given by:
$$\nu\sim \beta\sim 1/(d-2)\ \Rightarrow\ 
\eta_{\sigma}=4-d \,.\eqnd{\enu}$$ 
At leading order, for $N\to\infty$, $\nu$ has the same value as in the
non-linear $\sigma$-model.\par 
At leading order in the scaling limit the thermodynamical (or
effective) potential $V(\sigma)$ then becomes
$$ V(\sigma)= \ud t\sigma^2+(N'/d)C(d)|\sigma|^d .\eqnd\efpotGN $$
We note that, although in terms of the $\sigma$-field the model has a simple
Ising-like symmetry, the scaling equation of state for large $N$ is quite
different. \par
We read from the large $N$ action that at this order $\eta_{\psi}=0$.\par
Finally from the large $N$ action we can calculate the $\sigma$-propagator
at leading order. 
Quite generally, using the saddle point equation,  one finds for the inverse
$\sigma$-propagator in the massive phase:  
$$\eqalignno{\Delta^{-1}_{\sigma}(p)&=\Lambda^{d-4}\left({p^2\over
g^2}+{\lambda \over3g^4 }\sigma^2\right) \cr&\quad + {N' 
\over 2 (2\pi)^d }\left(p^2+4\sigma^2 \right)\int^\Lambda{\d^d q \over \left(
q^2+\sigma^2 \right) \left[(p+q)^2 +\sigma^2 \right]}.&\eqnd\eGNsigprop \cr}$$
We see that in the scaling limit $p,\sigma\to 0$, the integral yields the
leading contribution. Neglecting corrections to scaling we find that
the propagator vanishes for $p^2=-4\sigma^2$ which is 
just the $\bar\psi\psi$  threshold. Thus, in this limit,  $m_{\sigma}=2
m_{\psi}$ in all dimensions, a result consistent with $d=2$ exact value.\par
At the transition the propagator reduces to
$$\Delta_\sigma\sim {2 \over N' b(\varepsilon)  p^{d-2}},\eqnn
$$  
with (equation \econstb)
$$ b \left(\varepsilon \right)=-{\pi \over\sin(\pi d/2)} 
{\Gamma^2 (d/ 2) \over \Gamma (d-1)}N_d \,. \eqnn $$
The result is consistent with the value of $\eta_{\sigma}$ found above.\par
Let us finally note that the behaviour of the propagator at the critical
point, $\Delta_\sigma(p) \propto p^{2-d}$, implies for the field $\sigma$  
the canonical dimension $[\sigma]$ in the large $N$
expansion, for $2\le d\le 4$: 
$$[\sigma]=1\,. \eqnn $$ 
\medskip
{\it Corrections to scaling and the IR fixed point.}  The IR fixed point is
determined by demanding the cancellation of the leading corrections to
scaling. Let us thus consider the effective potential $V(\sigma)$. The
leading correction to scaling is proportional to
$$\left({\lambda\over4! g^4}-{N' a(d)\over 4}\right)\sigma^4 ,$$
($ a \left(\varepsilon \right)\sim{1 / 8\pi^{2}\varepsilon}
$).
Demanding the
cancellation of the coefficient of $\sigma^2$, we obtain a relation between
$\lambda$ and $g^2$
$$ g_*^4={\lambda_*\over 6N' a(d)}
={4\lambda_*\varepsilon \pi^2\over 3N'}+O\left(\varepsilon^2\right),$$
a result consistent with the results of the $\varepsilon$-expansion.\par 
In the same way it is possible to calculate the leading correction to the
$\sigma$-propagator \eGNsigprop. Demanding the cancellation of the leading
correction we obtain 
$${p^2\over g^2_*}+{\lambda_* \over3g_*^4}\sigma^2 -\ud N'
\left(p^2+4\sigma^2\right)a(d)=0\,.$$
The coefficient of $\sigma^2$ cancels from the previous relation and
the cancellation the coefficient of $p^2$ yields
$$g_*^2={2\over N'a(d)}
={16\pi^2\varepsilon\over N'}
+O\left(\varepsilon^2\right),$$    
in agreement with the $\varepsilon$-expansion for $N$ large.
\medskip
{\it  The relation to the GN model for dimensions $2\le d\le 4$.}
We have seen that the terms $(\partial_{\mu}\sigma)^2$ and $\sigma^4$ of the
large $N$ action which have a canonical dimension 4, are irrelevant in the IR
critical region for $d\le 4$.  We recognize a situation already encountered in
the $(\phib^2)^2$ field theory in the large $N$ limit. In the scaling region
it is possible to omit them and one then finds the action: 
$$S_N\left(\bar \psib, \psib,\sigma\right)=\int \d^d x \left[-
\bar\psib\cdot \left(\sla{\partial}+\sigma \right) \psib
+\Lambda^{d-4}{ m^2\over2g^2} \sigma^2 \right].\eqnd\eactGN $$
The integral over the $\sigma$ field can explicitly be performed and yields
the action of the GN model:
$$S_N\left(\bar \psib, \psib\right)=-\int\d^d x  \left[\bar\psib\cdot
\sla{\partial} \psib +{\Lambda^{4-d} \over 2m^2}g^2\left(\bar\psib\cdot \psib
\right)^2 \right].$$ 
The GN and GNY models are thus equivalent for the large
distance physics. In the GN model, in the large $N$ limit,  the $\sigma$
particle appears as a $ \bar\psi\psi$ boundstate at threshold.\par 
Conversely, it would seem that the GN model depends on a smaller number of
parameters than its renormalizable extension. Again this problem is only
interesting in four dimensions where corrections to scaling, i.e.\ to free
field theory, are important. However, if we examine the divergences of the
term $\tr\ln \left(\sla{\partial}+\sigma \right)$ in the effective action
\eactefGN\ relevant for the large $N$ limit, we find a local polynomial in
$\sigma$ of the form:
$$\int\d^4x\left[A \sigma^2(x)+B\left(\partial_{\mu}\sigma\right)^2 +C
\sigma^4(x)\right].$$
Therefore the value of the determinant can be modified by a local polynomial
of this form by changing the way the
cut-off is implemented: additional parameters, as in the case
of the non-linear $\sigma$-model, are hidden in the cut-off procedure.
Near two dimensions these operators can be identified with $(\bar\psi\psi)^2,
[\partial_{\mu} (\bar\psi\psi)]^2,  (\bar\psi\psi)^4$. It is clear that by
changing the cut-off procedure we change the amplitude of higher dimension
operators. These bare operators in the IR limit have a component on all
lower dimensional renormalized operators.  \par
Note finally that we could have added to the GNY model an explicit
breaking term linear in the $\sigma$ field, which becomes a fermion mass term
in the GN model, and which would have played the role of the magnetic field
of the ferromagnets. 
\subsection The large $N$ expansion

Using the large $N$ dimension of fields and power counting arguments one can
then prove that the $1/N$ expansion is renormalizable with arguments quite
similar to those presented in section \sssfivNRT. 
\medskip
{\it Alternative theory.} To prove that the large $N$
expansion is renormalizable one proceeds as in the case of the scalar theory
in section \sssfivNRT. One starts from a critical action with an additional
term quadratic in $\sigma$ which generates the large $N$ $\sigma$-propagator
already in perturbation theory
$$S(\psi,\bar\psi,\sigma)=\int \d^d x\left[-\bar\psi
(\sla{\partial}+\sigma)\psi +{1\over
2\vv^2}\sigma(-\partial^2)^{d/2-1}\sigma\right] .\eqnn $$ 
The initial theory  is recovered in the limit $\vv\to\infty$. One then
rescales $\sigma$ in $\vv \sigma$. The model is renormalizable without
$\sigma$ field renormalization because divergences generate only local 
counter-terms
$$S_\r(\psi,\bar\psi,\sigma)=\int \d^d x\left[-Z_\psi\bar\psi
(\sla{\partial}+\vv_\r Z_\vv\sigma)\psi +{1\over
2}\sigma(-\partial^2)^{d/2-1}\sigma\right] .\eqnn $$ 
RG equations follow 
$$\left[\Lambda {\partial \over \partial \Lambda}+
\beta_{\vv^2}(\vv){\partial \over \partial \vv^2}-{n\over
2}\eta_\psi(\vv)\right] \Gamma^{(l,n)}=0\,.\eqnn $$ 
Again the large $N$ expansion is obtained by first summing the bubble
contributions to the $\sigma$-propagator. We define
$$D(\vv)={2\over b(\varepsilon)}+N' \vv^2 .$$
Then the large $N$ $\sigma$ propagator reads
$$\left<\sigma\sigma\right>={2\over b(\varepsilon)D(\vv) p^{d-2}}.\eqnn $$
The solution to the RG equations can be written:
$$\Gamma^{(l,n)}(\tau p, \vv,\Lambda)=Z^{-n/2}(\tau)
\tau^{d-l-n(d-2)/2} \Gamma^{(l,n)}(p, \vv(\tau),\Lambda)  ,
\eqnn $$ 
with the usual definitions
$$\tau{\d \vv^2\over \d \tau}=\beta(\vv(\tau))\,,\quad \tau{\d \ln Z
\over \d \tau}=\eta_\psi(\vv(\tau))\, .$$
We are interested in the neighbourhood of the fixed point $\vv^2=\infty$.
Then the RG function $\eta(\vv)$ approaches the exponent 
$\eta$. The flow equation
for the coupling constant becomes:
$$\tau{\d \vv^2\over \d \tau}=\rho \vv^2 , \ \Rightarrow\ \vv^2(\tau)\sim
\tau^{\rho}. $$ 
We again note that a correlation function with $l$ $\sigma$ fields becomes
proportional to $\vv^l$. Therefore
$$\Gamma^{(l,n)}(\tau p, \vv,\Lambda)\propto 
\tau^{d-(1-\rho/2)l-n(d-2+\eta_\psi)/2} .
\eqnn  $$ 
We conclude
$$ d_\sigma=\ud(d-2+\eta_\sigma)=1-\ud\rho \ \Leftrightarrow\
\eta_\sigma=4-d-\rho \,.\eqnn   $$
\medskip
{\it RG functions at order $1/N$.}
A new generic integral is useful here
$${1\over(2\pi)^d}\int{\d^d q (\sla{p}+\sla{q})\over (p+q)^{2\mu} q^{2\nu}}=
\sla{p}
p^{d-2\mu-2\nu}{\Gamma(\mu+\nu-d/2)\Gamma(d/2-\mu+1)\Gamma(d/2-\nu)
\over (4\pi)^{d/2}\Gamma(\mu)\Gamma(\nu)\Gamma(d-\mu-\nu+1)}
\, . \eqnn $$
We first calculate the
$1/N$ contribution to the fermion two-point function at the critical point
(from a diagram similar to diagram \efigiii)
$$\Gamma^{(2)}_{\bar\psi\psi}(p)=i\sla{p}+{2i \vv^2\over
b(\varepsilon)D(\vv)(2\pi)^d} 
\int^\Lambda{\d^d q (\sla{p}+\sla{q})\over q^{d-2}(p+q)^2}.$$
We need the coefficient of $\sla{p}\ln\Lambda/p$.
Since we work only at one-loop order we again replace the $\sigma$
propagator $1/q^{d-2}$ by $1/q^{2\nu}$, and send the cut-off to
infinity. The residue of the pole at $2\nu=d-2$ gives the coefficient
of the term $\sla{p} \ln\Lambda$ and the finite part the $\sla{p}\ln p$
contribution. We find
$$\Gamma^{(2)}_{\bar\psi\psi}(p)=i\sla{p}+{2i \vv^2\over
b(\varepsilon)D(\vv)}N_d \left(d-2\over d\right)
\sla{p}\ln(\Lambda/p)\,, \eqnn $$
where $N_d$ is the loop factor \eintNcor{a}.
Expressing that the $\left<\bar \psi \psi\right>$ function satisfies
RG equations we immediately obtain the RG function $\eta_\psi(\vv)$ 
$$\eta_\psi(\vv)={\vv^2\over D(\vv)}{(d-2) \over d}X_1\,,\eqnn $$
where $X_1$ is given by equation \eXone.
We then calculate the function $\left<\sigma\bar\psi\psi\right>$ at
order $1/N$ 
$$\Gamma^{(3)}_{\sigma\bar\psi\psi}(p)=\vv+A_1 D^{-1}(\vv)\vv^3 \ln\Lambda
 \,,$$ 
with
$$A_1 =-{2\over b(\varepsilon) } N_d =-X_1 \,,$$
where $A_1$ corresponds to the diagram of figure \efigiv. The diagram
of figure \efigv~vanishes because the $\sigma$ 3-point function vanishes for
symmetry reasons. \par 
The $\beta$-function follows
$$\beta_{\vv^2}(\vv) ={4(d-1) \vv^4 \over d}X_1 D^{-1}(\vv)
 ,\eqnn $$
and thus
$$\rho={8(d-1)N_d\over d b(\varepsilon) N'}={4(d-1)\over d N'}X_1\,.$$
The exponents $\eta_{\psi}$ and $\eta_\sigma$ at order $1/N$, and thus
the corresponding dimensions of fields $d_\psi, d_\sigma$ follow
$$\eqalignno{\eta_{\psi}&= {(d-2)\over d}{X_1\over N'}= {(d-2)^2 \over
d}{\Gamma(d-1)\over 
\Gamma^3(d/2)\Gamma(2-d/2)N'}.&\eqnd\eetapsii \cr
2d_\psi&=d-1-{2(d-2)\over d}{X_1\over N'}\,. &\eqnn \cr}$$
For $d=4-\varepsilon$ we find $\eta_{\psi}\sim \varepsilon/N'$, result
consistent with \expeps\ for  $N$ large. For $d=2+\varepsilon$ instead one
finds $\eta_{\psi}\sim \varepsilon^2/2N'$, consistent with \expetaii.
The dimension $d_\sigma$ of the field $\sigma$ is
$$d_\sigma=\ud(d-2+\eta_\sigma)=1-{2(d-1)\over d N'}X_1 +O(1/N'{}^2)
.\eqnn $$ 

A similar evaluation of the $\left<\sigma^2\sigma\sigma\right>$ function
allows to determine the exponent $\nu$ to order $1/N$
$${1\over\nu}=d-2-{2(d-1)(d-2)\over dN'}X_1\,. \eqnn $$
Actually all exponents are known to order $1/N^2$ except $\eta_\psi$ which is
known to order $1/N^3$.
\section Other models with chiral fermions

Let us for completeness shortly examine two other models with chiral
fermions, in which large $N$ techniques can be applied, massless QED
and the $U(N)$ massless Thirring model.
\subsection Massless electrodynamics

Let us give another example with a structure different 
from a Yukawa-type theory. We now consider a model of 
$N$ charged massless fermion fields $\psib,\bar \psib$, 
coupled  through an abelian gauge field $A_{\mu}$ (massless QED):
$$ S \left(\bar \psib ,\psib,A _{\mu} \right) = \int \d^{d}x
\left[{\textstyle{ 1 \over 4e^2}}  F ^{2} _{\mu \nu}(x) -
\bar \psib (x) \cdot\left(\sla{\partial} + i\Abar \right) 
\psib(x) \right] . \eqnd\eactQEDN $$
This model possesses, in addition to the $U(1)$ gauge invariance, a 
chiral  $U(N)\times U(N)$ symmetry because the fermions are 
massless. Again the interesting question is whether the model exhibits in some
dimension $2\le d\le 4$ a spontaneous breaking of the chiral symmetry. 
\medskip
{\it Dimension $d=4-\varepsilon$.} 
In terms of the standard coupling constant $\alpha$:
$$\alpha\equiv {e^2\over 4\pi}\,, \eqnn $$
the RG $\beta$  function reads (taking  $\tr {\bf 1}=4$ in the space of
$\gamma$ matrices):  
$$\eqalignno{\beta(\alpha)&=-\varepsilon \alpha +{2 N\over 3\pi}\alpha^2+
+{N\over2\pi^2}\alpha^3-{N(22N+9)\over144\pi^3}\alpha^4 \cr
&-{1\over 64 \pi^4}N\left[\frac{616}{243}N^2+\left(\frac{416}{9}\zeta(3)
-\frac{380}{27}\right)N+23\right]\alpha^5+
O\left(\alpha^6\right). &\eqnd\ebetQED \cr} $$
The model is free at low momentum  in four dimensions. Therefore
no phase  transition is expected, at least for $e^2$ small enough. A
hypothetical  phase transition would rely on the existence on non-trivial
fixed points outside of the perturbative regime. \par
In the perturbative framework the model provides an example of the famous 
triviality  problem. For a generic effective coupling constant at
cut-off scale (i.e.~bare coupling), the effective coupling constant at
scale $\mu\ll\Lambda$ is given by
$$\alpha(\mu)\equiv {e^2(\mu)\over 4\pi}\sim {3\pi\over2N\ln(\Lambda/\mu)}.$$
This result can be used  to bound $N$.\par
In  $4-\varepsilon$ dimension, one instead finds a non-trivial  IR fixed point
corresponding  to a coupling constant:  
$$e^2_*=24 \pi^2 \varepsilon\Lambda^{\varepsilon}/N'\,,$$
($N'=N\tr{\bf 1}$) and  correlation functions have a scaling behaviour
at long distance. 
As we have discussed in the case of the $\phi^4$ field  theory, the effective
coupling constant at large distance becomes close to the IR fixed  point,
except when the  initial coupling constant is very small. 
\par
The RG function associated  with the field renormalization is also
known at order $\alpha^3$ but this is a non-physical quantity since gauge
dependent
$$\eta_\psi= \xi {\alpha\over2\pi}-{4N+3\over 16\pi^2}\alpha^2+{40 N^2 +
54N+27 \over 576\pi^3}\alpha^3+O\left(\alpha^4\right),$$
where the gauge is specified by a term $(\partial_\mu A_\mu)^2/2\xi$.
\subsection The large $N$ limit

To evaluate correlation functions for $N$ large, one first integrates over the
fermion fields and one obtains the  effective action: 
$$S \left(\bar \psi ,\psi,A _{\mu} \right)= \int \d^{d}x
\left[{\textstyle{ 1 \over 4e^2}}  F^{2}_{\mu \nu}(x) -N\tr\ln
\left(\sla{\partial} + i\Abar \right)\right] . \eqnd{\eactQEDN} $$
The  large  $N$ limit is taken with $e^2N$ fixed. Therefore, at leading order,
only $S_2(A_\mu)$, the quadratic term in $A_{\mu}$ in the expansion
of the fermion determinant, contributes. A short calculation yields
$$\left.\eqalign{S_2(A_\mu)&=-N' \int \d^d k\, A_{\mu}(k)
\left[k^2\delta_{\mu\nu}-k_{\mu}k_{\nu}\right] A_{\nu}(-k)K(k), \cr
{\rm with}\quad K(k)&={d-2\over4 (d-1)}\left[b(\varepsilon)k^{d-4}
-a(d)\Lambda^{d-4}\right]+O\left(\Lambda^{-2}\right),\cr}\right.
\eqnd{\eAAlN}$$
where $a(d)$ is a  regularization-dependent constant. \par
For $d<4$ the leading term in the IR region comes from the integral. 
The behaviour at small momentum of the vector field is modified, which 
confirms the existence of a non-trivial IR fixed point. The fixed point 
is found by demanding cancellation of the leading corrections to scaling
coming from $F_{\mu\nu}^2$ and the divergent part of the loop integral,
$$e_*^2={2(d-1)\over (d-2)a(d)}{\Lambda^{4-d}\over N'}.$$
However there is still no indication of chiral symmetry breaking. 
Power counting within the $1/N$ expansion confirms that the IR
singularities have been eliminated, because the large $N$ vector propagator 
is less singular than in perturbation theory. Of course this result
is valid only for $N$ large. Since the long range forces generated by the
gauge coupling have not been totally eliminated the problem remains open
for $d$ not close to four, or for $e^2$ not very small and $N$ finite.
Some numerical simulations indeed suggest a  chiral phase transition for $d=4$
and $d=3$, $N\le N_c \sim 3$. \par 
The exponents corresponding to the IR fixed point have been calculated up to
order $1/N^2$. At order $1/N$ ($X_1$ is defined  by equation \eXone)
$$\eqalign{\eta_\psi&=-{(d-1)^2(4-d)\over
d(d-2)} {X_1\over 4N}+O\left(1/N^2\right) \cr
\eta_m& =-{(d-1)^2\over
d(d-2)} {X_1\over N}+O\left(1/N^2\right) \cr
\beta'(\alpha^*)&= 4-d -{(d-3)(d-6)(d-1)^2(4-d)\over
d(d-2)} {X_1\over 4N}+O\left(1/N^2\right)  \,. \cr}$$
Finally note that in the  $d=2$  limit, the integral generates a contribution
$Ne^2/\pi k^2$ times the propagator of the free gauge
field
$$N'K(k)\mathop{\sim}_{d\to 2}{N\over 2\pi}{1\over k^2}. $$
As a direct  analysis of the $d=2$  case confirms, this corresponds
to a massive bound state, of mass squared $Ne^2/\pi$. However, for
generic values of the coupling constant, this mass is of the order of
the cut-off $\Lambda$. Only when $e$ is small with respect to the
microscopic scale, as one assumes in conventional renormalized
perturbation theory, does this mass correspond in the continuum limit
to a propagating particle.   
\medskip
{\it Two dimensions.}
As stated above we now assume that the dimensional quantity $e^2$ is small
in the microscopic scale. The model is then a simple
 extension  of the Schwinger model and can be exactly solved in the same
way. For $N=1$ the model exhibits the simplest example of a  chiral
anomaly, illustrates  the property of confinement and spontaneous chiral
symmetry breaking. For $N>1$ the situation is more subtle. The neutral
$\bar\psi\psi$ two-point function decays algebraically
$$\left< \bar{\psib}(x)\cdot\psib(x)\bar{\psib}(0)\cdot\psib(0) \right>
\propto x^{2/N-2},$$
indicating the presence of a massless mode and
$\left<\bar\psi\psi\right>=0$. Instead if we calculate the two-point function
of the composite operator ${\cal O}_N(x)$
$${\cal O}_N(x)=\prod_{i=1}^N \bar{\psi}_i(x)\psi_i(x),$$
we find
$$\left<{\cal O}_N(x){\cal O}_N(0)\right>\propto \ {\rm const.}\ .$$
We have thus identified an operator which has a non-zero expectation value.
As a consequence of the fermion antisymmetry, if we perform a transformation
under the group $U(N)\times U(N)$ corresponding to matrices
$U_+,U_-$, the operator is multiplied  by $\det U_+/\det
U_-$. The operator thus is invariant under the group $SU(N)\times
SU(N)\times U(1)$. Its non-vanishing expectation value is the sign of the
spontaneous breaking of the remaining $U(1)$ chiral group.

\subsection The $U(N)$ Thirring model

We now consider the model
$$S(\bar\psib,\psib)= - \int \d^d x\left[\bar\psi\left(\sla{\partial}+m_0
\right) \psi - \ud g J_{\mu}J_{\mu}\right] ,\eqnd{\eactThir} $$
where
$$J_{\mu}=\bar\psib\gamma_{\mu}\cdot\psib\, . \eqnn $$
The special case $N=1$ corresponds to the simple Thirring model. In two
dimensions it is then equivalent to a free massless boson field theory (with
mass term for fermions one obtains the sine--Gordon model). Both to bozonize
the model in $d=2$ and to study that large $N$ properties one introduces a
abelian gauge field $A_\mu$ coupled to the current $J_\mu$
$$\ud g  J_{\mu}J_{\mu} \longmapsto A_{\mu}^2/2g +iA_{\mu}J_{\mu}\, .
\eqnn $$
One then finds massive QED without the $F^2_{\mu\nu}$ term
$$S(A_{\mu},\bar\psi,\psi)= - \int \d^2 x\left[\bar\psi\left(\sla{\partial}+
i\Abar+ m_0 \right)\psi -A_{\mu}^2 /2g\right] .
\eqnd{\eactAmu} $$
If we integrate over the fermions, the fermion determinant generates
a  kinetic term for the gauge field. For $m_0=0$ we are thus in
situation very similar to massless QED, except that the gauge field is massive.
\beginbib

\nrf Nambu and Jona-Lasinio were the first to propose a mechanism based on
a four-fermion interaction with $U(1)$ chiral invariance to generate nucleon,
scalar and pseudo-scalar $\sigma,\pi $ masses\rf
Y. Nambu and G. Jona-Lasinio, {\it Phys. Rev.} 122 (1961) 345.
\nrf The difficulties connected with this approach (approximate treatment
of Dyson--Schwin\-ger equations without small parameter,  non renormalizable
theory with cut-off) have partially solved in \rf
K.G. Wilson, {\it Phys. Rev.} D7 (1973) 2911;
D.J. Gross and A. Neveu, {\it Phys. Rev.} D10 (1974) 3235.
\nrf In these articles the $1/N$ expansion was introduced and the
existence of IR fixed points pointed out. The semi-classical spectrum for
$d=2$ in the large $N$ limit of the GN model (with discrete chiral
invariance) was obtained from soliton  calculation in
\rf
R. Dashen, B. Hasslacher and A. Neveu, {\it Phys. Rev.} D12 (1975) 2443.
\nrf This study as well as some additional considerations concerning the
factorization of $S$ matrix elements at order $1/N$ have led to a
conjecture of the exact spectrum at $N$ finite\rf
A.B. Zamolodchikov and Al.B. Zamolodchikov, {\it Phys. Lett.} 72B (1978) 481;
M. Karowski and H.J. Thun, {\it Nucl. Phys.} B190 (1981) 61.
\nrf See also \rf
P. Forgacs, F. Niedermayer and P. Weisz, {\it Nucl. Phys.} B367 (1991)
123, 144.
\nrf The properties of the NJL model in two dimensions are discussed in
\rf
J.H. Lowenstein, {\it Recent Advances in Field Theory and Statistical
Mechanics}, Les Houches 1982, J.B. Zuber and R. Stora
eds., (Elsevier Science Pub., Amsterdam 1984). 
\nrf The thermodynamics of the GN and NJL models at all temperatures and
densities at $d=2$ for $N\to\infty$ are discussed in \rf
R.F. Dashen, S.K. Ma and R. Rajaraman, {\it Phys. Rev.} D11 (1975) 1499,
\nrf where the existence of instantons responsible of the symmetry
restoration is demonstrated. More recently a more complete analysis has
appeared in\rf
A. Barducci, R. Casalbuoni, M. Modugno and G. Pettini, R. Gatto, {\it Phys.
Rev.} D51 
(1995) 3042.
\nrf The large $N$ expansion in $d=3$ has been discussed in\rf
B. Rosenstein, B. Warr and S.H. Park, {\it Phys. Rev. Lett.} 62 (1989) 1433;
{\it Phys. Rep.} 205 (1991) 59; G. Gat, A. Kovner and B. Rosenstein,
{\it Nucl. Phys.} B385 (1992) 76.
\nrf The relation with the GNY model is discussed in \rf
A. Hasenfratz, P. Hasenfratz, K. Jansen, J. Kuti and Y. Shen, {\it Nucl.
Phys.} B365 (1991) 79;
J. Zinn-Justin, {\it Nucl. Phys.}  B367 (1991) 105.
\nrf Recent investigations at $d=4$ can be found in \rf
P.M. Fishbane and R.E. Norton, {\it Phys. Rev.} D48 (1993) 4924; B.
Rosenstein, H.L. Yu and A. Kovner, {\it Phys. Lett.} B314 (1993) 381.
\nrf A number of $1/N$ and $2+\varepsilon$ calculations have been reported
W. Wentzel, {\it Phys. Lett.} 153B (1985) 297; N.A. Kivel, A.S.
Stepanenko, A.N. Vasil'ev, {\it Nucl. Phys.} B424 (1994) 619;
J.A. Gracey, {\it Nucl. Phys.} B367 (1991) 657; {\it Z.
Phys.} C61 (1994) 115; 
{\it Int. J. Mod. Phys.} A9 (1994) 567 and 727, hep-th/9306107;
{\it Phys. Rev.} D50 (1994) 2840, hep-th/9406162; {\it Phys. Lett.} B342 (1995)
297. 
\nrf
A comparison in dimension three between numerical simulations of the GN
model and the $\varepsilon$-expansion at second order obtained from the GNY
model is reported in
\rf 
L. K\"arkk\"ainen, R. Lacaze, P.Lacock and B. Petersson, {\it Nucl. Phys.}
B415 (1994) 781, Erratum {\it ibidem} B438 (1995) 650; E. Focht, J. Jerzak and
J. Paul, {\it Phys. Rev.} D53 (1996) 4616. 
\nrf The models are also compared numerically in dimension two in
\rf
A.K. De, E. Focht, W. Franski, J. Jersak and M.A. Stephanow, {\it Phys. Lett.}
B308 (1993) 327.
\nrf For recent rigorous results see \rf
C. Kopper, J. Magnen and V. Rivasseau, {\it Comm. Math. Phys.} 169 (1995) 121.
\nrf A few references on the Schwinger model and its relation with
the confinement problem: \rf 
J. Schwinger, {\it Phys. Rev.} 128 (1962) 2425;
J.H. Lowenstein and J.A. Swieca, {\it Ann. Phys. (NY)} 68 (1971) 172;
A. Casher, J. Kogut and L. Susskind, {\it Phys. Rev.} D10 (1974) 732;
S. Coleman, R. Jackiw and L. Susskind, {\it Ann. Phys. (NY)} 93 (1975) 267;
S. Coleman, {\it Ann. Phys. (NY)} 101 (1976) 239.
\nrf A few references on Schwinger and Thirring models (which have also been
systematically investigated) for $N$ large
\rf
G. Parisi, {\it Nucl. Phys.} B100 (1975) 368; S. Hikami and T. Muta, {\it
Prog. Theor. Phys.} 57 (1977) 785; 
D. Espriu, A. Palanques-Mestre, P. Pascual and R. Tarrach, {\it Z. Phys} C13
(1982) 153; A. Palanques-Mestre and P. Pascual, {\it Comm. Math. Phys.} 95
(1984) 277; 
J.A. Gracey, {\it Nucl. Phys.} B414 (1994) 614; S.J. Hands, {\it Phys. Rev.}
D51 (1995) 5816; S.J. Hands, {\it Phys. Rev.} D51 (1995) 5816;
S. Hands, hep-lat/9806022. 
\nrf For the calculation of the QED RG $\beta$ function in the $\overline{\rm
MS}$  scheme see\rf 
S.G. Gorishny, A.L. Kataev and S.A. Larin, {\it Phys. Lett.} B194 (1987) 429;
B256 (1991) 81.
\nrf A recent simulation concerning the 3D Thirring model is reported in \rf
L. Del Dubbio, S.J. Hands, J.C. Mehegan, {\it Nucl. Phys.} B502 (1997)
 269,  hep-lat/9701016. 
\nrf Finally these analytic techniques have also been applied to the
supersymmetric extension of the non-linear $\sigma$ model and other models,
see for example \rf 
J.A. Gracey, {\it Nucl. Phys.} B352 (1991) 183; 
P.M. Ferreira, I. Jack, D.R.T. Jones, {\it Phys. Lett.} B399 (1997) 258, 
hep-ph/9702304; P.M. Ferreira, I. Jack, D.R.T. Jones, C.G. North, {\it Nucl.
Phys.} B504 (1997) 108, hep-ph/9705328. 

\endbib 
\vfill\eject
% V(\phi^2) and double scaling limit
% ************** math symbols ********************************************
\def\to{\rightarrow}
\def\half{\frac{1}{2}}
\def\del{\partial}
\def\phib{{\phi}}
\def\Vc{\Omega_c}
\newskip\tableskipamount \tableskipamount=10pt plus 4pt minus 4pt

\def\upar{\uparrow\kern-9.\exec1pt\lower.2pt \hbox{$\uparrow$}}

%%%%%%%%%%%%%%%%%%%%%%%%%%%%%%%%%%%%%%%%%%%%%%%%%%%%%%%%%

\section{The $O(N)$ vector model in the large $N$ limit:
multi-critical points and double scaling limit}

We now discuss the large $N$ limit of the general $N$-vector models
with one scalar field. To illustrate the method we study 
multi-critical points. Of particular interest are the subtleties
involved in the stability of the phase structure at critical dimensions.
 \sslbl\ssdblescal\par
Another issue involves the so-called {\it double scaling limit}. 
Statistical mechanical properties of random surfaces as well as randomly
branched polymers can be analyzed within the framework of large $N$ expansion.
In the same manner in which matrix models in their double scaling limit  
provide representations  of dynamically triangulated random
surfaces summed on different topologies,  
$O(N)$ symmetric vector models represent discretized branched polymers
in this limit, where
$N\to\infty$ and the coupling constant $g \to g_{c}$ in a correlated 
manner. The surfaces in the case of matrix models, and the randomly
branched polymers in the case of vector models are classified by the 
different topologies of their
Feynman graphs and thus by powers of $1/N$.  Though  matrix theories attract
most  attention, a detailed understanding of these theories exists
only for dimensions  $d \leq 1$. On the other hand,
in  many cases, the $O(N)$ vector models can be successfully studied  also  in
dimensions $d > 1$, and thus, provide us with intuition for the search 
for a possible description of quantum field theory
in terms of extended objects in four dimensions,
which is a long lasting problem in elementary 
particle theory.  

The double scaling limit in $O(N)$ vector quantum field theories 
reveals an interesting phase structure beyond $N\to \infty$ limit.
In particular, though the $ N  \to \infty $ multicritical  structure 
of these models is generaly well understood, there are certain cases where
it is still unclear which of the features survives at finite $N$, and to what
extent. One such problem is the multicritical behavior
of $O(N)$ models {\it at critical dimensions}.
Here, one finds that in the $N \to \infty$ limit, 
there exists a non-trivial UV fixed point, scale invariance
is spontaneously broken, and the one parameter family of ground
states contains a massive vector and a massless bound state, a Goldstone 
boson-dilaton.
However, since it is unclear whether this structure is likely to survive
for finite $N$ one would like to know whether 
it is possible to construct a local field theory of a massless 
dilaton via the double scaling limit, 
where all orders in
$1/N$ contribute. The double scaling limit is viewed as the limit at which the
attraction between the $O(N)$ vector quanta reaches a value at $g \to g_c$,
at which a massless bound state
is formed in the $N \to \infty$ limit, while the 
mass of the vector particle stays finite. In this
limit, powers of $1/N$ are compensated by IR singularities and thus all
orders in $1/N$ contribute.
\par
In section \sintQM~the double scaling limit for simple integrals 
and quantum mechanics is explained, introducing a formalism which will be
useful for field theory examples. \par
In section \stwodim~the special case of field theory in dimension two is
discussed.\par
In higher dimensions a new phenomenon arises: the possibility of a spontaneous
breaking of the $O(N)$ symmetry of the model, associated to the Goldstone
phenomenon. 

Before discussing a possible double scaling limit, the critical and
multicritical points of the $O(N)$ vector model are examined in section
\smulticr. In particular, a certain sign ambiguity that appears in the
expansion of the gap equation is noted, and related to the existence
of the IR fixed point in dimensions $2<d<4$ discussed in section
\sssEGRN. In section  \sboundst~we discuss the subtleties and
conditions for the existence of an $O(N)$ singlet massless bound state 
along with a small mass  $O(N)$ vector particle excitation. It is pointed out
that the correct massless effective field theory is obtained after the
massive $O(N)$ scalar is integrated out. Section \sdouble~is devoted to 
the double scaling limit with a particular emphasis on this limit in theories
at their critical dimensions. In section \sconclud~the main conclusions are
summarized. 
\subsection Double scaling limit: simple integrals and quantum mechanics

We first discuss $d=0$ and $d=1$ dimensions, dimensions in
which the matrix models has equally been solved. We however introduce a
general method, not required here, but useful in the general field theory
examples. \sslbl\sintQM  
\medskip
{\it The zero dimensional example.} 
Let us first consider the zero dimensional example. The partition function $Z$
is given by
$$\e^Z=\int\d^N\phib\exp\left[-NV\left(\phib^2\right)\right] .$$
The simplest method for discussing the large $N$ limit is of course to
integrate over angular variables. Instead we introduce two
new variables $\lambda,\rho$ and use the identity
$$\exp\left[-N V(\phib^{2})\right] \propto \int
\d\rho\,\d\lambda\exp\left\{-N\left[\half\lambda\left(\phib^{2}-\rho\right) +
V(\rho) \right]\right\}.\eqnd\egenideno $$
The integral over $\lambda$ is really a Fourier representation of
a $\delta$-function and thus the contour of integration runs parallel to the
imaginary axis. The identity
\egenideno\ transforms the action into a quadratic form in $\phib$.
Hence the integration over $\phib$ can be  performed and the
dependence in $N$ becomes explicit
$$\e^Z \propto 
 \int \d\rho\,\d\lambda\exp\left\{-N\left[-\half\lambda\rho + 
V(\rho)+\half \ln\lambda \right]\right\}. $$
The large $N$ limit is obtained by steepest descent. The saddle point is given
by 
$$V'(\rho)=\half\lambda\,,\qquad \rho=1/\lambda\,.$$
The leading contribution to $Z$ is proportional to $N$ and obtained by
replacing $\lambda,\rho$ by the saddle point value. 
The leading correction is obtained by expanding $\lambda,\rho$ around the
saddle point and performing the gaussian integration. It involves the
determinant $D$ of the matrix $\bf M$ of second derivatives
$$ {\bf M}=\pmatrix{-\half\lambda^{-2} & -\half \cr -\half &
V''(\rho) \cr} \,,\quad D=\det{\bf
M}=-\half\left(V''(\rho)/\lambda^2+\half\right).$$  
In the generic situation the resulting contribution to $Z$ is $-\half \ln D$.
However if the determinant $D$ vanishes the leading order integral is no
longer gaussian, at least for the degree of freedom which corresponds to the
eigenvector with vanishing eigenvalue. The condition of vanishing of the
determinant also implies that two solutions of the saddle point equation
coincide and thus corresponds to a surface in the space of the coefficients of
the potential $V$ where the partition function is singular.  \par
To examine the corrections to the leading large $N$ behaviour it remains
however possible to integrate over one of the variables by steepest descent.
At leading order this corresponds to solving
the saddle point equation for one
of the variables, the other being fixed. Here it is convenient to eliminate
$\lambda$ by the equation $\lambda=1/\rho$. One finds
$$\e^Z \propto
\int \d\rho\,\exp\left[-N\bigl(V(\rho)-\half\ln\rho\bigr)+O(1)\right].
$$ 
In the leading term we obviously recover
the result of the angular integration 
with $\rho=\phib^2$. For $N$ large the leading contribution arises from the
leading term in the expansion of $W(\rho)=V(\rho)-\half\ln\rho$ near the
saddle point: 
$$W(\rho)-W(\rho_s)\sim\frac{1}{n!}W^{(n)}(\rho_s)(\rho-\rho_s)^n.$$
The integer $n$ characterizes the nature of the critical point.
Adding relevant perturbations $\delta_k V$ of parameters $v_k$ to
the critical potential 
$$\delta_k V= v_k (\rho-\rho_s)^k ,\quad  1\le k\le n-2\, $$
(the term $k=n-1$ can always be eliminated by a shift of $\rho$)
we find the partition function at leading order for $N$
large in the scaling region: 
$$\e^{Z(\{u_k\})}\propto\int\d z\,\exp\left(-z^n-\sum_{k=1}^{n-2}u_k
z^k\right),$$ 
where $z\propto N^{1/n}(\rho-\rho_s)$ and 
$$u_k \propto N^{1-k/n}v_k $$
is held fixed
\medskip
{\it Quantum mechanics.} 
The method we have used above immediately generalizes to quantum mechanics,
although a simpler method involves solving the radial Schr\"odinger
equation. We consider the euclidean action
$$S(\phib)=N\int\d
t\,\left[\half\bigl(\d_t\phib(t)\bigr)^2+V(\phib^2)\right]. \eqnn $$ 
Note the unusual field normalization, the factor $N$ in front of the
action simplifying all expressions in the large $N$ limit.\par
To explore the large $N$ limit one has to take the scalar function $\phib^2$,
which self-averages, as a dynamical variable.
At each time $t$ we thus perform the transformation \egenideno. One
introduces two paths $\rho(t),\lambda(t)$ and writes
$$\eqalignno{&\exp\left[-N\int \d t\,V(\phib^{2
})\right]&\cr&\hskip10truemm \propto \int \left[\d\rho(t)\,\d\lambda(t)
\right] \exp\left\{-N\int \d t\left[\half\lambda\left(\phib^{2}-\rho\right) +
V(\rho) \right]\right\}.\hskip12truemm&\eqnd\egeniden\cr} $$
The integral over the path $\phib(t)$ is then gaussian and can be performed.
One finds 
$$e^Z=\int[\d\rho(t)\d\lambda(t)]\,\exp\left[-S_N(\lambda,\rho)\right]\,\eqnn
$$ 
with
$$S_N=N\int\d t\,\left[-\half \lambda\rho+V(\rho)\right]+\half \tr\ln
\bigl(-\d_t^2+\lambda(\cdot) \bigr).\eqnn $$
Again, in the large $N$ limit the path integral can be calculated by steepest
descent. The saddle points are constant paths solution of
$$V'(\rho)=\half\lambda\,,\qquad \rho={1\over2\pi}\int{\d\omega\over
\omega^2+\lambda}={1\over2\sqrt{\lambda}} \,,\eqnn $$
where $\omega$ is the Fourier energy variable conjugated to $t$.
Again a critical point is defined by the property that at least two
solutions to the saddle point equations coalesce. This happens when 
the determinant of the matrix of first derivatives of the equations vanishes:
$$\det\pmatrix{V''(\rho) & -\half \cr -\half & -{1\over 8
\lambda^{3/2}} \cr} = -{1\over
8\lambda^{3/2}}V''(\rho)-{1\over4}=0\,.\eqnn $$  
The leading correction to the saddle point contribution is given by a gaussian
integration. The result involves the determinant of the operator
second derivative of $S_N$. By Fourier transforming time the operator
becomes a tensor product of $2\times2$ matrices with determinant $D(\omega)$
$$D(\omega)=\det\pmatrix{V''(\rho) & -\half \cr -\half &- \ud B(\omega)\cr}
\ {\rm with}\ B(\omega)=
{1\over2\pi}\int{\d\omega'
\over(\omega'{}^2+\lambda)[(\omega-\omega')^2+\lambda]} \,.$$ 
Thus, the criticality condition is equivalent to $D(0)=0$. 
When the criticality condition is satisfied, the leading correction is no
longer given by steepest descent. Again, since at most one mode can be
critical, we can integrate over one of the path by steepest descent,
which means solving the saddle point equation for one function, the 
other being fixed.
While the $\rho$ equation remains local, the $\lambda$ is now non-local,
involving the diagonal matrix element of the inverse of the differential
operator $-\d^2_t+\lambda(t)$. We shall see in next section how
this problem 
can be overcome in general. A special feature of quantum mechanics, however,
is that the determinant can be calculated, after a simple change of variables.
We set
$$\lambda(t)=\dot s(t)+s^2(t), \eqnd\echgvar $$
in such a way that the second order differential operator factorizes
$$-\d^2_t+\lambda(t)=-\bigl(\d_t+s(t)\bigr)\bigl(\d_t-s(t)\bigr).
\eqnn $$
The determinant of a first order differential operator can be calculated
by expanding formally in $s$. Only the first term survives but the 
coefficient is ambiguous  
$$\tr\ln \bigl(1 -\d_t^{-1}s(\cdot)\bigr)=-\theta(0)\int\d t\,s(t).$$
A more refined analysis, which involves boundary conditions, is required to 
determine the ambiguous value $\theta(0)$ of step function. Here one finds
$$\ln\det\bigl(-\d^2_t+\lambda(\cdot)\bigr)=
\tr\ln\bigl(-\d^2_t+\lambda(\cdot)\bigr)=
\int\d t\, s(t) .\eqnn $$
The jacobian of the transformation \echgvar~contributes at higher order in
$1/N$ and can be neglected. Therefore the large $N$ action becomes
$$\eqalign{S_N &=N\int\d t\,\left[-\half (\dot s+s^2)\rho+V(\rho)+\half
s \right] \cr
&=N\int\d t\,\left[-\half \rho s^2 +\half s(\dot\rho+1)+V(\rho)\right].\cr}$$ 
We can now replace $s$ by the solution of a local saddle point equation
(or perform the gaussian integration, but neglect the determinant
which is of higher order): 
$${\delta S_N\over \delta s(t)}=0\ \Leftrightarrow\ -s\rho+\half
(\dot\rho+1)=0\,,$$ 
and find
$$S_N=N\int\d t\,\left[{\dot\rho^2\over8\rho}+{1\over8\rho}+
V(\rho)\right].\eqnn  $$
We recognize the action for the large $N$ potential at zero angular momentum
in the radial coordinate $\rho(t)=\phib^2(t)$. Critical points then are
characterized by the behaviour of the potential $W(\rho)$ 
$$W(\rho)=V(\rho)+{1\over8\rho}\,,$$
near the saddle point $\rho_s$
$$W(\rho)-W(\rho_s)\sim W^{(n)}(\rho_s){\left(\rho-\rho_s\right)^n\over
n!}\,.$$
At critical points the ground state energy, after subtraction of the classical
term which is linear in $N$, has a non-analytic contribution. To eliminate
$N$ from the action we set
$$t\mapsto t N^{(n-2)/(n+2)},\qquad \rho(t)-\rho_s\mapsto N^{-2/(n+2)}z(t). $$
We conclude that the leading correction to the energy levels is proportional
to $N^{-(n-2)/(n+2)}$. Note also that the scaling of time implies that higher
order time derivatives would be irrelevant, an observation which can be used
more  directly to expand the determinant in local terms, and will be important
in next section.\par
If we add relevant corrections to the potential
$$\delta_k V=v_k (\rho-\rho_s)^k ,\quad 1\le k\le n-2\,,$$
the coefficients $v_k$ must scale like
$$v_k\propto N^{2(k-n)/(n+2)}.$$
\subsection{The 2D $V(\phib \sp{2}) $ field theory in the double scaling
limit} 

In the first part we study the $O(N)$ symmetric
$V(\phib^2)$ field theory, where $\phib$ is $N$-component field, in the large
$N$ limit in dimension two because phase transitions occur in higher
dimensions, a problem which has to be considered separately. The action
is:\sslbl\stwodim 
$$S(\phib)= N\int\d^2 x \left\{\half \left[ \partial_{\mu} \phib (x)
\right]^{2} +V\left(\phib^2\right) \right\} ,\eqnd{\eactONgii}$$
where an implicit cut-off $\Lambda$ is always assumed below. Whenever the
explicit dependence in the cut-off will be relevant we shall assume a
Pauli--Villars's type regularization, i.e.\  the replacement in action
\eactONgii\ of $-\phib\del^2\phib$ by
$$-\phib\del^2 D(-\del^2/\Lambda^2)\phib\,,\eqnd\eDPV$$
where $D(z)$ is a positive non-vanishing polynomial with $D(0)=1$.\par
As before one introduces two fields
$\rho(x)$ and $\lambda(x)$ and uses the identity \egeniden.
The large $N$  action is then:
$$S_N=N\int\d^2 x \left[V(\rho)-\half \lambda
\rho \right]+\half N \tr\ln(-\Delta+\lambda ).\eqnd\eactefNgii $$
Again for $N$ large we evaluate the integral by steepest descent.
Since the saddle point value $\lambda$ is the
$\phib$-field mass squared, we set in general $\lambda=m^2$. With this
notation the two equations for the saddle point $m^2,\rho_s=\langle \phib^2
\rangle$ are: 
\eqna\esaddNgenii
$$\eqalignno{
V'(\rho_s)&=\half m^2\,,&\esaddNgenii{a}\cr
\rho_s&={1\over (2\pi)^2 }\int^{\Lambda} {\d^2 k \over k^2
+m^2}\,, & \esaddNgenii{b} \cr}$$
where we have used a short-cut notation
$${1\over (2\pi)^2 }\int^\Lambda {\d^2 k \over k^2
+m^2}\equiv {1\over (2\pi)^2 }\int{\d^2 k \over D(k^2/\Lambda^2) k^2
+m^2}\equiv B_1(m^2). \eqnn $$
For $m\ll\Lambda$ one finds
$$B_1(m^2)={1\over2\pi}\ln(\Lambda/m)+{1\over4\pi}\ln (8\pi K)
+O(m^2/\Lambda^2) , $$
where $K$ is a regularization dependent constant.
%\ln (8\pi C) &= \int_0^\infty\d s\left( {1\over
%D(s)}-\theta(1-s)\right). 
\par
As we have discussed in the case of quantum mechanics a critical point
is characterized by the vanishing at zero momentum of the  
determinant of second derivatives of the action at the saddle point. 
The mass-matrix has then a zero eigenvalue which, in field theory, corresponds
to the appearance of a new massless excitation other than $\phib$. In order to
obtain the effective action for this scalar massless mode we  must
integrate over one of the fields. In the
field theory case the resulting effective action can no longer be written in
local form.  
To discuss the order of the critical point, however, we only need the action
for space independent fields, and thus for example we can eliminate $\lambda$
using the $\lambda$ saddle point equation. 
\par
The effective $\rho$ potential $W(\rho)$ then reads
$$W(\rho)=V(\rho)-\half\int^{\lambda(\rho)}\d \lambda' \,\lambda'
{\del \over \del \lambda'} B_1(\lambda'),\eqnd\eWeffrho$$
where at leading order for $\Lambda$ large 
$$\lambda(\rho)=8\pi K \Lambda^2\e^{-4\pi\rho}. $$
The expression for the effective action in equation \eWeffrho~is
correct  for any $d$ and will be used  also in section \sdouble.
Here we have:
$$W(\rho)=V(\rho)+K\Lambda^2\e^{-4\pi\rho}=V(\rho)+\frac{1}{8\pi}m^2
\e^{-4\pi(\rho-\rho_s)}.$$
A multicritical point is defined by the condition
$$W(\rho)-W(\rho_s)=O\left((\rho-\rho_s)^n\right)\eqnd\eWcrit.$$
This yields the conditions:
$$V^{(k)}(\rho_s)=\half (-4\pi)^{k-1} m^2\quad{\rm for}\ 1\le k\le n-1\,.$$
Note that the coefficients $V^{(k)}(\rho_s)$ are the coupling
constants renormalized at leading order for $N$ large.
If $V(\rho)$ is a polynomial of degree $n-1$ (the minimal polynomial model)
the  multicritical condition in equation~\eWcrit~determines the
critical values of renormalized coupling constants as well as $\rho_s$
\par 
When the fields are space-dependent it is simpler to eliminate $\rho$ instead,
because the corresponding field equation: 
$$V'\bigl(\rho(x)\bigr)=\half \lambda(x). \eqnd\esadroge $$
is local. This equation can be solved by expanding $\rho(x)-\rho_s$ in a power
series in $\lambda(x)-m^2$:
$$\rho(x)-\rho_s= {1\over2 V''(\rho_s)}\bigl(\lambda(x)-m^2\bigr)
+O\left((\lambda-m^2)^2\right) . \eqnd\esolrola $$
The resulting action for 
the field $\lambda(x)$ remains non-local but because, as we
shall see,  adding powers of $\lambda$ 
as well as adding derivatives make
terms less relevant, only the few first terms of a local expansion of the
effective action will be important.\par
If in the local expansion of the determinant we keep only the two
first terms we obtain an action containing at leading order a kinetic
term proportional to $(\del_\mu\lambda)^2$ and the interaction
$(\lambda(x)-m^2)^n$:
$$S_N(\lambda) \sim N\int\d^2 x\left[{1\over96\pi
m^4}(\del_\mu\lambda)^2 + \frac{1}{n!}S_n \bigl(\lambda(x)-m^2)^n\right],$$
where the neglected terms are of order $(\lambda-m^2)^{n+1}$, $\lambda\del^4
\lambda$, and $\lambda^2\del^2\lambda$ and 
$$S_n=W^{(n)}(\rho_s)[2V''(\rho_s)]^{-n}=W^{(n)}(\rho_s)(-4\pi m^2)^{-n} .$$
Moreover we note that together with the cut-off $\Lambda$, $m$ now also
acts as a cut-off in the local expansion.\par
To eliminate the $N$ dependence in the action we have, as in the example
of quantum mechanics, to rescale both the field $\lambda-m^2$ and space:
$$\lambda(x)-m^2=\sqrt{48\pi}m^2 N^{-1/2}\varphi(x)\,,\quad x\mapsto
N^{(n-2)/4}x \,.
\eqnd\elamrescal $$
We find
$$S_N(\varphi)
\sim\int\d^2x\left[\half(\partial_\mu\varphi)^2+\frac{1}{n!} g_n
\varphi^n\right] .$$
In the minimal model, where the polynomial $V(\rho)$ has exactly degree $n-1$,
we find $g_n=6(48\pi)^{(n-2)/2}m^2$. 

As anticipated we observe that derivatives and powers of $\varphi$
are affected by negative powers of $N$, justifying a local expansion.
However we also note that the cut-offs ($\Lambda$  or the mass $m$) are now
also multiplied by $N^{(n-2)/4}$. Therefore the large $N$ limit also
becomes a large cut-off limit.
\medskip
{\it Double scaling limit.} The existence of a double scaling limit
relies on the existence of IR singularities due to the massless or small mass
bound state which can compensate the $1/N$ factors appearing in the large $N$
perturbation theory. \par 
We now add to the action relevant perturbations:
$$\delta_k V=v_k(\rho(x)-\rho_s)^k,  \quad 1\le k\le n-2. $$
proportional to $\int\d^2x(\lambda-m^2)^k$:
$$\delta_k S_N(\lambda)=N S_k \int\d^2x\,(\lambda-m^2)^k ,$$
where the coefficients $S_k$ are functions of the coefficients $v_k$.
After the rescaling \elamrescal~
$$\delta_k S_N(\varphi)=\frac{1}{k!} g_k
N^{(n-k)/2}\int\d^2x\,\varphi^k(x)  \quad 1\le k\le n-2  $$
%(the term $k=n-1$ can always be eliminated by a shift of $\varphi(x)$).
However, unlike quantum mechanics, it is not sufficient to scale the
coefficients $g_k$ with the power $N^{(k-n)/2}$ to obtain a finite scaling
limit. Indeed 
perturbation theory is affected by UV divergences, and we have just noticed
that the cut-off diverges with $N$. In two dimensions the nature of
divergences is very simple: it is entirely due to the self-contractions
of the interactions terms and only one divergent integral appears:
$$\left<\varphi^2(x)\right>={1\over4\pi^2}\int {\d^2 q\over q^2+\mu^2}\,,$$
where $\mu$ is the small mass of the bound state, required as an IR cut-off to
define perturbatively the double scaling limit.
We can then extract the $N$ dependence
$$\left<\varphi^2(x)\right>={1\over8\pi}(n-2)\ln N+O(1) .$$
Therefore the coefficients $S_k$ have also to cancel these UV divergences,
and thus have a logarithmic dependence in $N$ superposed to the natural
power obtained from power counting arguments. In general for any potential
$U(\varphi)$
$$U(\varphi)=:U(\varphi):+\left[\sum_{k=1}{1\over2^k
k!}\left<\varphi^2\right>^k \left(\partial\over\partial \varphi\right)^{2k}
\right]:U(\varphi):\,,$$
where $:U(\varphi):$ is the potential from which self-contractions have been
subtracted (it has been normal-ordered). For example for $n=3$
$$\varphi^3(x)=:\varphi^3(x):+3\left<\varphi^2\right> \varphi(x) ,$$
and thus the double scaling limit is obtained with the behaviour
$$N g_1+{1\over16\pi}\ln N g_3  \ {\rm fixed}\ .$$
For the example $n=4$ 
$$ g_1 N^{3/2}\quad {\rm and}\quad N g_2 +{g_4\over8\pi}\ln N \ {\rm fixed}\
.$$ 
\subsection{The $V(\phib \sp{2}) $ in the large $N$ limit: phase transitions}

In higher dimensions something new happens: the possibility of 
phase transitions associated with  spontaneous 
breaking the $O(N)$ symmetry. In the first part we thus
study the $O(N)$ symmetric $V(\phib^2)$ field theory, in the large $N$
limit to explore the possible phase transitions and identify the corresponding
multicritical points. The action is:\sslbl\smulticr
$$S(\phib)= N\int\d^d x \left\{\half \left[ \partial_{\mu} \phib (x)
\right]^{2} +V\left(\phib^2\right) \right\} ,\eqnd{\eactONg}$$
where, as above (equations \eqns{\eactONgii,\eDPV}), 
an implicit cut-off $\Lambda$ is always assumed below. \par
The identity 
\egeniden\ transforms the action into a quadratic form in $\phib$
and therefore the integration over $\phib$ can be  performed. It is convenient
however here to integrate only over 
$N-1$ components, to keep a component of the vector field, which we denote
$\sigma$, in the action. The large $N$ action is then:
$$S_N=N\int\d^d x \left[\half \left(\partial_\mu\sigma\right)^2+
V(\rho)+\half \lambda\left(\sigma^2- \rho\right)\right]+\half
(N-1)\tr\ln(-\Delta+\lambda ).\eqnd{\eactefNg}$$
\medskip
{\it The saddle point equations: the $O(N)$ critical point.}
Let us then write the saddle point equations for a general potential $V$. At
high temperature $\sigma=0$ and $\lambda$ is the
$\phib$-field mass squared. We thus set in general $\lambda=m^2$. With this
notation the three saddle point equations are: 
\eqna\esaddNgen
$$\eqalignno{m^2\sigma&=0\,, & \esaddNgen{a} \cr
V'(\rho)&=\half m^2\,,&\esaddNgen{b}\cr
\sigma^2&=\rho-{1\over (2\pi)^d }\int^{\Lambda} {\d^d k \over k^2
+m^2}\,. & \esaddNgen{c} \cr}$$
In the ordered phase $\sigma\ne0$ and thus $m$ vanishes. 
Equation \esaddNgen{c} has a solution only for $\rho>\rho_c$,
$$\rho_c={1\over (2\pi)^d }\int^\Lambda {\d^d k \over k^2}\,,\ \Rightarrow\
\sigma=\sqrt{\rho-\rho_c}\,.$$
Equation \esaddNgen{b} which reduces to $V'(\rho)=0$ then yields the critical
temperature. Setting
$V(\rho)=U(\rho)+\half r\rho$, we find
$$r_c=-2 U'(\rho_c).$$
To find the magnetization critical exponent $\beta$ we need the relation
between the $r$ and $\rho$ near the critical point. \par
In the disordered phase, $\sigma=0$, equation \esaddNgen{c} relates $\rho$ to
the $\phib$-field mass $m$. 
For $m\ll\Lambda$, $\rho$ approaches $\rho_c$,
and the relation becomes (equation \eintasym):
$$\rho-\rho_c =-C(d)
m^{d-2}+a(d)m^2\Lambda^{d-4}
+O\left(m^d\Lambda^{-2}\right)
+O\left(m^4\Lambda^{d-6}\right). \eqnd\edmum  $$
For $2<d<4$ (the situation we shall assume below except when stated otherwise)
the $ O\left(m^d\Lambda^{-2}\right)$  from the
non-analytic part dominates the corrections to the leading part of this
expression. 
For $d=4$ instead 
$$\rho-\rho_c=\frac{1}{8\pi^2} m^2\left(\ln m/\Lambda+\ {\rm const.}\right),$$
and for $d>4$ the analytic contribution dominates and 
$$\rho-\rho_c\sim a(d)m^2\Lambda^{d-4}.$$
The constant $C(d)$ is universal (equation \eintNcor{b}).
The constant  $a(d)$, which also appears in equation \eintasym, instead
depends on the cut-off procedure, 
and is given by
$$a(d)= {1\over (2\pi)^d}\int{\d^d k\over k^4}\left(1-{1\over
D^2(k^2)} \right)  .  \eqnd\eadef $$
\medskip
{\it Critical point.} In a generic situation $V''(\rho_c)=U''(\rho_c)$ does
not vanish. We thus find in the low temperature phase
$$t=r-r_c\sim -2 U''(\rho_c)(\rho-\rho_c) \ \Rightarrow\ \beta=\half\,.
\eqnd\erminrc$$
This is the case of an ordinary critical point. Stability implies
$V''(\rho_c)>0$ so that $t<0$.\par
At high temperature, in the disordered phase, the $\phib$-field mass $m$ is
given by $2U'(\rho)+ r=m^2$ and thus, using \edmum, at leading order
$$t\sim 2U''(\rho_c)C(d)m^{d-2},$$
in agreement with the result of the normal critical point. Of course the
simplest realization of this situation is to take $V(\rho)$ quadratic, and we
recover the $(\phib^2)^2$ field theory.
\smallskip
{\it The sign of the constant $a(d)$.} A comment concerning the non-universal
constant $a(d)$ defined in \edmum\ is here in order because, while its
absolute value is irrelevant, its sign plays a role in the discussion of
multicritical points. Actually the relevance of this sign to
the RG properties of the large $N$ limit of the simple
$(\phib^2)^2$ field theories has already mentioned (section \sssEGRN).
For the simplest Pauli--Villars's type regularization we have $D(z)>1$ and
thus $a(d)$ is finite and positive in dimensions $2 < d < 4$, but this clearly
is not a universal feature.\par
A new situation arises if we can adjust a parameter of the
potential in such a way that $U''(\rho_c)=0$. This can be achieved only if the
potential $V$ is at least cubic. We then expect a tricritical
behavior. Higher critical points can be obtained when more derivatives vanish.
We shall examine the general case though, from the point of view of real phase
transitions, higher order critical points are not interesting because $d>2$
for continuous symmetries and mean-field behavior is then obtained for
$d\ge 3$. The analysis will however be useful in the study of  double scaling
limit.\par
Assuming that the first non-vanishing derivative is $U^{(n)}(\rho_c)$,
we expand further equation \esaddNgen{b}. In the ordered low temperature
phase we now find
$$t=-{2\over (n-1)!}U^{(n)}(\rho_c)(\rho-\rho_c)^{n-1},\ \Rightarrow\
\sigma\propto (-t)^\beta,\quad \beta={1\over2(n-1)}\,  ,  \eqnd\ecritbeta $$
which leads to the exponent $\beta$ expected in the mean field approximation
for such a multicritical point.
We have in addition the condition $U^{(n)}(\rho_c)>0$.\par
In the high temperature phase instead
$$m^2=t+ (-1)^{n-1}{2\over (n-1)!}U^{(n)}(\rho_c)C^{n-1}(d)m^{(n-1)(d-2)}.
\eqnd\emsq$$
For $d>2n/(n-1)$ we find a simple mean field behavior, as expected since we
are above the upper-critical dimension . \par
For $d<2n/(n-1)$ we find a peculiar phenomenon, the term in the r.h.s.\ is
always dominant, but depending on the parity of $n$ the equation has solutions
for $t>0$ or $t<0$. For $n$ even, $t$ is positive and we find
$$m\propto t^\nu,\qquad \nu={1\over(n-1)(d-2)},\eqnd\emoft$$
which is a non mean-field behavior below the critical dimension.
However for $n$ odd (this includes the tricritical point) $t$ must be
negative, 
in such a way that we have now two competing solutions at low temperature.
We have to find out which one is stable. We shall verify below that only the
ordered phase is stable, so that the correlation length of the $\phib$-field
in the high temperature phase remains always finite. Although these dimensions
do not correspond to physical situations because $d<3$ the result is
peculiar and inconsistent with the $\varepsilon$-expansion. 
\par
For $d=2n/(n-1)$ we find a mean field behavior without logarithmic
corrections, provided one condition is met:
$${2\over(n-1)!}U^{(n)}(\rho_c)C^{n-1}\left(2n/(n-1)\right)<1\,,\qquad
C(3)=1/(4\pi).\eqnd\etriccond$$
We examine, as an example, in more details the tricritical point below.
We will see that the special point
$${2\over(n-1)!}U^{(n)}(\rho_c)C^{n-1}\left(2n/(n-1)\right)=1\,,
\eqnd\eendtric$$
has several peculiarities. 
In what follows we call $\Vc$ this special value of $U^{(n)}(\rho_c)$.
\medskip
{\it Discussion.} In the mean field approximation the function
$U(\rho)\propto\rho^n$ is not bounded from below for $n$ odd, however $\rho=0$
is the minimum because by
definition $\rho\ge0$. Here instead we are in the situation where $U(\rho)
\sim (\rho-\rho_c)^n$ but $\rho_c$ is positive. Thus this extremum of the
potential is likely to be unstable for $n$ odd. To check the global
stability requires further work. The question is whether such multicritical
points can be studied by the large $N$ limit method.
\par
Another point to notice concerns renormalization group: The $n=2$ example is
peculiar in the sense that the large $N$ limit exhibits a non-trivial IR fixed
point. For higher values of $n$ no coupling renormalization arises in the
large $N$ limit and the IR fixed point remains pseudo-gaussian. We are in a
situation quite
similar to usual perturbation theory, the $\beta$ function can only be
calculated perturbatively in $1/N$ and the IR fixed point is outside the
perturbative regime.
\medskip
{\it Local stability and the mass matrix.} 
The matrix of the general second partial derivatives of the effective action
is:
$$N\pmatrix{p^2+m^2 & 0 & \sigma \cr
0 & V''(\rho) & -\half \cr
\sigma & -\half & -\frac{1}{2}B_\Lambda(p,m)  \cr}  ,  \eqnd\ematrix$$
where $B_\Lambda(p,m)$ is defined in \ediagbul.\par
We are in position to study the local stability of the critical points.
Since the integration contour for $\lambda=m^2$ should be parallel to the
imaginary axis, a necessary condition for stability is that the determinant
remains negative.
\smallskip
{\it The disordered phase.} Then $\sigma=0$ and thus we have only to study
the $2\times2$ matrix $\bf M$ of the $\rho,m^2$  subspace. Its determinant
must remain negative which implies
$$\det{\bf M}<0\ \Leftrightarrow\
2V''(\rho)B_\Lambda(p,m)+1>0\,.\eqnd\estable $$
For Pauli--Villars's type regularization the function 
$B_\Lambda(p,m)$ is decreasing
so that this condition is implied by the condition at zero momentum
$$\det{\bf M}<0\ \Leftarrow\ 2V''(\rho)B_\Lambda(0,m)+1>0\,.$$
For $m$ small we use equation \eBLamze~and
%find
%$$B_2(0;m^2)=\half(d-2)C(d)m^{d-4}-a(d)\Lambda^{d-4}+O\left(m^{d-2}\right),
%\eqnd\eBmzero$$
at leading order the condition becomes:
$$ C(d)(d-2)m^{d-4}V''(\rho)+1>0\,.$$
This condition is satisfied by a normal critical point since $V''(\rho_c)>0$.
For a multicritical point, and taking into account equation \edmum\ we find:
$$(-1)^n {d-2\over (n-2)!}C^{n-1}(d)m^{n(d-2)-d}V^{(n)}(\rho_c)+1>0\,.
\eqnd\multicr $$ 
We obtain a result consistent with our previous analysis: For $n$ even it is
always satisfied. For $n$ odd it is always satisfied above the critical
dimension and never below. At the upper-critical dimension we find a condition
on the value of $V^{(n)}(\rho_c)$ which we recognize to be identical to
condition \etriccond\ because then $2/(n-1)=d-2$.
\smallskip
{\it The ordered phase.} Now $m^2=0$ and the determinant $\Delta$ of the
complete matrix is:
$$-\Delta>0\ \Leftrightarrow\
2V''(\rho)B_\Lambda(p,0)p^2+p^2+4V''(\rho)\sigma^2>0\,.\eqnd\eorder$$
We recognize a sum of positive quantities, and the condition is always
satisfied. Therefore in the case where there is a competition with a
disordered saddle point only the ordered one can be stable.
\subsection The scalar bound state

In this section we study the limit of stability in the disordered phase
($\sigma=0$). This is a problem which only arises when $n$ is odd, the first
case being provided by the tricritical point.\sslbl\sboundst\par
The mass-matrix has then a zero eigenvalue which corresponds to the appearance
of a new massless excitation other than $\phib$. Let us denote by $\bf M$ the
$\rho,m^2$ $2\times2$ submatrix. Then
$$\det{\bf M}=0\ \Leftrightarrow\ 2V''(\rho)B_\Lambda(0,m)+1=0\,.$$
In the two-space the corresponding eigenvector has components 
$(\half,V''(\rho))$.
\medskip
{\it The small mass $m$ region.} 
In the small $m$ limit the equation can be rewritten in terms of the
constant $C(d)$ defined in \eintasym:
$$ C(d)(d-2)m^{d-4}V''(\rho)+1=0\,.\eqnd\ephitwoa $$
Equation \ephitwoa\ tells us that $V''(\rho)$ must be small. We are thus
close to a multicritical point. Using the result of the stability analysis
we obtain
$$(-1)^{n-1}{d-2\over (n-2)!}C^{n-1}(d)m^{n(d-2)-d}V^{(n)}(\rho_c)=1\,.
\eqnd\estabil$$
We immediately notice that this equation has solutions only for $n(d-2)=d$,
i.e.\ at the critical dimension. The compatibility then fixes the value of
$V^{(n)}(\rho_c)$. We again find the point \eendtric, $V^{(n)}(\rho_c)=\Vc$.
If we take into account the leading correction to the small $m$ behavior
we  find  instead:
$$V^{(n)}(\rho_c)\Vc^{-1}-1\sim(2n-3){a(d)\over C(d)}\left({m\over\Lambda}
\right)^{4-d}.\eqnd\estabil  $$
This means that when $a(d)>0$ there exists a small region
$V^{(n)}(\rho_c)>\Vc$ where the vector
field is massive with a small mass $m$ and the bound-state massless. The
value $\Vc$ is a fixed point value.%% the other case???
\medskip
{\it The scalar field at small mass.} We want to extend the analysis to a
situation where the scalar field has a small but non-vanishing mass $M$ and
$m$ is still small. The goal is in particular to explore the neighbourhood of
the special point \eendtric.
Then the vanishing of the determinant of $\bf M$ implies
$$1+2V''(\rho)B_\Lambda(iM,m)=0\,.\eqnd\emassless$$
Because  $M$ and $m$ are small, this equation still implies
that $\rho$ is close to a point $\rho_c$ where $V''(\rho)$ vanishes.
Since reality imposes $M<2m$, it is easy to verify that this equation
has also solutions for only the critical dimension. Then
$$V^{(n)}(\rho_c)f(m/M)=\Vc\,,\eqnd\eOmegaC$$
where we have set:
$$f(z)=\int_0^1\d x\left[1+(x^2-1)/(4z^2)\right]^{d/2-2},\qquad \half<z\,.
\eqnd\efofz$$
In three dimensions it reduces to:
$$f(z)=z\ln\left({2z+1\over 2z-1}\right)\,.$$
$f(z)$ is a decreasing function which diverges for $z=\half$ because $d\le3$.
Thus we find solutions in the whole region $0<V^{(n)}(\rho_c)<\Omega_c$,
i.e.\ when the multicritical point is locally stable.\par
Let us calculate the propagator near the pole. We find the matrix $\bf \Delta$
$${\bf \Delta}={2\over G^2}\left[N\left.{\d B_\Lambda(p,m)\over \d
p^2}\right \vert_{p^2=-M^2}\right]^{-1}{1\over p^2+M^2}\pmatrix{1 &G
\cr G  &  G^2 \cr},\eqnd\edeltapole$$
where we have set
$$ G={2(-C)^{n-2}W^{(n)} \over  (n-2)! } m^{4-d}\,. $$
For $m/M$ fixed the residue goes to zero with $m$ as 
$m^{d-2}$ because the derivative of $B$ is of the order of $m^{d-6}$. 
Thus the bound-state decouples on the multicritical line.
\medskip
{\it The scalar massless excitation: general situation.}
Up to now we have
explored only the case where both the scalar field and the vector field
propagate. Let us now relax the latter condition, and examine what happens
when $m$ is no longer small. The condition $M=0$ then reads
$$2V''(\rho_s)B_\Lambda(0,m)+1=0\,   $$
together with
$$m^2=2V'(\rho_s),\qquad \rho_s= {1\over (2\pi)^d }\int^{\Lambda}
{\d^d k \over k^2 +m^2}\,. \eqnd\GapEq $$
An obvious remark is: there exist solutions only for $V''(\rho_s)<0$, and
therefore the ordinary critical line can never be approached. In terms
of the function $F(z)$ 
$$\Lambda^{d-2}F(m^2/\Lambda^2)= {1\over (2\pi)^d }\int^{\Lambda}
{\d^d k \over k^2 +m^2}\equiv {1\over (2\pi)^d }\int
{\d^d k \over k^2D(k^2) +m^2}\eqnn $$
and thus
$$F(z)=N_d \int_0{k^{d-1}\d k \over k^2D(k^2)+z}\, , \eqnd\eFofz$$
the equations can be rewritten
$$\rho_s=\Lambda^{d-2} F(z),\quad z=2V'(\rho_s)\Lambda^{-2},\quad
2\Lambda^{d-4} V''(\rho_s)F'(z)=1\,.$$
The function $F(z)$ in Pauli--Villars's regularization is a decreasing
function.
In the same way $-F'(z)$ is a positive decreasing function.\par
The third equation is the condition for the two curves corresponding to the
two first ones become tangent. For any value of $z$ we can find potentials and
thus solutions. Let us call $z_s$ such a value and specialize to cubic
potentials. Then
$$\rho_s=\Lambda^{d-2}F(z_s)~~,$$ 
$$\quad V(\rho)=V'(\rho_s)(\rho-\rho_s)+\half
V''(\rho_s)(\rho-\rho_s)^2+
\frac{1}{3!}V^{(3)}(\rho_s)(\rho-\rho_s)^3,\eqnd\Vofrho $$
which yields a two parameter family of solutions. For $z$ small we see that
for $d<4$ the potential becomes proportional to $(\rho-\rho_c)^3$.

\subsection Stability and double scaling limit

In order to discuss in more details the stability issue and the double scaling
limit we 
now construct the effective action for the scalar bound state. We consider
first only the massless case. We only need the action in the IR limit, and in
this limit we can integrate out the vector field and the second massive
eigenmode. \sslbl\sdouble
\medskip
{\it Integration over the massive modes.} 
As we have already explained in section \stwodim~we can integrate over one of
the fields, the second being fixed, and we need only the result at leading
order. Therefore we replace in the functional integral
$$\e^Z=\int [\d\rho\d\lambda]\exp\left[-\frac{N}{2}\tr \ln(-\del^2 + \lambda)
+N\int \d^dx\left(- V(\rho) +\half \rho \lambda\right)\right],\eqnd\eZ$$
one of the fields by the solution of the field
equation. It is useful to first discuss the effective potential of the
massless mode. This requires calculating the action only for constant fields.
It is then simpler to eliminate $\lambda$. We 
assume in this section that $m$ is small (the vector propagates). 
For $\lambda\ll\Lambda$ the $\lambda$-equation reads ($d<4$)
$$\rho-\rho_c=-C(d)\lambda^{(d-2)/2}.  \eqnd\erholambda $$
It follows that the resulting potential $W(\rho)$, obtained from
equation \eWeffrho~is 
$$W(\rho)=V(\rho)+{d-2\over 2d (C(d))^{2/(d-2)}}(\rho_c-\rho)^{d/(d-2)}.
\eqnd\effUrho$$
In the sense of the double scaling limit the criticality conditions are
$$W(\rho)=O\bigl((\rho-\rho_s)^n\bigr).$$
It follows
$$V^{(k)}(\rho_s)=- \half C^{1-k}(d){\Gamma\bigl(k-d/(d-2)\bigr)\over
\Gamma\bigl(-2/(d-2)\bigr)}m^{d-k(d-2)}\quad 1\le k\le n-1\, .$$
For the potential $V$ of minimal degree we find
$$W(\rho)\sim \frac{1}{2 n!} C^{1-n}(d){\Gamma\bigl(n-d/(d-2)\bigr)\over
\Gamma\bigl(-2/(d-2)\bigr)}m^{d-n(d-2)}(\rho-\rho_s)^n .$$
\medskip
{\it The double scaling limit.}
We recall here that quite generally one verifies that a non-trivial double
scaling limit may exist only if the resulting field theory of the massless
mode is 
super-renormalizable, i.e.~below its upper-critical dimension $d=2n/(n-2)$,
because perturbation theory has to be IR divergent. Equivalently, to
eliminate $N$ from the critical theory, one has to rescale
$$\rho-\rho_s\propto N^{-2\theta}\varphi\,,\quad x\mapsto x N^{(n-2)\theta}
\quad {\rm with}\ 1/\theta=2n-d(n-2),$$ 
where $\theta$ has to be positive.
\par
We now specialize to dimension three, since $d<3$ has already been
examined, and the expressions above are valid  only for $d<4$.
The normal critical point ($n=3$), which leads to a $\varphi^3$ field theory, 
and can be obtained for a quadratic potential $V(\rho)$  (the $(\phib^2)^2$)
has been discussed elsewhere. We thus concentrate on the next critical
point $n=4$ where the minimal potential has degree three.
\medskip %####
{\it The $d=3$ tricritical point.}
The potential $W(\rho)$ then becomes
$$W(\rho)=V(\rho)+\frac{8\pi^2}{3}(\rho_c-\rho)^3. \eqnd\effWthreeD$$
If the potential $V(\rho)$ has degree larger than three, we obtain
after a local expansion and a rescaling of fields,
$$\rho-\rho_s= ({-1\over 32\pi^2 \rho_c})
(\lambda-m^2)\propto \varphi/N\,,\quad x\mapsto Nx\,,\eqnd\rescal$$
a simple super-renormalizable $\varphi^4(x)$ field theory. 
If we insist
instead that the initial theory should be renormalizable, then we remain with
only one candidate, the renormalizable $(\phib^2)^3$ field theory, also
relevant for the tricritical phase transition with $O(N)$ symmetry breaking.
Inspection of the potential $W(\rho)$ immediately shows a remarkable feature:
Because the term added to $V(\rho)$ is itself a polynomial of degree three,
the critical conditions lead to a potential $W(\varphi)$ which vanishes
identically. This result reflects the property that 
the two saddle point equation  ($\del S/\del \rho = 0 ~,~  
\del S/\del \lambda = 0$ in equations~\esaddNgen{}) are proportional 
and thus have a continuous one-parameter family of solutions. This results in
a flat effective potential for $\varphi(x)$.
The effective action for $\varphi$ depends only on the derivatives
of $\varphi$, like in the $O(2)$ non-linear $\sigma$ model.\par
We conclude that no non-trivial double scaling limit can be obtained in this
way.
In three dimensions with a $(\phib^2)^3$ interaction we can generate at
most a normal critical point $n=3$, but then a simple $(\phib^2)^2$
field theory suffices.

The ambiguity of the sign of $a(d)$ discussed in section \smulticr~has an
interesting appearance in $d= 3$ in the small $m^2$ region.   
If one keeps the
extra term proportional to $a(d)$ in equation \effUrho~we have
$$W(\rho)=V(\rho)+{8\pi^2 \over 3}(\rho_c-\rho)^3
+{a(3)\over\Lambda} 4\pi^2 (\rho_c-\rho)^4.$$
Using now equation \erholambda~and, as mentioned in section \sboundst, 
the fact that in the small $m^2$ region the potential is proportional to
$(\rho-\rho_c)^3$ we can solve for $m^2$.
Since $m^2>0$ the appearance of a phase with small mass depends
on the sign of $a(d)$. Clearly this shows a
non-commutativity of the limits of $m^2/\Lambda^2 \to 0 $ and $ N \to \infty$.
The small  $m^2$ phase can be reached by a special tuning  and
cannot be reached with an improper sign of $a(d)$.
Calculated in this way, $m^2$ can be made proportional to the deviation  of
the  coefficient of $\rho^3$ in $V(\rho)$ from its critical  value $16\pi^2$. 
\subsection Conclusions 

This is a study of several subtleties in the phase structure of  $O(N)$ vector
models around multicritical points of odd and even orders.
One of the main topics is the understanding of the multicritical behavior
of these models  at their critical dimensions and the effective field theory
of the $O(N)$-singlet bound state obtained in the $N \to \infty$, ~$g \to g_c$
correlated limit. It is pointed out that the integration over massive $O(N)$
singlet modes is essential in order to extract the correct effective field
theory of the small mass scalar excitation. 
After performing this integration, it has been established here that
the double scaling limit of $(\phib^2)^K$ vector model 
in its critical dimension $d=2K/(K-1)$
results in a theory of a free massless $O(N)$ singlet bound state.  
This fact is a consequence of the existence of 
flat directions at the scale invariant multicritical point 
in the effective action. In contrast to the case $d < 2K/(K-1)$
where IR singularities compensate powers of $1/N$ in the double scaling 
limit,  at  $d=2K/(K-1)$ there is no such compensation and only a 
noninteracting effective field theory  of the massless bound state is left.
\sslbl\sconclud\par
Another interesting issue in this study is the ambiguity of the
sign of $a(d)$.
The coefficient of $m^2\Lambda^{d-4}$ denoted by $a(d)$ in the expansion
of the gap equation in equations \esaddNgen{c} and \edmum~seems to have a
surprisingly important role in the approach to the continuum limit ($\Lambda^2
\gg m^2$). The existence of an IR fixed point  at $g \sim O(N^{-1}),$ as seen
in the $\beta$ function for the unrenormalized coupling constant (section
\sssEGRN), depends on the sign of $a(d)$. Moreover, %as seen in section ???  
the existence of a phase with a small mass $m$ for the $O(N)$ vector quanta
and a massless $O(N)$ scalar depends also on the sign of
$a(d)$. It may very well be that the importance of 
the sign of $a(d)$ is a mere reflection of the limited  coupling constant
space used to described the model. This is left here as an open question that
deserves a detailed renormalization group or lattice simulation study in the
future. 
\beginbib

\nrf The last section is taken from \rf
Galit~Eyal, Moshe~Moshe, Shinsuke~Nishigaki and Jean~Zinn-Justin,
{\it Nucl. Phys.} B470 (1996) 369, hep-th/9601080.
\nrf For a review on matrix models and double scaling limit see \rf
 P. Di Francesco, P. Ginsparg and J. Zinn-Justin,
{\it Phys. Rep.} 254 (1995) 1.
\nrf The $d=1$ matrix problem is discussed in \rf
P. Ginsparg and J. Zinn-Justin, {\it Phys. Lett.} B240 (1990) 333;
E. Br\'ezin, V. A. Kazakov, and Al. B. Zamolodchikov,
{\it Nucl. Phys.} B338 (1990) 673;
G. Parisi, {\it Phys. Lett.} B238 (1990) 209, 213;
{\it Europhys. Lett.} 11 (1990) 595; 
D. J. Gross and N. Miljkovic, {\it Phys. Lett.} B238 (1990) 217.
\nrf Previous references on the double scaling limit in vector models
include\rf
A. Anderson, R.C.
Myers and V. Perival, {\it Phys. Lett.} B254 (1991) 89, {\it Nucl. Phys.}
B360 (1991) 463;
S. Nishigaki and T. Yoneya, {\it Nucl. Phys.} B348 (1991) 787;
P. Di Vecchia, M. Kato and N. Ohta, {\it Nucl. Phys.} B357 (1991) 495;
J. Zinn-Justin, {\it Phys. Lett.} B257 (1991) 335;
P. Di Vecchia, M. Kato and N. Ohta, {\it Int. J. Mod. Phys.}A7,
(1992)1391; T. Yoneya, {\it Prog. Theo. Phys. Suppl.} 92, 14 (1992).
\nrf References on tricritical behaviour, dilaton...include
\rf
W.A. Bardeen, M. Moshe, M. Bander, {\it Phys. Rev. Lett.} 52 (1984) 1188;
F. David, D.A. Kessler and H. Neuberger, {\it Phys. Rev. Lett.} 53
(1984) 2071, {\it Nucl. Phys.} B257 [FS14] (1985) 695;
D.A. Kessler and H. Neuberger, {\it Phys. Lett.} 157B (1985) 416;
P. Di Vecchia and M. Moshe, {\it Phys.~Lett.~}B300 (1993) 49;
H.~J.~Schnitzer, {\it Mod.~Phys.~Lett.~}A7 (1992) 2449.

\endbib

\bye